\documentclass[fleqn,usenatbib]{mnras}

\usepackage{newtxtext,newtxmath}

\usepackage[T1]{fontenc}
\usepackage{ae,aecompl}

\usepackage{subfig}
\usepackage{tikz}
\usepackage{standalone}

\usepackage{graphicx}	
\usepackage{amsmath}	
\usepackage{amssymb}	
\usepackage{booktabs}

\def\beq{\begin{eqnarray}}
\def\eeq{\end{eqnarray}}

\newcommand{\av}[1]{\langle{#1\rangle}} 
\let\vec\mathbf

\newcommand{\resub}[1]{#1}

\numberwithin{equation}{section}

\title[Small-Scale Power Spectrum \& Bispectrum Estimators]{Computing the Small-Scale Galaxy Power Spectrum and Bispectrum in Configuration-Space}

\author[O.\,H.\,E. Philcox \& D.\,J. Eisenstein]{
Oliver H.\,E. Philcox$^{1,2}$\thanks{E-mail: \href{mailto:ohep2@alumni.cam.ac.uk}{ohep2@alumni.cam.ac.uk}}
and Daniel J. Eisenstein$^{2}$
\\
$^{1}$Department of Astrophysical Sciences, Princeton University, Princeton, NJ 08544, USA\\
$^{2}$Center for Astrophysics | Harvard \& Smithsonian, 60 Garden St., MA 02138, USA\\
}

\date{Accepted 2019 November 26. Received 2019 October 20; in original form 2019 June 19}

\pubyear{2019}

\begin{document}
\label{firstpage}
\pagerange{\pageref{firstpage}--\pageref{lastpage}}
\maketitle

\begin{abstract}
We present a new class of estimators for computing small-scale power spectra and bispectra in configuration-space via weighted pair- and triple-counts, with no explicit use of Fourier transforms. Particle counts are truncated at $R_0\sim 100h^{-1}\,\mathrm{Mpc}$ via a continuous window function, which has negligible effect on the measured power spectrum multipoles at small scales. This gives a power spectrum algorithm with complexity $\mathcal{O}(NnR_0^3)$ (or $\mathcal{O}(Nn^2R_0^6)$ for the bispectrum), measuring $N$ galaxies with number density $n$. Our estimators are corrected for the survey geometry and have neither self-count contributions nor discretization artifacts, making them ideal for high-$k$ analysis. Unlike conventional Fourier transform based approaches, our algorithm becomes more efficient on small scales (since a smaller $R_0$ may be used), thus we may efficiently estimate spectra across $k$-space by coupling this method with standard techniques. We demonstrate the utility of the publicly available power spectrum algorithm by applying it to BOSS DR12 simulations to compute the high-$k$ power spectrum and its covariance. In addition, we derive a theoretical rescaled-Gaussian covariance matrix, which incorporates the survey geometry and is found to be in good agreement with that from mocks. Computing configuration- and Fourier-space statistics in the same manner allows us to consider joint analyses, which can place stronger bounds on cosmological parameters; to this end we also discuss the cross-covariance between the two-point correlation function and the small-scale power spectrum.
\end{abstract}




\section{Introduction}\label{sec: intro}
Along with the two-point correlation function (2PCF), the galaxy power spectrum $P(\vec k)$ is the most commonly used tool in the analysis and interpretation of large cosmological surveys. Simply from considering the angle-averaged power spectrum monopole, we can constrain a variety of effects, most notably Baryon Acoustic Oscillations (BAO) on scales close to $100h^{-1}\mathrm{Mpc}$ \citep[e.g.][]{2005ApJ...633..560E,2014MNRAS.441...24A,2016MNRAS.460.4188G,2016MNRAS.460.4210G,2017MNRAS.464.3409B,2017MNRAS.470.2617A}. This encodes a wealth of cosmological information regarding the Universe's expansion history and composition, allowing high precision measurements of the Hubble expansion parameter and the relation between angular diameter distance and redshift.

Additional constraints are obtained by considering the dependence of the power spectrum on $\mu$, the cosine of the angle between the galaxy separation vector and the line-of-sight (LoS). The anisotropic power spectrum $P(k,\mu)$ has been measured by various surveys \citep[e.g.][]{2006PhRvD..74l3507T,2008PThPh.120..609Y,2011MNRAS.415.2876B,2014MNRAS.444.1400N,2015MNRAS.453L..11B,2015PhRvD..92h3532S,2017JCAP...07..002H} and allows us to probe a variety of phenomena including redshift space distortions (RSD; \citealt{1987MNRAS.227....1K}) and the Alcock-Paczynski (AP; \citealt{1979Natur.281..358A}) effect. The former arises from peculiar velocities biasing the conversion between redshift and comoving distance, whilst the latter is due to an incorrect assumed fiducial cosmology giving anisotropy artifacts in the clustering dataset. Both effects are useful in cosmological analyses, e.g. by placing constraints on the growth of structure through the $\sigma_8$ parameter \citep[e.g.][]{2017MNRAS.466.2242B}, and allowing tests of General Relativity on the largest scales \citep[e.g.][]{2008Natur.451..541G}. In anisotropic analyses, we must be aware of wide-angle effects, which arise if the varying LoS between pairs is not fully accounted for \citep{1998ApJ...498L...1S,2004ApJ...614...51S}, although these tend to be subdominant on small scales \citep{2012MNRAS.420.2102S,2015MNRAS.447.1789Y}. The varying LoS makes Fourier methods more difficult to apply; this is ameliorated using LoS approximations such as in \citet{2006PASJ...58...93Y} and \citet{2011MNRAS.415.2876B}, which are included in most modern analyses.  

A further tranche of cosmological information is provided by the bispectrum, particularly with regards to the Universe's non-Gaussianity \citep[e.g.][]{1999ApJ...517..531S,2006PhRvD..74b3522S}. As our library of cosmological survey data grows greater, the bispectrum will become increasingly important, giving a complementary probe of many parameters. Whilst a number of analyses have included this information \citep[e.g.][]{2001PhRvL..86.1434F,2001ApJ...546..652S,2002MNRAS.335..432V,2015MNRAS.451..539G,2015MNRAS.452.1914G}, it is still seldom used compared to the power spectrum, though \citet{2017MNRAS.465.1757G} and \citet{2018MNRAS.478.4500P} provide exciting examples of its utility. In the first paper, the bispectrum is combined with the power spectrum giving improved constraints on RSD and AP parameters, whilst the second provides the first measurement of bispectrum BAO features at high significance. In general, however, the bispectrum is far less understood than the power spectrum and its computation is significantly more expensive.

Conventional algorithms (e.g. \citealt{1994ApJ...426...23F,2006PASJ...58...93Y,2012PhRvD..86f3511F,2013PhRvD..88f3512S,2015PhRvD..92h3532S,2015MNRAS.453L..11B,2019MNRAS.484..364S}) measure the power spectrum and bispectrum through Fast Fourier Transforms (FFTs), transforming the measured density fields of galaxy and random-position catalogs to accurately constrain large-scale clustering (corresponding to small wavevector $k$). An integral part of this approach is the discretization of galaxies onto a regular grid in configuration-space; the bias produced by this approximation can be reduced by using a finer grid. Measuring spectra on particularly small scales (large $k$) is difficult however, since the grid density must scale as $k^{-1}$, requiring significant computational power. In addition, Fourier-transform based methods suffer on small-scales from \resub{the inclusion of self-counts}, leading to \resub{high-$k$ spectra approximately consistent with white-noise}. For the power spectrum, this strictly affects only the monopole ($\ell = 0$) and is usually subtracted off in the Poissonian limit, though recent analyses have found some evidence for more complex behavior \citep{2016MNRAS.460.4188G}, \resub{due to non-Poissonian sampling of the underlying density field and discretization effects}. Many future galaxy surveys will focus on creating catalogs with far higher number density that previously used; this will pose new challenges for cosmology since we will require efficient methods to compute small-scale power spectra and bispectra, and accurate models to compare them to. The latter requirement is particularly non-trivial; at small $k$,we move deeply into the non-linear, and non-Gaussian, regime, where perturbation theory becomes less useful thus we must look for alternative methods, e.g. using simulations. Our primary goal in this work is to propose a method to assist with the first problem.

The principal idea of this paper is that the power spectrum $P(\vec k)$ (defined as the Fourier transform of the 2PCF) can be written as a configuration-space pair-count across the survey, weighted by $e^{i\vec k\cdot\vec r}$ for pair separation vector $\vec r$. This is fully analogous to standard 2PCF estimators, and our method can be seen as an extension of treatments that Fourier transform the binned correlation function \citep{2001MNRAS.325.1389J,2016ApJ...833..287L}, though avoiding errors from configuration-space binning and introducing a controlled apodization scheme. Averaging over angle, the multipole powers $P_\ell(k)$ can be similarly written as pair-counts, instead weighting by $j_\ell(kr)L_\ell(\vec r\cdot\vec x)$ for pair mid-point $\vec x$, spherical Bessel function $j_\ell$ and Legendre polynomial $L_\ell$. To measure the true power, we must consider all particle pairs separated by arbitrarily large distances, which is computationally expensive. Instead, we propose to truncate the pair-count at some $R_0\sim 100h^{-1}\mathrm{Mpc}$ via a smooth window $W(r;R_0)$. This allows us to quickly measure the convolved power $\left[P\ast \widetilde{W}\right](\vec k)$, which is shown to be highly consistent with the true power on small-scales ($\gtrsim 0.2h\,\mathrm{Mpc}^{-1}$). As noted in \citet{2016ApJ...833..287L}, by basing our estimator on the \citet{1993ApJ...412...64L} 2PCF formalism, it is naturally corrected for the survey geometry, avoiding the need for later deconvolution, and has no contribution from \resub{self-counts}. The estimator is highly efficient at small $k$ (where $R_0$ can be small) and may be paired with traditional FFT-based techniques to measure the power spectrum on all scales. Furthermore, it is highly applicable to N-body simulations, due to the avoidance of large, memory-intensive meshes and wide availability of fast pair-counting routines. The algorithm may be simply extended to the bispectrum, which instead involves weighted triple-counts, again truncated at $R_0$. Computing spectra and correlation functions on the same footing naturally allows joint analyses, and a strong motivation for this work is that it enables us to robustly compute \textit{cross-covariance} matrices between configuration- and Fourier-space statistics, following techniques developed for the correlation functions \citep{2016MNRAS.462.2681O,2019MNRAS.487.2701O,rascalC,3pcfCov}. This will be of paramount importance in future analyses, where we must use multiple statistics in concert to obtain strongest constraints on cosmological parameters.

We apply the method to mock galaxy catalogs appropriate to the Baryon Oscillation Spectroscopic Survey \citep[BOSS; ][]{2013AJ....145...10D,2015ApJS..219...12A,2017MNRAS.470.2617A} Data Release 12 \citep[DR12; ][]{2015ApJS..219...12A} dataset of Sloan Digital Sky Survey III \citep[SDSS-III; ][]{2011AJ....142...72E} to produce power spectrum estimates up to $k = 1h^{-1}\mathrm{Mpc}$ which are reasonably well fit by simple non-linear RSD models. We additionally compute theoretical covariance matrices following a similar prescription to \citet{2016MNRAS.462.2681O}, with non-Gaussianity incorporated via a simple shot-noise rescaling. These are shown to fit well with the high-$k$ monopole and quadrupole covariances and may be used to produce fast estimates of the covariances without requiring a large suite of mocks, as well as to compute cross-covariances between the 2PCF and the power spectrum. \resub{The code used in this paper has been made publicly available with extensive documentation.\footnote{HIPSTER: HIgh-k Power SpecTrum EstimatoR (\href{https://HIPSTER.readthedocs.io}{HIPSTER.readthedocs.io})}}

The outline of this paper is as follows. We begin by introducing our configuration-space power spectrum estimators for both isotropic and anisotropic power in Sec.\,\ref{sec: ideal-power-estimators}, before we consider the effect of pair-separation window functions and the survey geometry in Secs.\,\ref{sec: windows}\,\&\,\ref{sec: survey-correction-integrals}. Theoretical covariance matrices are discussed in Sec.\,\ref{sec: power-cov}, before we compute both the power and its covariance from mock galaxy catalogs in Sec.\,\ref{sec: data}. We end with a discussion of generalization to the bispectrum in Sec.\,\ref{sec: bispectrum-estimators} and conclude in Sec.\,\ref{sec: conclusion}. Appendices \ref{appen: aniso-power-deriv} to \ref{appen: bispectrum-deriv} contain mathematical derivations and useful results linking the 2PCF and power spectrum multipoles.





\section{Idealized Power Spectrum Estimators}\label{sec: ideal-power-estimators}
We begin by presenting estimators for the power spectrum in the form of weighted pair-counts over all galaxies in a survey of arbitrary geometry. Sec.\,\ref{subsec: power-unbinned} introduces our approach in continuous form before we consider isotropic and Legendre binning in Secs.\,\ref{subsec: power-iso} and \ref{subsec: power-aniso} respectively. The \resub{exclusion of self-counts (and hence shot-noise)} in our estimators is discussed in Sec.\,\ref{subsec: shot-noise}. 

\subsection{Unbinned Power Integrals}\label{subsec: power-unbinned}
The first step in our analysis is to define the power spectrum, $P(\vec k)$, as the Fourier transform of the (unbinned) redshift-space anisotropic two-point correlation function (2PCF) $\xi(\vec r)$. This differs from standard approaches, which define the power as the average of the Fourier-transformed overdensity field \citep[e.g.][]{1994ApJ...426...23F,2015MNRAS.453L..11B}, and is useful since we can account for non-uniform survey geometries and window-function effects in the estimator directly, obviating the need for deconvolution, or the comparison of the power spectrum with window-convolved models \citep[e.g.][]{2006PASJ...58...93Y,2017MNRAS.464.3409B,2017MNRAS.464.3121W}. In addition, we do not need to grid the particles, making this approach ideal for high-$k$ power estimation. Following the FKP \citep{1994ApJ...426...23F} pair-counting estimator, we define
\beq\label{eq: power_as_2PCF_FT}
P(\vec k) = \mathcal{F}\left[\xi(\vec r)\right](\vec k) = \mathcal{F}\left[\frac{NN(\vec r)}{RR(\vec r)}\right](\vec k) \equiv \int d^3\vec r\,\frac{NN(\vec r)}{RR(\vec r)}e^{i\vec k\cdot\vec r}
\eeq
with $N = D-R$ for data and random counts $D$ and $R$ respectively.\footnote{\resub{Throughout this work we assume the random counts to be reweighted such that the mean weighted number densities of data and random points agree. The random counts are thus invariant to the exact number of randoms used.}} The 2PCF is simply the ratio of data-minus-random and random pair-counts, here taken as a continuous (unbinned) functions of position $\vec r$. These are defined via
\beq
    NN(\vec r) &=& \int d^3\vec r_i\, d^3\vec r_j\,n(\vec r_i)n(\vec r_j)w(\vec r_i)w(\vec r_j)\delta(\vec r_i)\delta(\vec r_j)\delta^{(D)}(\vec r-[\vec r_i-\vec r_j])\\\nonumber
    RR(\vec r) &=& \int d^3\vec r_i\, d^3\vec r_j\,n(\vec r_i)n(\vec r_j)w(\vec r_i)w(\vec r_j)\delta^{(D)}(\vec r-[\vec r_i-\vec r_j])
\eeq
where $n$, $w$ and $\delta$ are the (continuous) survey number density, weights and overdensities, and $\delta^{(D)}$ is a Dirac delta-function.\footnote{We do not explicitly include the dependence on the LoS (given by $\tfrac{1}{2}(\vec r_i+\vec r_j)$); our expressions are thus implicitly integrated over this.} These reduce to the standard pair-count forms \citep[e.g.][]{1993ApJ...412...64L,2016MNRAS.462.2681O} when integrated over some $\vec r$-bin, which replaces the Dirac deltas with binning functions $\Theta^a(\vec r)$ which are unity for $\vec r$ in bin $a$ and zero else. For an ideal uniform unbounded survey of volume $V$ (where the weight $w$ and uncorrelated number density $n$ are independent of position), the $RR$ counts in bin $a$ with center $\vec r_a$ and width $\delta\vec r_a$ are simply
\beq
    RR^a_\mathrm{ideal} &\approx& RR(\vec r_a)\delta\vec r \approx \int d^3\vec r_i\,d^3\vec r_j\,n(\vec r_i)n(\vec r_j)w(\vec r_i)w(\vec r_j)\delta(\vec r_i)\delta(\vec r_j)\Theta^a(\vec r_i-\vec r_j)\\\nonumber
    &=& V(nw)^2 \int d^3\vec r_{ij}\, \Theta^a(\vec r_{ij}) = V(nw)^2\delta \vec r_a
\eeq
for $\vec r_{ij} = \vec r_i-\vec r_j$, noting that this becomes exact in the limit of thin-bins. Following \citet{2007MNRAS.376.1702P}, \citet{2009MNRAS.393..297P}, \citet{2010ApJ...718.1224X} and \citet{3pcfCov}, we introduce a survey-correction function $\Phi$, \resub{which allows us to incorporate the geometric effects arising from non-uniformities and survey boundaries by relating the ideal and true pair-counts, via the definition}
\beq\label{eq: true-RR-definition}
    RR(\vec r)\equiv \frac{V\overline{(nw)^2}}{\Phi(\vec r)}
\eeq
where an overbar indicates averaging with respect to the survey volume. In $r,\mu$-coordinates\footnote{Note that we restrict to $\mu\in [0,1)$ identifying $-\mu$ with $\mu$, giving an extra factor of 2.} (where $\arccos{\mu}$ is the angle between the pair separation vector and the local LoS), for thin bins centered at $r_a,\mu_b$ with $r\in[r_{a,\mathrm{min}},r_{a,\mathrm{max}}]$, we obtain volume $\delta \vec r = 4\pi/3\left(r_{a,\mathrm{max}}^3-r_{a,\mathrm{min}}^3\right)\delta \mu_b$. An estimator for $\Phi$ is \resub{given by the ratio of pair counts in a given bin};
\beq\label{eq: phi_r_mu_defn}
    \hat\Phi(r_a,\mu_b) = \frac{4\pi V\overline{(nw)^2}\left(r_{a,\mathrm{max}}^3-r_{a,\mathrm{min}}^3\right)\delta\mu_b}{3RR^a_b} 
\eeq
where $RR^a_b$ is the true $RR$ pair-count in this bin computed via exhaustive pair-counting e.g. using \texttt{corrfunc}\footnote{\href{https://corrfunc.readthedocs.io}{corrfunc.readthedocs.io}} \citep{2017ascl.soft03003S}. The function $\Phi$ is expected to be smooth and close to unity, with deviations arising from non-uniformities in the survey and the finite domain. In practice, Eq.\,\ref{eq: phi_r_mu_defn} is used to compute values of $\Phi$ for each bin which are then fit to smooth functions to define $\Phi(\vec r)=\Phi(r,\mu)$ and hence a functional form for $RR(\vec r)$. Since $\Phi$ is expected to be smooth we do \textit{not} require excessively large random catalogs to compute this. The impact of the survey-correction function on the measured power spectrum is discussed in Sec.\,\ref{sec: survey-correction-integrals}.

Following these definitions, $P(\vec k$) becomes
\beq\label{eq: full-power-estimator}
    P(\vec k) &=& \int d^3\vec r\, \frac{NN(\vec r)\Phi(\vec r)}{V\overline{(nw)^2}}e^{i\vec k\cdot \vec r} = \frac{1}{V\overline{(nw)^2}}\int d^3\vec r\,d^3\vec r_i\,d^3\vec r_j\,n(\vec r_i)n(\vec r_j)w(\vec r_i)w(\vec r_j)\delta(\vec r_i)\delta(\vec r_j)\delta^{(D)}(\vec r-[\vec r_i-\vec r_j])\Phi(\vec r)e^{i\vec k\cdot \vec r}\\\nonumber
    &=& \frac{1}{V\overline{(nw)^2}}\int d^3\vec r_i\,d^3\vec r_j\,n(\vec r_i)n(\vec r_j)w(\vec r_i)w(\vec r_j)\delta(\vec r_i)\delta(\vec r_j)\Phi(\vec r_i-\vec r_j)e^{i\vec k\cdot(\vec r_i-\vec r_j)}
\eeq
integrating over the Delta function in the final line. This can be compared to the standard FKP power spectrum estimator \citep{1994ApJ...426...23F};
\beq\label{eq: FKP_estimator}
    P_\mathrm{FKP}(\vec k) &=& \frac{1}{I}\int d^3\vec r_i\,d^3\vec r_j\,n(\vec r_i)n(\vec r_j)w(\vec r_i)w(\vec r_j)\delta(\vec r_i)\delta(\vec r_i)e^{i\vec k\cdot(\vec r_i-\vec r_j)}-P_\mathrm{shot}(\vec k)\\\nonumber
    I &=& \int d^3\vec r\,n^2(\vec r)w^2(\vec r) \equiv V\overline{(nw)^2}
\eeq
in our notation.\footnote{Our $n(\vec r)\delta(\vec r)$ function may be identified with $n_g(\vec r)-\alpha n_s(\vec r)$ in \citet{1994ApJ...426...23F}, for $\alpha=1$. We also assume the continuous limit, such their $\bar n(\vec r)$ is equal to our $n(\vec r)$.} Note that, unlike the FKP integral, our $P(\vec k)$ does \textit{not} include a \resub{Poissonian} shot-noise term, as discussed in Sec.\,\ref{subsec: shot-noise}. \resub{In addition, it does not contain the correction function $\Phi$, with normalization provided simply by the idealized pair counts $V\overline{(nw)^2}$; this means that the FKP estimator is not corrected for the survey geometry, requiring deconvolution in post-processing (or convolution of power spectrum models with the Fourier transform of the window function).} In the ideal periodic survey limit, the survey correction factor $\Phi$ is unity everywhere thus the two estimators agree (up to the shot-noise term). By introducing $\Phi$ into our estimator, we are able to account for window-function effects directly in our power spectrum estimator. \resub{This is discussed in detail in Sec.\,\ref{sec: survey-correction-integrals}.}

One effect that we do not include in our $P(\vec k)$ integral is that of the `integral constraint' \citep[e.g.][]{1991MNRAS.253..307P,2014MNRAS.443.1065B}. In typical analyses, we assume that the average density of our survey matches that of the Universe, i.e. that $P(\vec 0) = 0$. Due to the existence of modes larger than the survey which modulate the local average density, this biases our power spectrum estimate, giving an underestimate of the large-scale power (for modes where $2\pi/k$ is comparable to the survey size $L$). Since we are only interested in $k\gg 1/L$ in this analysis, the effect may be safely neglected.

\subsection{Isotropic Estimator}\label{subsec: power-iso}
The isotropic power in a $|\vec k|$-bin $a$ is given as an angular integral over Eq.\,\ref{eq: full-power-estimator};
\beq\label{eq: isotropic_estimator}
    P^a &=& \frac{1}{V_\mathrm{shell}}\int d^3\vec k\,\Theta^a(|\vec k|)P(\vec k) = \frac{1}{V\overline{(nw)^2}V_\mathrm{shell}}\int d^3\vec k\,\Theta^a(|\vec k|)\int d^3\vec r_i\,d^3\vec r_j\,n(\vec r_i)n(\vec r_j)w(\vec r_i)w(\vec r_j)\delta(\vec r_i)\delta(\vec r_j)\Phi(\vec r_i-\vec r_j)e^{i\vec k\cdot(\vec r_i-\vec r_j)}
\eeq
where $V_\mathrm{shell}$ is the volume of the shell defined by the binning function $\Theta^a(|\vec k|)$. 
This is similar to the 2PCF integrals introduced in \citet{2016MNRAS.462.2681O}, except that we bin in $\vec k$-space rather than $\vec r$-space with an $e^{i\vec k\cdot(\vec r_i-\vec r_j)}$ kernel. The $\vec k$-dependence is simplified by defining the kernel
\beq
    A^a(\vec r_i-\vec r_j) &\equiv& \frac{1}{V_\mathrm{shell}}\int d^3\vec k\,\Theta^a(|\vec k|)e^{i\vec k\cdot(\vec r_i-\vec r_j)}\\\nonumber
    &=& \frac{1}{V_\mathrm{shell}}\int k^2\,dk\,d\phi_k\,d\mu_k\, \Theta^a(k)e^{ik|\vec r_i-\vec r_j|\measuredangle[\vec k\,,\vec r_i-\vec r_j]}
\eeq
\resub{where the polar and azimuthal angles of $\vec k$ are denoted by $\arccos{\mu_k}$ and $\phi_k$ respectively.} For simplicity, we define $\measuredangle[\vec x\,,\vec y]$ as the cosine of the angle between vectors $\vec x$ and $\vec y$, i.e. $\hat{\vec x}\cdot\hat{\vec y}$, denoting unit vectors with hats. Assuming the binning function $\Theta^a(k)$ to have unit support for $k\in[k_{a,\mathrm{min}},k_{a,\mathrm{max}}]$ we obtain
\beq\label{eq: iso-power-pair-count-kernel}
    A^a(\vec r_i-\vec r_j) &=& \frac{2\pi}{V_\mathrm{shell}}\int k^2dk\,d\mu_k\,\Theta^a(k)e^{ik|\vec r_i-\vec r_j|\resub{\mu_k}}
    = \frac{3}{k_{a,\mathrm{max}}^3-k_{a,\mathrm{min}}^3} \int_{k_{a,\mathrm{min}}}^{k_{a,\mathrm{max}}} k^2 dk\,j_0(k|\vec r_i-\vec r_j|)\\\nonumber
    &=& \frac{3}{k_{a,\mathrm{max}}^3-k_{a,\mathrm{min}}^3}\left[\frac{k^2j_1(k|\vec r_i-\vec r_j|)}{|\vec r_i-\vec r_j|}\right]_{k_{a,\mathrm{min}}}^{k_{a,\mathrm{max}}} \approx j_0(k_a|\vec r_i-\vec r_j|),
\eeq
where $j_n$ is the $n-$th order spherical Bessel function of the first kind, choosing the $\vec k$-space polar axis to be aligned along $\vec r_i-\vec r_j$ such that $\measuredangle[\vec k,\vec r_i-\vec r_j] = \mu_k$. In the thin-bin limit, $V_\mathrm{shell}\approx 4\pi k_a^2\Delta k$ and we may use the approximate $j_0$ solution. The kernel depends only on the pair separation and the chosen binning, with wider bins leading to a greater phase shift between the $j_1$ functions and hence smaller oscillations at large $k$. This gives the configuration-space isotropic power spectrum integral
\beq\label{eq: power_real_space_estimator_continuous}
    P^a = \frac{1}{V\overline{(nw)^2}}\int d^3\vec r_i d^3\vec r_j n(\vec r_i)n(\vec r_j)w(\vec r_i)w(\vec r_j)\delta(\vec r_i)\delta(\vec r_j)A^a(\vec r_i-\vec r_j) \Phi(\vec r_i-\vec r_j).
\eeq
Practically this may be estimated by pair-counting, using
\beq\label{eq: iso_power_pair_counting}
    \hat P^a = \frac{\widetilde{DD}^a - 2\,\widetilde{DR}^a+\widetilde{RR}^a}{V\overline{(nw)^2}}
\eeq
with the modified pair-counts defined via
\beq\label{eq: DD,DR,RR power estimator}
    \widetilde{XY}^a &=& \sum_{i\in X}\sum_{j\in Y,\,i\neq j\text{ if }X=Y}w_iw_jA^a_{ij}\Phi_{ij}
\eeq
for $X,Y\in\{D,R\}$, where $i$ and $j$ are drawn from fields $X$ and $Y$ respectively (excluding \resub{$i=j$} self-counts and with the random counts rescaled to ensure that the galaxy and random fields have the same mean number density). This corresponds to the discrete limit of Eq.\,\ref{eq: power_real_space_estimator_continuous}, with number densities being replaced by summations over delta functions centered at each particle position. Notably, there is no restriction on the $|\vec r_i-\vec r_j|$ separations appearing here, since all contribute to a given $k$-mode. In practice we are limited by (a) the survey size and (b) computation time, requiring the inclusion of a pair-separation window function, as discussed in Sec.\,\ref{sec: windows}. It is also interesting to note that this formalism does not explicitly require $\hat P^a$ to be positive, unlike the FKP estimator (which depends on $|\widetilde{N}(\vec k)|^2$). This is not found to be an issue in practice.

\subsection{Legendre Moments of the Anisotropic Power Spectrum}\label{subsec: power-aniso}
Analogous to \citet{2015MNRAS.453L..11B}, we can define the Legendre moments of the anisotropic power spectrum from our $P(\vec k)$ form (Eq.\,\ref{eq: full-power-estimator}), with the estimator in multipole $\ell$ being
\beq\label{eq: legendre_power_estimator_inital}
    P^a_\ell &=& \frac{2\ell+1}{V\overline{(nw)^2}V_\mathrm{shell}}\int d^3\vec r_i d^3\vec r_j d^3\vec k\,\Theta^a(|\vec k|) n(\vec r_i)n(\vec r_j)w(\vec r_i)w(\vec r_j)\delta(\vec r_i)\delta(\vec r_j)\Phi(\vec r_i-\vec r_j)\times e^{i\vec k\cdot(\vec r_i-\vec r_j)} L_\ell(\measuredangle[\vec k\,,\tfrac{1}{2}(\vec r_i+\vec r_j)])
\eeq
for Legendre polynomial $L_\ell$, evaluated at the angle between the $\vec k$-vector and the mid-point of $\vec r_i$ and $\vec r_j$. This is simply the integral of $P(\vec k)$ weighted by Legendre polynomials, switching the order of integration such that we integrate over $\vec k$ before $\vec r_i+\vec r_j$. Note that this choice of angle between $\vec k$ and the galaxy pair is not unique; various options exist which differ by $\mathcal{O}(\theta^2)$ for survey opening angle $\theta$ \citep{2015arXiv151004809S}. The impacts of this are negligible for $k\gtrsim0.1h\,\mathrm{Mpc}^{-1}$ \citep{2015MNRAS.452.3704S} and are hence not considered in this paper. Note that we again neglect any contribution \resub{from self-counts} (cf.\,Sec.\,\ref{subsec: shot-noise}).

As before, consider the kernel function
\beq\label{eq: aniso-power-kernel-init}
    A_\ell^a(\vec r_i\,,\vec r_j) \equiv \frac{2\ell+1}{V_\mathrm{shell}}\int d^3\vec k\,\Theta^a(|\vec k|)e^{i\vec k\cdot\vec u}L_\ell(\measuredangle[\vec k, \vec x])
\eeq
denoting the local LoS vector and separation vector by $\vec x = (\vec r_i+\vec r_j)/2$ and $\vec u = \vec r_i - \vec r_j$ respectively. In appendix \ref{appen: aniso-power-deriv}, this is shown to have the approximate form (for $\Delta k\ll k_a$)
\beq\label{eq: simple_aniso_kernel}
    A_\ell^a(\vec r_i\,,\vec r_j) \approx (-1)^{\ell/2} (2\ell+1) L_\ell(\vec x\cdot \vec u) j_\ell(k_au),
\eeq
with the full solution expressible either as a generalized hypergeometric function or in terms of spherical Bessel functions and the Sine integral, both of which are stated in appendix \ref{appen: aniso-power-deriv}. The latter form is adopted in this paper for computational efficiency. With this kernel, the anisotropic power spectrum integral becomes
\beq\label{eq: aniso_power_estimator_final}
    P_\ell^a = \frac{1}{V\overline{(nw)^2}}\int d^3\vec r_i\, d^3\vec r_j\, n(\vec r_i)n(\vec r_j)w(\vec r_i)w(\vec r_j)\delta(\vec r_i)\delta(\vec r_j)A^a_\ell(\vec r_i-\vec r_j)\Phi(\vec r_i-\vec r_j)
\eeq
analogous to the anisotropic 2PCF estimators in Legendre bins \citep{3pcfCov}, just with a different kernel function. As for the isotropic case (Eq.\,\ref{eq: iso_power_pair_counting}) this may be computed by pair-counting, replacing $A^a$ with $A^a_\ell$ in Eqs.\,\ref{eq: DD,DR,RR power estimator}. It is important to note that our estimators do \textit{not} require the \citet{2006PASJ...58...93Y} approximation, which approximates the LoS as $\vec r_i$ rather than $(\vec r_i+\vec r_j)/2$ (although they could be easily derived in this approximation). However, since all major LoS angle approximations (including that of \citealt{2006PASJ...58...93Y}) have an $\mathcal{O}(\theta^2)$ error, this is not a significant benefit.

\subsection{Self-Counts \& Shot-Noise}\label{subsec: shot-noise}
We here briefly discuss the omission of \resub{self-counts} in the above estimators, \resub{and their relation to shot-noise}. For a discrete \resub{Poisson-sampled} galaxy survey with galaxies (random positions) located at $\{\vec r^{(g)}_i\}$ ($\{\vec r^{(r)}_i\}$) for $i=1,2,...,N_g$ ($i=1,2,...,N_r$), we can write the number densities as a sum over Dirac delta functions
\beq
    n_g(\vec r) &=& \sum_{i=1}^{N_g}\delta^{(D)}(\vec r-\vec r^{(g)}_i), \qquad
    n_r(\vec r) = \frac{N_g}{N_r}\sum_{i=1}^{N_r}\delta^{(D)}(\vec r-\vec r^{(r)}_i)
\eeq
(ignoring weights and renormalizing the random number density to match the galaxies), which have discrete Fourier transforms
\beq
    \widetilde{n}_g(\vec k) = \frac{1}{N_g}\sum_{i=1}^{N_g} e^{i\vec k\cdot \vec r_i^{(g)}}, \qquad \widetilde{n}_r(\vec k) = \frac{N_g}{N_r^2}\sum_{i=1}^{N_r}e^{i\vec k\cdot\vec r_i^{(r)}}.
\eeq
Using conventional methods, the power spectrum estimator is essentially given by $|\widetilde{n}_g(\vec k)|^2 - |\widetilde{n}_r(\vec k)|^2$ with
\beq\label{eq: n-g-Fourier-sq}
    |\widetilde{n}_g(\vec k)|^2 = \frac{1}{N_g^2}\sum_{i=1}^{N_g}\sum_{j=1}^{N_g}e^{i\vec k\cdot (\vec r_i^{(g)}-\vec r_j^{(g)})} = \frac{1}{N_g^2}\sum_{i=1}^{N_g}\sum_{j\neq i}e^{i\vec k\cdot(\vec r_i^{(g)}-\vec r_j^{(g)})} + \frac{1}{N_g}
\eeq
(with an analogous expression for the random positions), separating out the $i=j$ self-count term and ignoring the effects of discretization. \resub{In this limit, the self-count term is simply the familiar $N_g^{-1}$ shot-noise.} Additionally, the self-counts may be thought of as the Fourier transform of the zero-separation 2PCF $\xi(\vec 0)$, which results in a white-noise monopole $\overline{nw^2}/\overline{(nw)^2}$ (using $\Phi(\vec 0)=1$) that reduces to $n^{-1}$ shot-noise for a uniform \resub{Poissonian} survey \citep[e.g.][]{1994ApJ...426...23F,2015MNRAS.453L..11B,2016MNRAS.460.4188G}.

\resub{In realistic contexts, galaxy sampling is not expected to be exactly Poissonian, since there exists an `exclusion radius' around each central galaxy, below which we cannot sample another galaxy. This results in the self-count term being more complex in form, requiring consideration of non-Poissonian `shot-noise' in Fourier-transform based approaches \citep[e.g.][]{2016MNRAS.460.4188G}.}\footnote{\resub{We should additionally include the effects of discretization, which may be partially ameliorated by better particle assignment schemes, e.g.\,\citet{2008ApJ...687..738C}.}} In our approach, we estimate Eq.\,\ref{eq: n-g-Fourier-sq} by counting pairs directly in configuration-space, but do not include any self-counts, i.e. we assert $i\neq j$ in the modified pair-counts (Eq.\,\ref{eq: DD,DR,RR power estimator}) and only include pairs with $|\vec r_i-\vec r_j|>0$. \resub{By avoiding self-counts, we ensure that our method correctly computes the power spectrum as the Fourier transform of the correlation function (with no need to consider shot-noise, discretization effects or sampling).}

\section{Pair-Separation Window Functions in the Power Spectrum}\label{sec: windows}
In practice, the above expressions for the power spectrum are difficult to compute, since the pair-count integrals (Eqs.\,\ref{eq: power_real_space_estimator_continuous}\,\&\,\ref{eq: aniso_power_estimator_final}) have non-trivial contributions from pairs separated by large distances, requiring every pair in the survey to be counted. To ameliorate this, we introduce a pair-separation window function $W(\vec r_i-\vec r_j;R_0)$, considering only pairs up to some truncation radius $R_0$. This greatly expedites computation, yet has the effect of convolving $P(\vec k)$ with the window Fourier transform $\widetilde{W}$ as discussed in Sec.\,\ref{subsec: window_convolution}. In Sec.\,\ref{subsec: pair-separation-forecast}, this is shown to have minimal effect on the small-scale power spectrum.

\subsection{Choice of Window Function}
\begin{figure}%
    \centering
    \subfloat[Window Function, $W(|\vec r|; R_0)$]{{\includegraphics[width=0.5\textwidth]{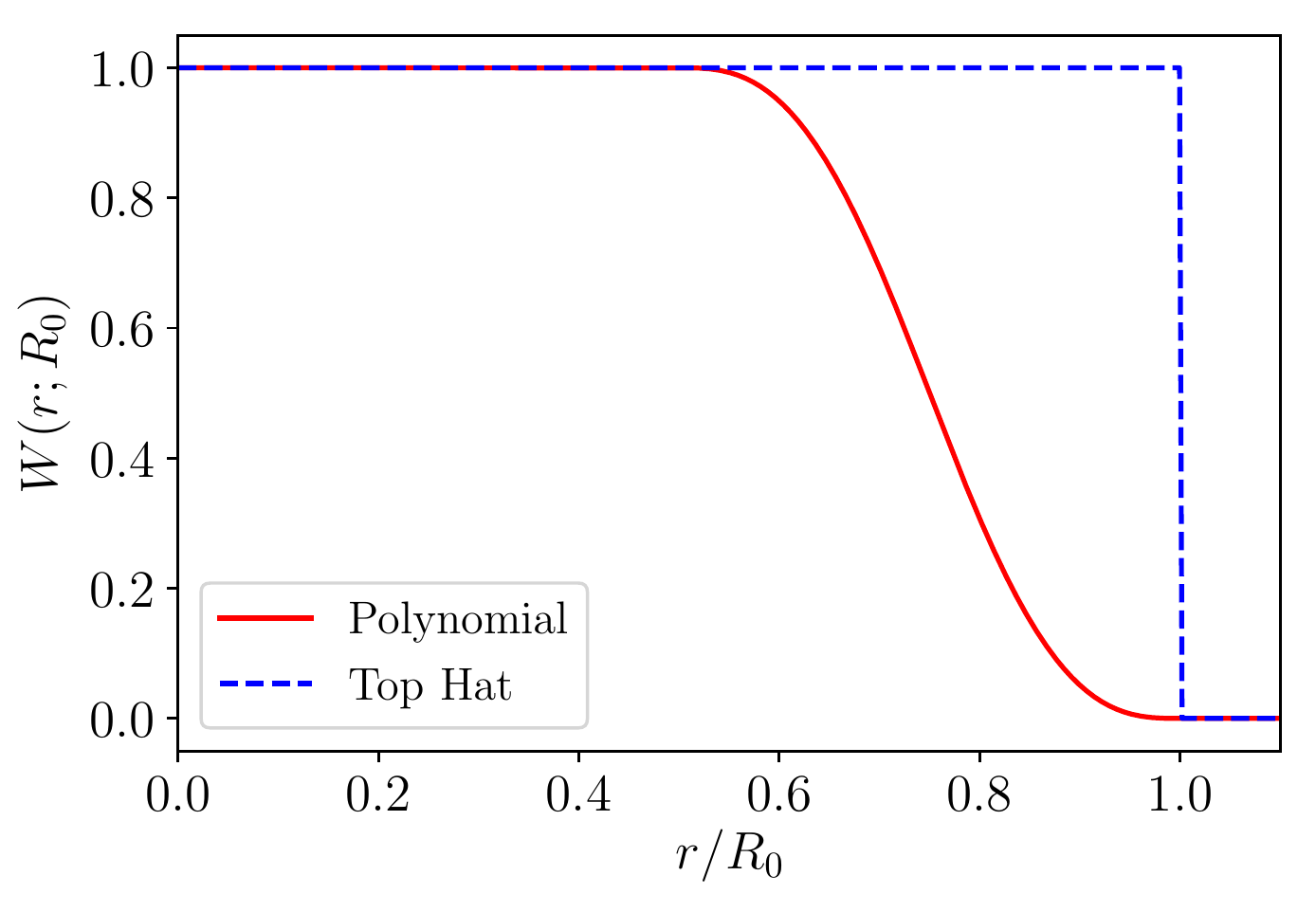} }}%
    \subfloat[Fourier Transformed Window $\widetilde{W}(|\vec k|; R_0)$]{{\includegraphics[width=0.5\textwidth]{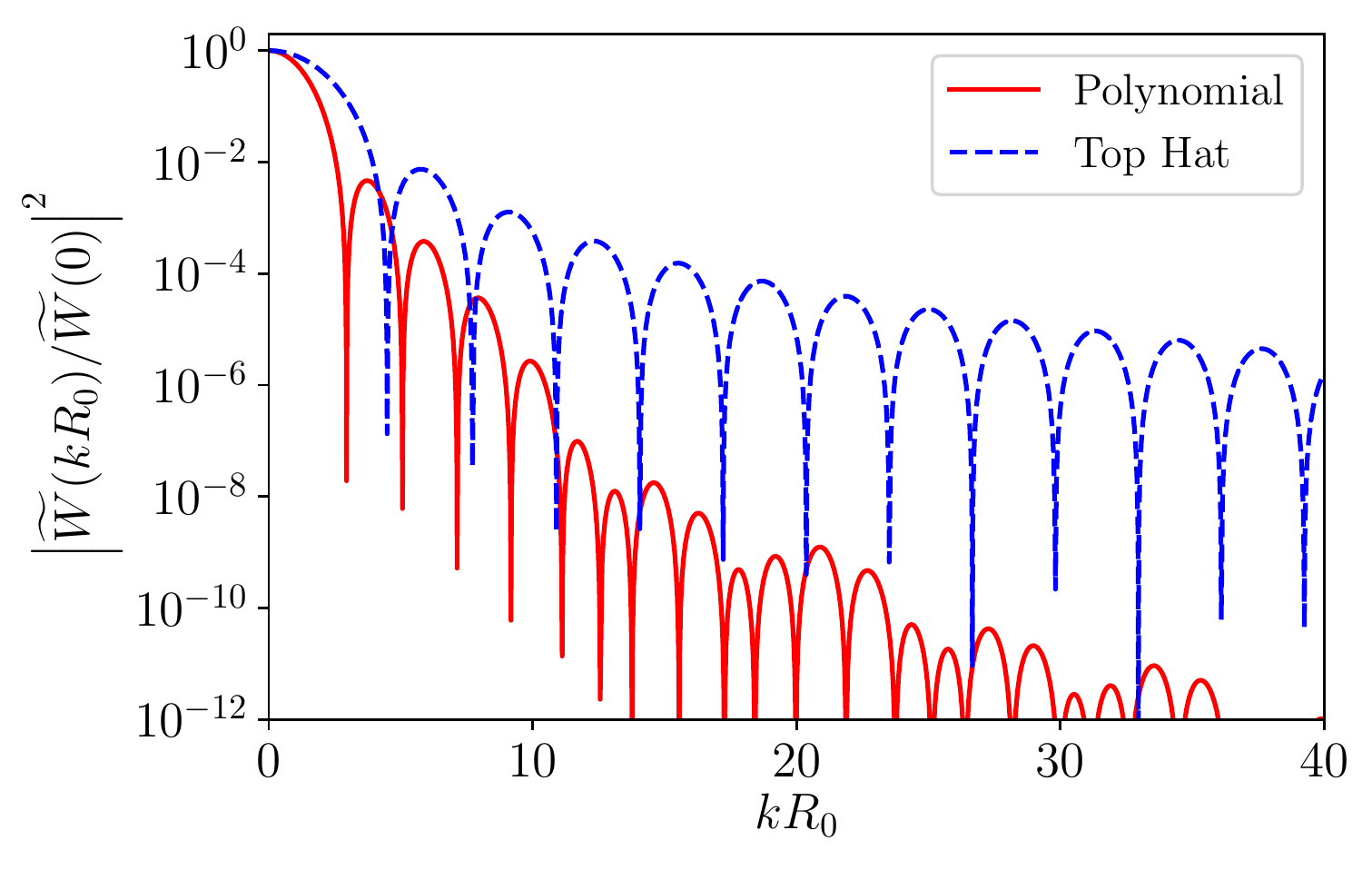}} }%
    \caption{Real and Fourier-space plots of the (radial) pair-separation window function $W_{R_0}$. Here, $R_0$ is the truncation scale of the window (typically $R_0\sim 100h^{-1}\mathrm{Mpc}$), and we plot both the utilized polynomial window (Eq.\,\ref{eq: window_defn}) and a spherical top-hat function. The Fourier space window functions are computed from Hankel transforms of $W(\vec r; R_0)$ and are here displayed in power-space, normalized by the $|\vec k|=0$ value to aid comparison. We note that the polynomial power spectrum exhibits much stronger fall-off with $|\vec k|$ than the top-hat window, due to its smoothness up to the second radial derivative.}
    \label{fig: window_plots}
\end{figure}

The simplest form for $W$ is that of a spherical top-hat filter, yet this gives large unwanted oscillations in Fourier space, significantly affecting the measured power. We here adopt a piecewise-continuous spherical window function
\beq\label{eq: window_defn}
    W(\vec r; R_0) &\equiv& f\left(\frac{|\vec r|}{R_0}\right)\\\nonumber
    f(x) &=& \begin{cases} 1 & \text{if } 0\leq x < 1/2 \\ 1-8\left(2x - 1 \right)^3 + 8 \left(2x - 1\right)^4& \text{if }1/2\leq x<3/4\\ -64\left(x-1\right)^3 - 128\left(x-1\right)^4 & \text{if } 3/4\leq x< 1\\ 0 & \text {else,}\end{cases}
\eeq
with coefficients chosen to ensure that the function, along with its first and second derivatives, are continuous at the break-points (necessary to avoid broad wings in the Fourier spectrum). The real- and power-space parts of $W(\vec r; R_0)$ are plotted in Fig.\,\ref{fig: window_plots} alongside a top-hat window for comparison. By construction, our piecewise-continuous polynomial function has a smooth fall-off with respect to the radial coordinate, but is large for most $r=|\vec r|<R_0$. The power spectrum of the polynomial window is significantly narrower than that of the top-hat function, with stronger decay with respect to $k$ resulting from the greater smoothness in real-space. This leads to reduced bias in the small-scale galaxy power spectrum measurement. In addition, this apodizes the survey-correction factor $\Phi(r,\mu)$, both increasing its smoothness and giving zero weight to separations at large $r$ where $\Phi$ departs strongly from unity.

Using this window function, the anisotropic power spectrum integral becomes
\beq\label{eq: windowed_aniso_power_estimator}
    P^a_\ell(R_0) &=& \frac{1}{V\overline{(nw)^2}}\int d^3\vec r_i\,d^3\vec r_j\,n(\vec r_i)n(\vec r_j)w(\vec r_i)w(\vec r_j)\delta(\vec r_i)\delta(\vec r_j)A^a_\ell(\vec r_i-\vec r_j)\resub{\Phi(\vec r_i-\vec r_j)}W(\vec r_i-\vec r_j; R_0)
\eeq
with $\ell=0$ giving the isotropic estimator. This converges to the true power, $P_\ell^a$, in the limit $R_0\rightarrow\infty$. This may again be computed using pair-counting, where the pair-counts are now defined as
\beq\label{eq: windowed_pair_counts}
    \widetilde{XY}^a_{R_0} = \sum_{i\in X}\sum_{j\in Y,i\neq j\text{ if }X=Y}w_iw_jA^a_{ij}\Phi_{ij}W_{ij}^{R_0}
\eeq
for $X,Y\in[D,R]$, drawing $i$ from $X$ and $j$ from $Y$. Here $A^a_{ij} \equiv A^a(\vec r_i-\vec r_j)$ and $W^{R_0}_{ij}\equiv W(\vec r_i-\vec r_j; R_0)$. 

\subsection{Power Spectrum Modification}\label{subsec: window_convolution}
The windowed 3D power spectrum, $P(\vec k; R_0)$, may be written using Eq.\,\ref{eq: full-power-estimator} as
\beq\label{eq: iso_power_convolution}
    P(\vec k;R_0) &=& \frac{1}{V\overline{(nw)^2}}\int d^3\vec r_i\,d^3\vec r_j\, n(\vec r_i)n(\vec r_j)w(\vec r_i)w(\vec r_j)\delta(\vec r_i)\delta(\vec r_j)\,e^{i\vec k\cdot (\vec r_i-\vec r_j)}\Phi(\vec r_i-\vec r_j)W(\vec r_i-\vec r_j;R_0)\\\nonumber
    &=& \frac{1}{V\overline{(nw)^2}}\int d^3\vec r_i\,d^3\vec r_j\,n(\vec r_i)n(\vec r_j)w(\vec r_i)w(\vec r_j)\delta(\vec r_i)\delta(\vec r_j)\Phi(\vec r_i-\vec r_j)\int \frac{d^3\vec p}{(2\pi)^3} e^{i(\vec k-\vec p)\cdot (\vec r_i-\vec r_j)}\widetilde{W}(\vec p; R_0)\\\nonumber
    &=& \int \frac{d^3\vec p}{(2\pi)^3}\, P(\vec k-\vec p) \widetilde{W}(\vec p; R_0) = \frac{1}{(2\pi)^3}\left[P \ast \widetilde W_{R_0}\right](\vec k) = \mathcal{F}\left[\xi(\vec r)W(\vec r; R_0)\right](\vec k).
\eeq
Here we have expressed $W(\vec r_i-\vec r_j; R_0)$ in terms of its Fourier transform, noting that this gives a simple convolution of the true power, $P(\vec k)$, with the Fourier-transformed window function $\widetilde W_{R_0}(\vec k)$. For the isotropic case, integrating over $\Omega_k$ gives 
\beq\label{eq: binned_convolved_iso_power}
    P(k;R_0) &=& \frac{1}{(2\pi)^3}\int \frac{d\Omega_k}{4\pi}\left[P\ast \widetilde{W}_{R_0}\right](\vec k) 
\eeq
which may be written in terms of the \textit{isotropic} 2PCF $\xi(r)$ and the (radial) $W(\vec r; R_0)$ function as
\beq\label{eq: iso_power_window_mod}
    P(k; R_0) &=& \int \frac{d\Omega_k}{4\pi}\int d^3\vec r\, e^{i\vec k\cdot\vec r}\xi(\vec r)W(\vec r; R_0) = \frac{1}{2}\int d\mu_k \int d^3\vec r\,e^{ikr\mu_k}\xi(\vec r)W(\vec r; R_0)\\\nonumber
    &=& \int_0^\infty r^2dr\,\int_{-1}^1 d\mu_r\,\int_0^{2\pi}d\phi_r\,j_0(kr)\xi(r,\mu_r)W(r; R_0) = 4\pi\int_0^\infty r^2dr\,\xi(r)W(r;R_0)j_0(kr) 
\eeq
(before radial binning), where $j_0$ is a zeroth-order spherical Bessel function of the first kind. Here we have aligned $\vec k$ with $\vec r$ in the first line (such that $\vec k\cdot\vec r = kr\mu_k$) and noted that $\int_{-1}^1d\mu_r \xi(r,\mu_r) = \xi(r)$. (This is also derived as the $\ell=0$ case of the result of appendix \ref{appen: power-to-2pcf-multipoles} for $A(\vec r)=\xi(\vec r)W(\vec r; R_0)$, $\widetilde{A}(\vec k) = P(\vec k; R_0)$.) This simple form will be used below to assess the effects of the polynomial window on the measured power spectrum.

For the anisotropic power spectrum multipoles, we adopt a similar approach, computing the angular integrals of $P(\vec k)$ weighted by Legendre polynomials. Notably, these depend on the LoS of a particular pair of points, $\vec x = \tfrac{1}{2}(\vec r_i+\vec r_j)$, thus we must integrate over $\Omega_k$ \textit{before} averaging over the LoS of galaxy pairs. This is made possible by noting that the quantities $\xi(\vec r)W(\vec r; R_0)$ appearing in Eq.\,\ref{eq: iso_power_convolution} are strictly the integrals of $\xi(\vec r; \vec x)W(\vec r; R_0)$ over $\vec x$, where $\xi(\vec r; \vec x)$ is the local 2PCF measurement for galaxies at mid-point $\vec x$. Using the linearity of Fourier transforms to switch the order of integration we obtain
\beq\label{eq: power-aniso-multipoles}
    P_\ell(k; R_0) &=& (2\ell+1)\int\frac{d^3\vec x}{V}\int \frac{d\Omega_k}{4\pi} \mathcal{F}\left[\xi(\vec r; \vec x)W(\vec r; R_0)\right]L_\ell(\hat{\vec k}\cdot\hat{\vec x})\\\nonumber
    &=& \frac{1}{(2\pi)^3}\int \frac{d^3\vec x}{V}\left\{(2\ell+1)\int \frac{d\Omega_k}{4\pi}\left[P\ast\widetilde{W}\right](\vec k)L_\ell(\hat{\vec k}\cdot\hat{\vec x})\right\} = \frac{1}{(2\pi)^3}\left\langle{}\left[P\ast \widetilde{W}\right]_\ell\right\rangle(k)
\eeq
thus the power spectrum multipoles are simply related to the multipoles of the convolution of $P(\vec k)$ and $\widetilde{W}(\vec k)$ (with angle brackets indicating averaging over the LoS). Although this result has been derived for a LoS angle convention (choosing $\vec x$ to be the mid-point of the two galaxies), it applies equally well to other conventions, e.g. the \citet{2006PASJ...58...93Y} approximation, setting $\vec x = \vec r_j$ or the flat-sky approximation, with $\vec x = \mathrm{const.}$ 

Analogous to the isotropic case, $P_\ell(k; R_0)$ may be expressed in terms of the 2PCF multipoles $\xi_\ell(r)$, noting that the power spectrum multipoles are those of the Fourier transform of $\xi(\vec r)W(\vec r; R_0)$. In appendix \ref{appen: power-to-2pcf-multipoles}, we show that the configuration- and Fourier-space multipoles of a function $A(\vec r)$ are related by
\beq
    \widetilde{A}_\ell(k) = 4\pi i^\ell \int_0^\infty r^2dr\,A_\ell(r)j_\ell(kr)
\eeq
(applicable both for fixed and moving lines-of-sight); setting $A(\vec r) = \xi(\vec r)W(\vec r; R_0)$ and hence $\widetilde{A}(\vec k) = P(\vec k; R_0)$ implies
\beq\label{eq: aniso_power_window_mod}
    P_\ell(k;R_0) = 4\pi i^\ell \int_0^\infty r^2dr\,\xi_\ell(r)W(r; R_0)j_\ell(kr)
\eeq
since $\left[\xi(\vec r)W(\vec r; R_0)\right]_\ell = \xi_\ell(r)W(r;R_0)$, due to $W$ being isotropic. Notably, there is no multipole mixing induced by our pair-separation window function.

\subsection{Assessing the Impact of the Pair-Separation Limit $R_0$}\label{subsec: pair-separation-forecast}
From Eqs.\,\ref{eq: iso_power_window_mod}\,\&\,\ref{eq: aniso_power_window_mod} we can predict the effects of the window function on measurements of the power spectrum multipoles for various $k$. Here, we assume an ideal infinite survey (such that $n,w$ are constant and $\Phi(\vec r) = 1$ for all $\vec r$), using a smooth 2PCF input derived from Hankel transforms of a matter power spectrum computed in \texttt{CAMB}\footnote{\href{https://camb.info}{camb.info}} \citep{Lewis:2002ah}. For simplicity, we assume a linear power spectrum at redshift $z = 0.57$ with the cosmology $\{\Omega_b = 0.048, \Omega_\Lambda = 0.71, h = 0.7\}$, but add Kaiser redshift space distortions (RSD) to create a non-trivial quadrupole and hexadecapole. We adopt Kaiser parameter $\beta = 0.774$ and bias $b=2$ to emulate the BOSS DR12 CMASS sample \citep{2016MNRAS.460.4188G}. No Finger-of-God (FoG) effects are included in this 2PCF since they are more difficult to model. Although the model is clearly not representative of the true $P_\ell(k)$, we expected it to be a good predictor of pair-separation window function biases.

\begin{figure}
    \centering
    \includegraphics[width=0.95\textwidth]{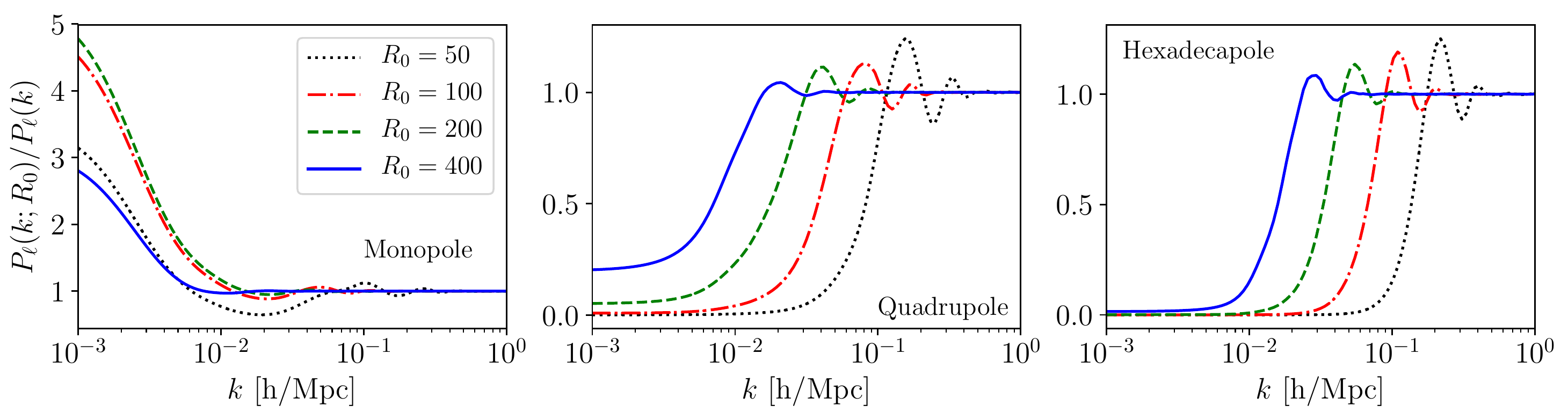}
    \includegraphics[width=\textwidth]{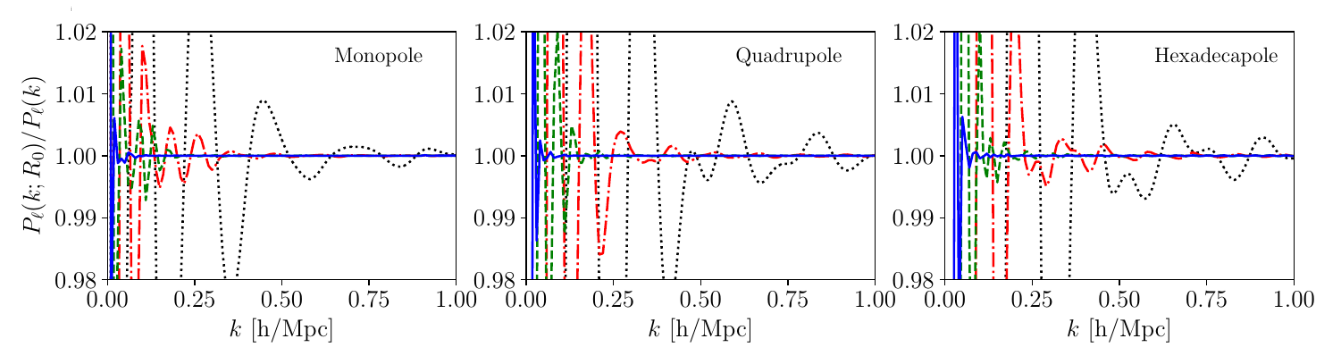}
    \caption{Comparison of windowed to true power spectrum moments using the polynomial window function (Eq.\,\ref{eq: window_defn}), computed for various multipoles and truncation scales, $R_0$ (in $h^{-1}\mathrm{Mpc}$). The upper and lower plots show the same data on logarithmic and linear scales respectively. Forecasts are obtained for a uniform unbounded survey (with correction function $\Phi=1$ everywhere), utilizing a smooth two-point correlation function derived from the linear matter power spectrum at $z = 0.57$ (from \texttt{CAMB}), with anisotropies included solely via the Kaiser redshift-space distortion (RSD) prescription, noting that we expect only weak dependence on the exact form of the power spectrum. For all multipoles, the windowed power matches the true power to high accuracy at large $k$, but we observe larger deviations at small $k$, especially for higher multipoles. For $R_0\gtrsim100h^{-1}\mathrm{Mpc}$, we achieve sub-percent accuracy with the windowed power spectrum on scales $k>0.2h\,\mathrm{Mpc}^{-1}$.}
    \label{fig: aniso_power_broadening_log}
\end{figure}

In Fig.\,\ref{fig: aniso_power_broadening_log} we plot the ratio of windowed to true anisotropic power as a function of $k$ for three sets of multipoles $\ell$ and truncation radii, obtained from numerical integration of Eq.\,\ref{eq: aniso_power_window_mod}. Notably there is a significant overestimate (underestimate) of monopole (quadrupole and hexadecapole) power on large scales (small $k$) when using the window function, but the ratio converges to unity at large $k$. Broader window functions are seen to have a smaller impact on the windowed power spectrum, as expected, and the impact is seen to be more significant at higher multipoles. For $R_0\gtrsim100h^{-1}\mathrm{Mpc}$, we obtain sub-percent agreement between windowed and true power for $k\gtrsim0.2h\,\mathrm{Mpc}^{-1}$, indicating that our estimator will work well in this regime. In addition, we note that these plots do not include $k$-space binning; using wide $k$ bins will reduce the oscillatory behaviour of $P_\ell(k; R_0)/P_\ell(k)$ with respect to $k$ and hence give improve the accuracy with which power can be measured.

\section{Dependence on Survey Geometry}\label{sec: survey-correction-integrals}
The survey-correction function $\Phi(\vec r) = RR_\mathrm{ideal}(\vec r)/RR(\vec r)$ (Eq.\,\ref{eq: true-RR-definition}) allows us to assess the impact of the survey geometry on the measured power spectrum. First, recall that our (boundary corrected) power spectrum estimator is derived simply from the Fourier transform of the \citet{1993ApJ...412...64L} estimator, i.e.
\beq
    P(\vec k) = \mathcal{F}\left[\frac{NN(\vec r)}{RR(\vec r)}\right](\vec k)
\eeq
(cf.\,Eq.\,\ref{eq: power_as_2PCF_FT}, here ignoring the pair-count window). If we had instead used the FKP estimator (Eq.\,\ref{eq: FKP_estimator}), we would have measured
\beq
    P^\mathrm{uncorr}(\vec k) = \mathcal{F}\left[\frac{NN(\vec r)}{RR_\mathrm{ideal}(\vec r)}\right] = \mathcal{F}\left[\frac{NN(\vec r)}{RR(\vec r)\Phi(\vec r)}\right] =  \mathcal{F}\left[\xi(\vec r)\Phi^{-1}(\vec r)\right] = \frac{1}{(2\pi)^3} \left[P\ast\mathcal{F}[\Phi^{-1}]\right](\vec k)
\eeq
applying the convolution theorem and noting that the factor $I = V\overline{(nw)^2}$ appearing in the denominator of the FKP estimator is simply the idealized survey pair-counts. The uncorrected power spectrum is thus a convolution of the corrected power with the Fourier transform of the reciprocal correction function. With the inclusion of the pair-wise window function, both $P^\mathrm{uncorr}(\vec k)$ and $P(\vec k)$ are further convolved with $\widetilde{W}(\vec k; R_0)$. 

To investigate this, we consider the expansion of $P^\mathrm{\,uncorr}$ into Legendre multipoles, utilizing the result of appendix \ref{appen: transformed-multipole-power}, which states that, for an arbitrary function $\omega(\vec r)$, the multipoles of $\mathcal{P}(\vec k) = \mathcal{F}\left[\xi(\vec r)\omega(\vec r)\right]$ are given by
\beq
    \mathcal{P}_\ell(k) &=& (2\ell+1) \sum_{\ell_1,\ell_2} \left(\begin{array}{ccc}
         \ell_1 & \ell_2 & \ell \\
          0 & 0 & 0
    \end{array}\right)^2\mathcal{G}_{\ell;\ell_1,\ell_2}(k)\\\nonumber
    \mathcal{G}_{\ell;\ell_1,\ell_2}(k) &=& 4\pi i^\ell\int_0^\infty r^2dr\,\,j_\ell(kr)\xi_{\ell_1}(r)\omega_{\ell_2}(r).
\eeq
in terms of a Wigner $3j$ symbol \citep[Sec.\,34.2]{nist_dlmf}, with the summations over $\ell_1$ and $\ell_2$ restricted to those values allowed by the $3j$ selection rules. This applies to both constant and moving lines-of-sight. Here $\omega(\vec r) = \Phi^{-1}(\vec r)W(\vec r; R_0)$ (reintroducing the pair-separation window function) giving
\beq\label{eq: pow-uncorr}
    P^\mathrm{uncorr}_\ell(k;R_0) &=& (2\ell+1)\sum_{\ell_1\ell_2}\left(\begin{array}{ccc}
         \ell & \ell_1 & \ell_2 \\
         0 & 0 & 0\\
    \end{array}\right)^2\mathcal{H}_{\ell;\ell_1,\ell_2}(k;R_0)\\\nonumber
    \mathcal{H}_{\ell;\ell_1,\ell_2}(k;R_0) &=& 4\pi i^{\ell}\int_0^\infty r^2dr\,\xi_{\ell_1}(r)[\Phi^{-1}]_{\ell_2}(r)W(r;R_0)j_\ell(kr)
\eeq
which may be compared with Eq.\,\ref{eq: aniso_power_window_mod}, which defines $P_\ell(k;R_0)$. This clearly shows how ignoring the survey correction function $\Phi$ introduces bias and multipole mixing into the windowed power spectrum.

\begin{figure}
    \centering
    \includegraphics[width=0.5\textwidth]{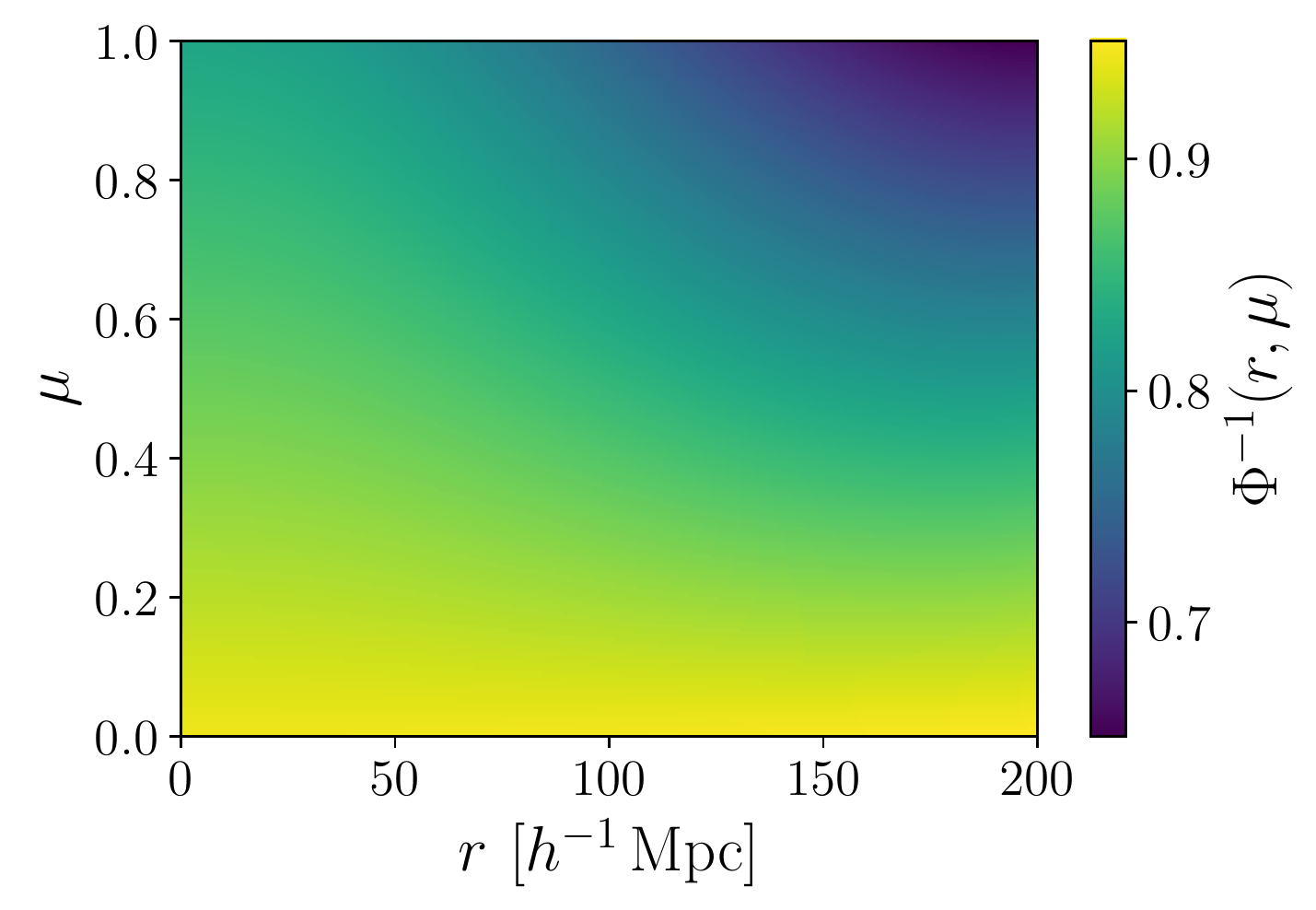}
    \caption{Reciprocal of the survey-correction function $\Phi(\vec r)$ for the BOSS DR12 CMASS-N survey \citep{2013AJ....145...10D} as a function of distance and angle of a particle pair from the LoS. $\Phi^{-1}$ is the ratio of the true $RR$ pair-counts (computed from exhaustive pair-counting with \texttt{corrfunc} \citep{2017ascl.soft03003S}) to those expected for a uniform, infinite survey, and deviations from unity show the effects of the survey geometry. As expected, deviations are largest for widely separated pairs oriented along the LoS. This plot is generated by fitting smooth splines to the Legendre moments of the measured grid of $\Phi$ values.}
    \label{fig: inverse-phi-plot}
\end{figure}

To illustrate this, we consider the corrected and uncorrected power using a survey correction function, $\Phi$, appropriate for the BOSS DR12 CMASS-N dataset \citep{2013AJ....145...10D}. This is computed by measuring the true $RR$ pair-counts in 400 radial and 100 angular bins via exhaustive pair-counting using $\texttt{corrfunc}$ and comparing these to the expected value, given the normalization of $V\overline{(nw)^2} \approx 53h^3\,\mathrm{Mpc}^{-3}$. The observed values of $\Phi^{-1}$ (defined by Eq.\,\ref{eq: phi_r_mu_defn}) are then fit with spline curves in Legendre multipole space, giving a smooth $\Phi(r,\mu)$ representation. In Fig.\,\ref{fig: inverse-phi-plot}, we show $\Phi^{-1}$ as a function of spatial position. As expected, $\Phi^{-1}\rightarrow1$ on small radial scales (when few pairs cross the survey boundary), but $\Phi^{-1}$ is reduced at larger $r$, especially for pairs parallel to the LoS $\mu\approx 1$, since these pairs are more likely to cross the survey boundary. $\Phi^{-1}(r,\mu)$ is clearly observed to be smooth here. 

\begin{figure}
    \centering
    \includegraphics[width=\textwidth]{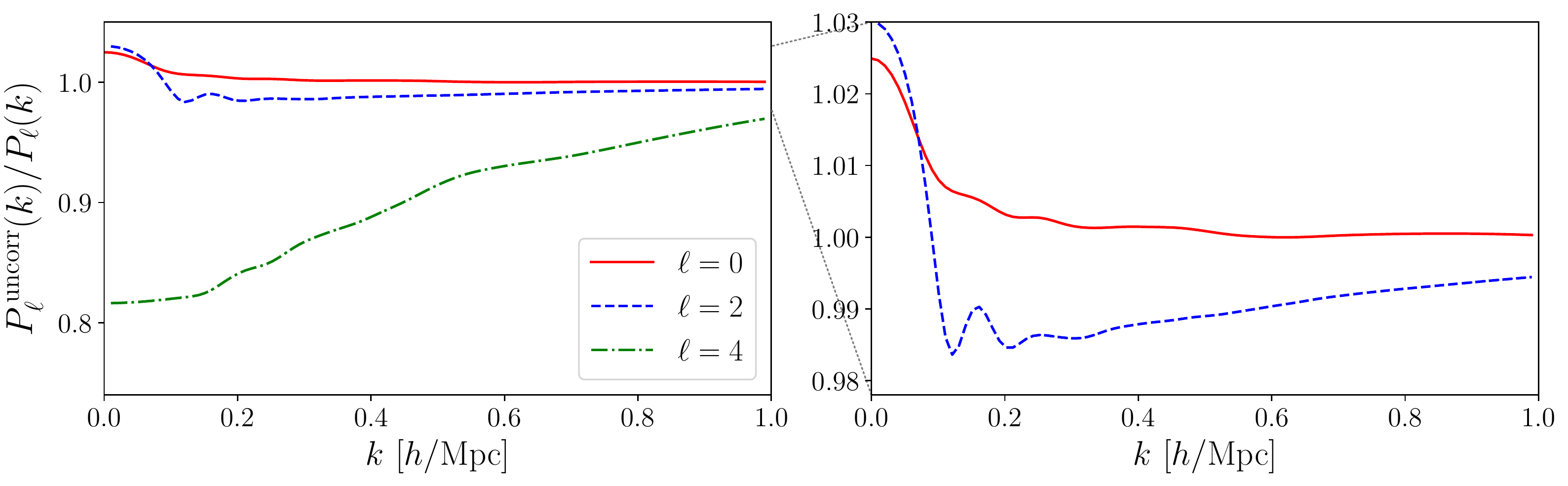}
    \caption{Comparison of the survey-geometry uncorrected and corrected measured power spectrum multipoles for simulated observations of the BOSS DR12 CMASS-N region, using a mock correlation function computed from a linear power spectrum in \texttt{CAMB} \citep{Lewis:2002ah} with Kaiser RSD anisotropies added. We note that the high-$k$ uncorrected power spectrum is equivalent to that used in \citet{2015MNRAS.453L..11B} (or \citet{1994ApJ...426...23F} for the monopole) without any deconvolution. The two plots are identical, but with different scales. These are derived from Eqs.\,\ref{eq: pow-uncorr}\,\&\,\ref{eq: aniso_power_window_mod}, using a polynomial pair-separation window function truncated at $200h^{-1}\mathrm{Mpc}$, with the inclusion of the reciprocal survey correction factor $\Phi^{-1}$ (shown in Fig.\,\ref{fig: inverse-phi-plot}) in the uncorrected power spectrum accounting for the lack of consideration of the survey-geometry. To account for small errors (at the few percent level) in the normalization of $\Phi$, the curves are normalized by their power at $k=2$ (where the geometry should have no significant effect). We note significant deviations from unity on small scales, especially for higher multipoles. Fig.\,\ref{fig: qpm_no_phi_comparison} gives the analogous plot applied to full simulated galaxy catalogs.}
    \label{fig: phi-correction-plot}
\end{figure}

Using a smooth mock 2PCF derived from a linear power spectrum with RSD added via the Kaiser prescription (as in Sec.\,\ref{subsec: pair-separation-forecast}), we may compute the ratio of uncorrected to corrected power via Eqs.\,\ref{eq: pow-uncorr}\,\&\,\ref{eq: aniso_power_window_mod}, utilizing the Legendre multipoles of $\Phi^{-1}$ (up to $\ell=6$) used to create Fig.\,\ref{fig: inverse-phi-plot}. Fig.\,\ref{fig: phi-correction-plot} shows this, with numerical integration used to evaluate the power integrals for various $k$ and multipoles $\ell\in[0,2,4]$. On the large scales probed here, we note only small effects on the isotropic power spectrum ($\ell=0$), at the sub-percent level for $k\gtrsim0.2h\,\mathrm{Mpc}^{-1}$. For higher multipoles, the difference between the corrected and uncorrected power is more striking, with the quadrupole (hexadecapole) measurements differing at the $1\%$ ($20\%$) level for $k\approx 0.2h\,\mathrm{Mpc}^{-1}$. The large difference in the measured hexadecapole is as expected since the hexadecapole is intrinsically a very small signal. It is clear, therefore, that proper consideration of the survey window functions is needed even on small scales. Standard Fourier transform approaches must do this via deconvolution with the survey mask power spectrum; our method here differs since it is performed directly as the power is estimated, not in post-processing. Although the results above are presented for a specific linear power spectrum, we do not expect them to have strong dependence on the exact power spectrum form.

\section{Power Spectrum Covariances}\label{sec: power-cov}
One motivation for estimating power spectra in configuration-space is the ability to generate survey-dependent covariance matrices using the techniques developed for correlation function covariances. In addition, computing the power spectrum and 2PCF in the same manner means that we can compute cross-covariances between statistics, facilitating robust joint analyses that will be important in upcoming galaxy surveys. We here discuss the form and implementation of these covariances, noting that they are fully analogous to the 2PCF and 3PCF covariances first presented in \citet{2016MNRAS.462.2681O} and \citet{3pcfCov}, with the addition of pair-wise window functions and a modified kernel.

\subsection{Theoretical Covariance Estimators}

\subsubsection{Autocovariance of $P_\ell(k)$}
The covariance matrix of the windowed anisotropic power spectrum for an arbitrary survey geometry may be derived directly from our estimators (Eq.\,\ref{eq: windowed_aniso_power_estimator}). We begin by writing the power spectrum estimator in discrete form;
\beq\label{eq: windowed_aniso_power_discrete}
    \hat{P}_\ell^a = \frac{1}{V\overline{(nw)^2}}\sum_{i\neq j}n_in_jw_iw_j\delta_i\delta_jA^a_{\ell,ij}\Phi_{ij}W_{ij}^{R_0}
\eeq
where the summations are over small volumes containing at most one galaxy and subscripts denote the function arguments, e.g. $\delta_i \equiv \delta(\vec r_i)$ and $\Phi_{ij} \equiv \Phi(\vec r_i-\vec r_j)$. Defining the covariance in radial bins $a,b$ and Legendre moments $p,q$ as
\beq
    \operatorname{cov}(\hat P^a_p\,,\,\hat P^b_q) \equiv \av{\hat P^a_p\hat P^b_q}-\av{\hat P^a_p}\av{\hat P^b_q}
\eeq
(suppressing the $R_0$ truncation radius argument henceforth) we can insert Eq.\,\ref{eq: windowed_aniso_power_discrete} to yield
\beq
    \operatorname{cov}(\hat P^a_p\,,\,\hat P^b_q) = \frac{1}{\left(V\overline{(nw)^2}\right)^2}\sum_{i\neq j}\sum_{k\neq l}n_in_jn_kn_lw_iw_jw_kw_lA_{p,ij}^a A_{q,kl}^b\Phi_{ij}\Phi_{kl}W^{R_0}_{ij}W^{R_0}_{kl}\left[\av{\delta_i\delta_j\delta_k\delta_l}-\av{\delta_i\delta_j}\av{\delta_k\delta_l}\right].
\eeq
As in \citet{2016MNRAS.462.2681O}, the double summation may be expanded into three terms involving two, three and four points in space and we can use Wick's theorem with the shot-noise contraction approximation $\delta_i^2 \approx (1+\delta_i)/n_i$ to evaluate the expectation terms arising. This gives the full covariance;
\beq\label{eq: power-cov-matrix-full}
    \operatorname{cov}(\hat P^a_p\,,\,\hat P^b_q) &=& {}^4C^{ab}_{pq} + {}^3C^{ab}_{pq} + {}^2C^{ab}_{pq}\\\nonumber
    {}^4C^{ab}_{pq} &=& \frac{1}{\left(V\overline{(nw)^2}\right)^2}\sum_{i\neq j\neq k\neq l}n_in_jn_kn_lw_iw_jw_kw_l A_{p,ij}^a A_{q,kl}^b\Phi_{ij}\Phi_{kl}W^{R_0}_{ij}W^{R_0}_{kl}\left(\xi^{(4)}_{ijkl}+2\xi_{ik}\xi_{jl}\right)\\\nonumber
    {}^3C^{ab}_{pq} &=& 4\times \frac{1}{\left(V\overline{(nw)^2}\right)^2}\sum_{i\neq j\neq k}n_in_jn_kw_i(w_j)^2w_k A_{p,ij}^a A_{q,jk}^b\Phi_{ij}\Phi_{jk}W^{R_0}_{ij}W^{R_0}_{jk}\left(\zeta_{ijk}+\xi_{ik}\right)\\\nonumber
    {}^2C^{ab}_{pq} &=& 2\times\frac{1}{\left(V\overline{(nw)^2}\right)^2}\sum_{i\neq j}n_in_j(w_iw_j)^2A_{p,ij}^aA_{q,ij}^b\left(\Phi^{}_{ij}W^{R_0}_{ij}\right)^2\left(1+\xi_{ij}\right)
\eeq
where $\zeta_{ijk}$ and $\xi^{(4)}_{ijkl}$ are the connected three- and four-point correlation functions (3PCF and 4PCF) respectively. These can be converted into integral expressions by replacing $\sum_i\rightarrow\int d^3\vec r_i$, $X_i\rightarrow X(\vec r_i)$ and $X_{ij}\rightarrow Y(\vec r_i-\vec r_j)$ for variables $X,Y$. In previous work \citep{2016MNRAS.462.2681O,2019MNRAS.487.2701O,rascalC,3pcfCov}, we have included a shot-noise rescaling parameter, $\alpha$, to encapsulate non-Gaussianity in our model by rescaling the two- and three-point covariance integrals (or summations) and neglecting any non-Gaussian contributions. Note that our power estimator focuses on small-scales, where we expect non-Gaussianity to dominate, thus it is not immediately apparent whether our covariance model will be sufficient. In addition, we have not explicitly accounted for the effects of power spectrum modes larger than the survey volume on the covariance, which can be shown to be non-negligible at large $k$ \citep[e.g.][]{2012JCAP...04..019D}. With the above approximations, our estimator becomes
\beq\label{eq: power-cov-matrix-gaussian}
    \operatorname{cov}(\hat P^a_p\,,\,\hat P^b_q) &=& {}^4C^{ab}_{pq} + \alpha \times {}^3C^{ab}_{pq} + \alpha^2 \times {}^2C^{ab}_{pq}\\\nonumber
\eeq
with identical definitions for the two-, three- and four-point matrices, except dropping the 3PCF and 4PCF terms. Techniques with which to compute these covariances efficiently are discussed in Sec.\,\ref{subsec: power-cov-implementation}.

\subsubsection{Cross Covariance of $P_\ell(k)$ and $\xi_\ell(r)$}
We may similarly derive the cross-covariance matrices between the $P_\ell(k)$ multipoles and the anisotropic 2PCF, assuming the latter to be binned in Legendre multipoles (as in \citealt{3pcfCov}). For Legendre multipoles $p$ and $q$ and radial bins $a$ and $b$ (in configuration- and Fourier-space respectively), we may define 
\beq
    \operatorname{cov}(\hat P^a_p\,,\,\xi^b_q) = \av{\hat P^a_p\hat\xi^b_q}-\av{\hat P^a_p}\av{\xi^b_q}.
\eeq
where the 2PCF estimator is given by 
\beq\label{eq: 2PCF in Legendre Moments}
    \hat\xi_\ell^a = \frac{2\ell+1}{\frac{4\pi}{3}V\overline{(nw)^2}\left(r_{a,\mathrm{max}}^3-r_{a,\mathrm{min}}^3\right)}\sum_{i\neq j}n_in_jw_iw_j\Theta^{a}(r_{ij})\Phi(r_a,\mu_{ij})L_\ell(\mu_{ij})\delta_i\delta_j
\eeq
(cf.\,\citet{3pcfCov}), where we note that $\mu_{ij}$ is the cosine of the angle between the LoS and the mid-point of $\vec r_i$ and $\vec r_j$ and $\Phi$ takes the radial value at the bin-center \textit{not} $r_{ij}$. Analogous to the $P_\ell(k)$ autocovariance, we can define the Gaussian cross-covariance matrix (with shot-noise rescaling $\alpha$) as 
\beq
    \operatorname{cov}(\hat P^a_p\,,\,\xi^b_q) &=& {}^4C^{ab}_{pq} + \alpha \times {}^3C^{ab}_{pq} + \alpha^2 \times {}^2C^{ab}_{pq}\\\nonumber
    {}^4C^{ab}_{pq} &=& \frac{2q+1}{\frac{4\pi}{3}V\overline{(nw)^2}\left(r_{a,\mathrm{max}}^3-r_{a,\mathrm{min}}^3\right)}\sum_{i\neq j\neq k\neq l}n_in_jn_kn_lw_iw_jw_kw_l A_{p,ij}^a \Phi(r_{ij},\mu_{ij})W^{R_0}_{ij}\Theta^b(r_{kl})L_q(\vec \mu_{kl})\Phi(r_b,\mu_{kl})\left(2\xi_{ik}\xi_{jl}\right)\\\nonumber
    {}^3C^{ab}_{pq} &=& 4\times\frac{2q+1}{\frac{4\pi}{3}V\overline{(nw)^2}\left(r_{a,\mathrm{max}}^3-r_{a,\mathrm{min}}^3\right)}\sum_{i\neq j\neq k}n_in_jn_kw_i(w_j)^2w_k A_{p,ij}^a \Phi(r_{ij},\mu_{ij})W^{R_0}_{ij} \Theta^b(r_{jk})L_q(\vec \mu_{jk})\Phi(r_b,\mu_{jk})\left(\xi_{ik}\right)\\\nonumber
    {}^2C^{ab}_{pq} &=& 2\times\frac{2q+1}{\frac{4\pi}{3}V\overline{(nw)^2}\left(r_{a,\mathrm{max}}^3-r_{a,\mathrm{min}}^3\right)}\sum_{i\neq j}n_in_j(w_iw_j)^2A_{p,ij}^a \Phi(r_{ij},\mu_{ij})W^{R_0}_{ij} \Theta^b(r_{ij})L_q(\vec \mu_{ij})\Phi(r_b,\mu_{ij})\left(1+\xi_{ij}\right)
\eeq
(noting that there is a $(2p+1)$ factor absorbed in the kernel $A_{p,ij}^a$). We note the inclusion of radial binning functions here as well as Legendre polynomials and survey-correction factors. This may be computed by pair-, triple- and quad-counting, with a quad giving contributions to a single 2PCF radial bin, but all $k$-space bins and Legendre multipoles.

The above auto and cross-covariances may be generalized to the multi-field case, giving the cross-covariances between auto- and cross-power spectra of multiple tracer galaxies. The cross-spectrum estimators are defined in a similar fashion to Eq.\,\ref{eq: iso_power_pair_counting}, except that we now include pair-counts from two data and random fields (e.g. replacing modified $DD$ counts with modified $D^SD^T$ counts for fields $S,T$). The covariances will take a similar form to those in \citet[][Sec.\,6]{rascalC} and \citet[][Sec.\,3.3]{3pcfCov}, with the inclusion of the $A^a_\ell$ kernel functions and pair-separation windows $W^{R_0}$. They may be computed in a similar fashion to the single-field case.

\subsection{Implementation of the Gaussian Power Covariance Matrix}\label{subsec: power-cov-implementation}
In practice, the survey-geometry-dependent covariance integrals (Eqs.\,\ref{eq: power-cov-matrix-full}) are difficult to compute, even in the fully Gaussian (yet non-linear) limit. Previously, integrals have been computed using the \texttt{RascalC} code \citep{rascalC},\footnote{\href{http://RascalC.readthedocs.io}{RascalC.readthedocs.io}} selecting sets of four particles in space and adding them to the relevant bins. Unlike for 2PCF covariances, each chosen quad will now contribute to all $k$-bins and Legendre moments, implying that an entire matrix must be computed for every quad drawn. This is grossly inefficient, especially given that we must compute generalized hypergeometric functions or Sine integrals for each $k$-bin for $\ell>0$, and we here consider an alternative solution.

Considering first the two-point matrix ${}^2C^{ab}_{pq}$, we note that all $k$-dependence arises from the kernel functions $A_{p,ij}^a$ and $A_{q,ij}^b$, which depend only on $r_{ij}$ and $\mu_{ij}$, and further, that the two dependencies are separable. Denoting the reduced kernel as $\mathcal{A}_p^a(|\vec r_i-\vec r_j|)$, we define
\beq
    &\,& A_p^a(\vec r_i,\vec r_j) \equiv (2p+1)L_p(\hat{\vec x}\cdot\hat{\vec u})\mathcal{A}_p^a(u)\\\nonumber
    &\Rightarrow& \mathcal{A}_p^a(u) = \frac{3(-1)^{p/2}}{u^3\left(k_{a,\mathrm{max}}^3-k_{a,\mathrm{min}}^3\right)}\left[D_p(k_{a,\mathrm{max}}u)-D_p(k_{a,\mathrm{min}}u)\right]
\eeq
where $\vec u = \vec r_i-\vec r_j$ and $\vec x = (\vec r_i+\vec r_j)/2$ as before. The two-point covariance matrix may thus be written as a weighted expectation of $\mathcal{A}^a_p\mathcal{A}^b_q$ over $u=|\vec r_i-\vec r_j|$;
\beq\label{eq: cov2-pdf-expression}
    {}^2C_{pq}^{ab} = \int_0^{R_0}du\,\mathcal{A}^a_p(u)\mathcal{A}^b_q(u)\times {}^2\Omega_{pq}(u).
\eeq
with the integrand critically having only finite support. Here ${}^2\Omega_{pq}(u)$ is the two-point integral including the Legendre polynomial factors, integrated over all dimensions except $|\vec r_i-\vec r_j|$. This is simply an unnormalized PDF for $u$ and is expected to be smooth. Estimates for ${}^2\Omega_{pq}(u)$ in some configuration-space bins $\{c\}$ with width $\{\Delta r_c\}$ may be determined from the two-point integral (here expressed in discrete form)
\beq\label{eq: omega2-def}
    {}^2\Omega^c_{pq}(R_0) = 2\times\frac{(2p+1)(2q+1)}{\left(V\overline{(nw)^2}\right)^2}\sum_{i\neq j}n_in_j(w_iw_j)^2L_p(\mu_{ij})L_q(\mu_{ij})\left(\Phi^{}_{ij}W^{R_0}_{ij}\right)^2\left(1+\xi_{ij}\right)\times \frac{\Theta^{c}_{ij}}{\Delta r_c}
\eeq
where the radial binning function $\Theta^c$ picks out values of $|\vec r_i-\vec r_j|$ in the bin $c$. This may be computed simply via pair-counting (and fit to a smooth function), since each pair only adds $(\ell_\mathrm{max}/2+1)^2$ Legendre bins (up to a maximum moment $\ell_\mathrm{max}$) rather than $(\ell_\mathrm{max}/2+1)^2\times n_k^2$ for a total of $n_k$ $k$-space bins. The two-point covariance matrix determination thus reduces to estimating a one-dimensional a $k$-space binning independent function ${}^2\Omega_{pq}(u;R_0)$ and reconstructing the covariance in post-processing.

We may derive analogous expressions for the three- and four-point covariances, noting that the $k$-dependence is now a function of two configuration-space separations, each limited to the region $[0,R_0]$. We similarly obtain
\beq\label{eq: cov34-pdf-expression}
    {}^3C_{pq}^{ab} = \int_0^{R_0} du_1\int_0^{R_0}du_2\,\mathcal{A}^a_p(u_1)\mathcal{A}^b_q(u_2)\times {}^3\Omega_{pq}(u_1,u_2;R_0)\\\nonumber
    {}^4C_{pq}^{ab} = \int_0^{R_0}du_1\int_0^{R_0}du_2\,\mathcal{A}^a_p(u_1)\mathcal{A}^b_q(u_2)\times {}^4\Omega_{pq}(u_1,u_2;R_0)
\eeq
where the (two-dimensional) ${}^3\Omega$ and ${}^4\Omega$ unnormalized PDFs may be found from estimates in configuration-space bins $c,d$;
\beq\label{eq: Omega34def}
    {}^3\Omega^{cd}_{pq}(R_0) = 4\times\frac{(2p+1)(2q+1)}{\left(V\overline{(nw)^2}\right)^2}\sum_{i\neq j\neq k}n_in_jn_kw_i(w_j)^2w_kL_p(\mu_{ij})L_q(\mu_{jk})\Phi^{}_{ij}\Phi^{}_{jk}W^{R_0}_{ij}W^{R_0}_{jk}\left(\xi_{ik}\right)\times \frac{\Theta^{c}_{ij}\Theta^d_{jk}}{\Delta r_c\Delta r_d}\\\nonumber
    {}^4\Omega^{cd}_{pq}(R_0) = \frac{(2p+1)(2q+1)}{\left(V\overline{(nw)^2}\right)^2}\sum_{i\neq j\neq k\neq l}n_in_jn_kn_lw_iw_jw_kw_lL_p(\mu_{ij})L_q(\mu_{kl})\Phi^{}_{ij}\Phi^{}_{kl}W^{R_0}_{ij}W^{R_0}_{kl}\left(2\xi_{ik}\xi_{jl}\right)\times \frac{\Theta^{c}_{ij}\Theta^d_{kl}}{\Delta r_c\Delta r_d}.
\eeq
Note that this is fully analogous to the Legendre-binned 2PCF covariance of \citet{3pcfCov}; the only difference is the inclusion of pair-separation windows $W(r_{ij};R_0)$ and the promotion of $\Phi$ to be a function of both $r$ and $\mu$ (for 2PCF covariances in Legendre bins, $\Phi$ is averaged over the desired radial bin). These expressions may be computed via pair-, triple- and quad-counting, with only a small modification to \texttt{RascalC}. Notably this is far easier to compute than the full power spectrum covariances, since the latter require a large matrix to be computed for each quad selected.

Unlike previous 2PCF covariances, we must sample the $\Omega$ matrices down to $u = 0$; this requires a minor change to the sampling strategies discussed in \citet{rascalC}. To do this, we multiply the previously used $\xi$-weighted importance sampling probability kernels by a factor $(1+5a^2/(a+r)^2)$ for cell separation $r$ and width $a$, and use a $(r+a)^{-3}$ kernel for $i-j$ separations. This allows for efficient sampling down to the minimum pair-separation in the random catalogs utilized.

\section{Application to Simulated Data}\label{sec: data}
In this section we apply the algorithms described above to simulated data to justify our analysis. We begin with a brief note on the choice of $R_0$ before considering the small-scale power estimator and its covariance in Secs.\,\ref{subsec: aniso-power-dat} and \ref{subsec: QPM-cov-matrices} respectively.

\subsection{Choice of Truncation Radius and Binning-Widths}\label{subsec: data-binning}
Two important hyperparameters in the power spectrum estimators are the truncation scale and binning widths. When performing pair-counts across a catalog with $N$ members, the total of operations scales as $N\times nV_{R_0}$ for number density $n$ and $V_{R_0} = 4\pi R_0^3/3$ since $nV_{R_0}$ is the number of secondary particles located within distance $R_0$ of a given primary. The computation time hence scales as $R_0^3$, although we obtain more accurate results with a larger truncation radius (Sec.\,\ref{subsec: pair-separation-forecast}\,\&\,Fig.\,\ref{fig: aniso_power_broadening_log}). In this section, we consider $R_0 = 50h^{-1}\mathrm{Mpc}$ and $100h^{-1}\mathrm{Mpc}$, with the latter giving sub-percent accuracies in $P_\ell(k;R_0)$ for $k\gtrsim0.25h\,\mathrm{Mpc}^{-1}$ at reasonable computation times.

As noted in Sec.\,\ref{subsec: window_convolution}, the pair-separation window function has the effect of convolving the true power with the Fourier transformed window $\widetilde{W}(k;R_0)$, which has characteristic scale $k\sim 3/R_0$ (Fig.\,\ref{fig: window_plots}b), which effectively sets the binning scale $\Delta k\gtrsim 3/R_0$. Using smaller $\Delta k$ will result in significant correlations between neighbouring bins, making the covariance matrix less diagonal and harder to invert, thus setting a minimum scale for $\Delta k$. In the ideal (unwindowed) limit, the monopole survey covariance matrix takes the form 
\beq\label{eq: ideal-covariance}
    C^{ab}_\mathrm{ideal} = \frac{(2\pi)^3}{V}\left[\frac{2P^2(k_a)}{4\pi k_a^2\Delta k}\delta_{ab}+\bar{T}_{ab}\right]
\eeq
\citep{1999ApJ...527....1S}, with a diagonal term proportional to $P_i^2$ and a non-Gaussian off-diagonal term depending on the reduced trispectrum $\bar{T}_{ab}$, which is the quadrilateral trispectrum $T(\vec k,-\vec k,\vec k',-\vec k')$ averaged over $\vec k,\vec k'$ in bins $a,b$. The non-Gaussian off-diagonal term gives non-negligible correlations between different $k$-bins at large $k$, and is found to be largely insensitive to the binning used \citep{2017MNRAS.466..780M} thus does not affect our choice of $\Delta k$.

Although in this paper we use linear binning in $k$-space with a fixed $R_0$, in future analyses it may be more convenient to use logarithmic bins in $k$-space. In this case one should use a $k$-dependent truncation radius $R_0$ to reduce correlations between bins. Assuming a binning of $\Delta \log_{10}k = \beta$, we require $R_0(k)\gtrsim 3\times \left[\log(10)\beta k\right]^{-1}$ in this instance. Using a $k$-dependent binning is somewhat less efficient (and thus not adopted here), since we must still sample all pairs up to the maximum truncation radius $R_0(k_\mathrm{min})$, although we would no longer need to compute the contributions to all $k$-space bins for a given pair, only those with $r<R_0(k)$ for pair-separation $r$. 

\subsection{Anisotropic Power Estimates}\label{subsec: aniso-power-dat}
\subsubsection{Experimental Methodology}
To show the utility of the above estimators for small-scale power spectrum estimation we apply them to Quick Particle Mesh \citep[QPM,][]{2014MNRAS.437.2594W} mock galaxy simulations, which emulate the NGC CMASS dataset \citep{2013AJ....145...10D} of the BOSS SDSS-III survey. Each simulation contains the positions and FKP weights of $\sim 640000$ galaxies, and are converted into a Cartesian coordinate space assuming the cosmology $\{\Omega_m=0.29,\Omega_k=0,w_\Lambda=-1\}$ \citep{2018MNRAS.477.1153V}. In addition, we use a set of 32292068 random particle positions (approximately 50 more randoms than galaxies) computed for the same survey geometry, which allow evaluation of the $\widetilde{DR}$ and $\widetilde{RR}$ pair-counts, as in correlation function estimation.

Since the method of this paper has been shown to work well at high-$k$, we principally use linear binning with $R_0 = 100h^{-1}\mathrm{Mpc}$, $\Delta k = 0.05h\,\mathrm{Mpc}^{-1}$ (consistent with Sec.\,\ref{subsec: data-binning}) and $k\in [0.05,1]$ for $\ell\in\{0,2,4\}$ (noting that higher multipoles are expected to be small and difficult to measure) giving a total of $19\times 3 = 57$ bins. To assess the effects of the truncation scale on the measured power, we additionally consider power spectra with $R_0 = 50h^{-1}\mathrm{Mpc}$, $\Delta k = 0.1h\,\mathrm{Mpc}^{-1}$ and $k\in [0.1,1]$, giving a total of $9\times 3 = 27$ bins. The survey correction function $\Phi$ (shown in Fig.\,\ref{fig: inverse-phi-plot}) is found from standard pair-counting using \texttt{corrfunc} as in Sec.\,\ref{sec: survey-correction-integrals}. Multipoles of the inverse function $\Phi^{-1}$ were found to be well described by polynomials of quadratic order; the coefficients of these are stored to allow the modified pair-counting algorithm to compute $\Phi$ for each set of $(r,\mu)$ separations.

To compute the modified pair-counts for two fields $X$ and $Y$ (defined by Eq.\,\ref{eq: windowed_pair_counts}), we first sort the particles into a cuboidal grid with cells of size $\sim 20\,\mathrm{Mpc}/h$. It is important to note that this is purely to allow us to efficiently find particles up to some truncation scale; we do \textit{not} discretize the particles into a regular grid of coordinates. For each primary grid-cell, we iterate over all secondary grid-cells at separation $\lesssim R_0$ (including the primary), and for each, evaluate the contributions to all pair-count bins from each pair of particles found therein (excluding self-counts). To allow fast computation of the kernel functions $D_\ell(k|\vec r_i-\vec r_j|)$ (Eq.\,\ref{eq: D_function_simple}), we pre-compute this function for $10^5$ values of $k|\vec r_i-\vec r_j|\in[0,k_\mathrm{max}R_0]$ and use linear interpolation throughout the pair-counting.

$\widetilde{DD}$ and $\widetilde{DR}$ are found via exhaustive pair-counting of the data and random fields described above. For the modified $\widetilde{RR}$ counts, we instead partition the random particles into 50 random disjoint subsets and compute the pair-counts for each set with itself, before co-adding the sets of counts (following \citet{1993ApJ...412...64L}, \citet{2015MNRAS.454.4142S} and \citet{2019arXiv190501133K}, which show this to be the most efficient method to compute random counts at fixed computational cost). We hence obtain $50$ times the number of $\widetilde{DR}$ and $\widetilde{RR}$ counts as for $\widetilde{DD}$, thus we expect any effects arising from the finite number of random positions to be largely subdominant. The full pair-counting takes $\sim 30$ minutes for the $\widetilde{RR}$ and $\widetilde{DR}$ counts on a modern 20-core machine (and is \resub{easily} parallelized). \resub{The utility of our method is seen via comparison to FFT-based approaches. These have computation time scaling as $N_\mathrm{grid}\log N_\mathrm{grid} \sim k_\mathrm{max}\log k_\mathrm{max}$ (for $N_\mathrm{grid}$ cells along each axis of the discretization grid) which becomes large on small scales. In contrast, the pair counts are computed \textit{faster} at higher $k$, since we may use a smaller truncation radius $R_0$. Note however that the pair counting work scales as $\mathcal{O}(Nn)$ for $N$ particles of number density $n$, whilst the FFT-based approaches have a weaker scaling with the sample density. Initial testing has shown our algorithm to be faster for BOSS-like surveys at mildly non-linear wavenumbers.}


\subsubsection{Analysis for a single QPM Mock}

\begin{figure}
\centering
\begin{minipage}[t]{.48\textwidth}
\vspace{0pt}
  \centering
  \includegraphics[width=0.95\textwidth]{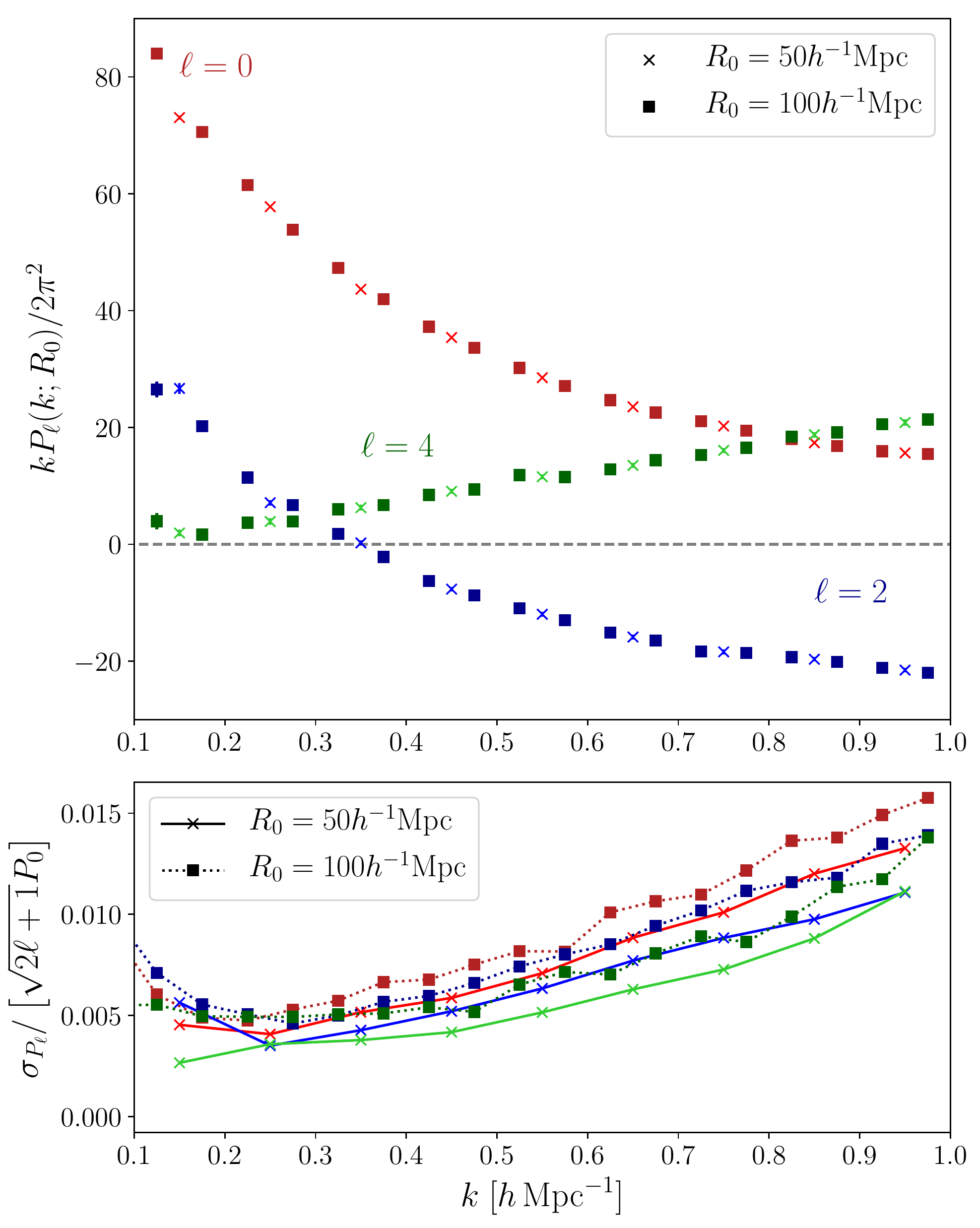}
  \caption{\textbf{Upper}: Small-scale power spectrum multipoles (upper plot) computed for $\ell \in \{0,2,4\}$ for a single QPM mock using the configuration-space power spectrum estimator (Eq.\,\ref{eq: windowed_pair_counts}). We display results counting pairs of particles up to truncation radii of $R_0 = 50h^{-1}\mathrm{Mpc}$ and $100h^{-1}\mathrm{Mpc}$ in crosses and squares using the linear bins of $\Delta k = 0.1h\,\mathrm{Mpc}^{-1}$ and $0.05h\,\mathrm{Mpc}^{-1}$ respectively to minimize cross-talk between bins. We note that the quadrupole spectrum becomes negative for $k\gtrsim0.3h\,\mathrm{Mpc}^{-1}$ due to the Finger of God (FoG) effect. We caution that, although the two datasets appear highly consistent here, the error from truncation at $R_0 = 50h^{-1}\mathrm{Mpc}$ swamps the statistical error on small scales (cf.\,Fig.\,\ref{fig: qpm_R0_comparison}). 
  \textbf{Lower}: Standard deviations of the above $P_\ell(k)$ measurements for both truncation radii, computed from the variance of 200 QPM mock power spectrum measurements. These are rescaled by the expected ideal isotropic power spectrum scaling of $\sigma_{P_\ell}(k)\propto \sqrt{2\ell+1}P_0(k)$ and are not normalized by $\sqrt{N_\mathrm{mocks}}$, i.e. they represent the error on a single survey measurement. The variances decrease slightly with $R_0$ due to the greater binning width, since this includes more $k$-modes (which are uncorrelated in the Gaussian limit) and leads to greater cancellation between the two $D_\ell(kr)$ kernels (Eq.\,\ref{eq: D_function_simple}).}
    \label{fig: qpm_power_spectrum}
\end{minipage}%
\hfill
\begin{minipage}[t]{.48\textwidth}
\vspace{0pt}
  \centering
  \includegraphics[width=\textwidth]{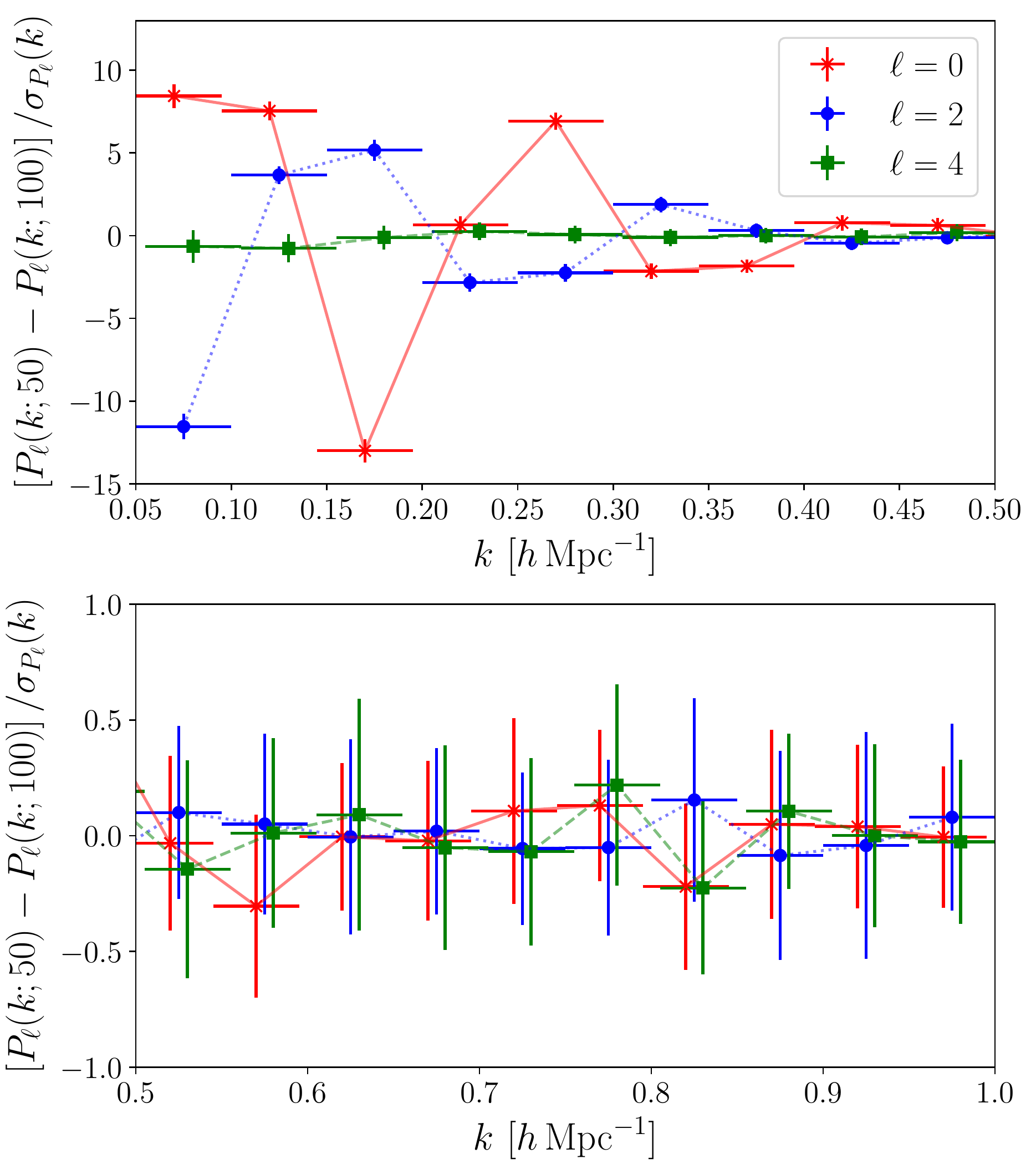}
    \caption{Difference between measured QPM power spectrum multipoles (as in Fig.\,\ref{fig: qpm_power_spectrum}) using pair separation truncation radii of $R_0 = 50h^{-1}\mathrm{Mpc}$ and $R_0 = 100h^{-1}\mathrm{Mpc}$ in units of the power spectrum error $\sigma_{P_\ell}(k)$ for $R_0 = 100h^{-1}\mathrm{Mpc}$. Both datasets are computed using linear $k$-space binning of $\Delta k = 0.05h\,\mathrm{Mpc}^{-1}$ (indicated by horizontal error bars) to allow comparison and vertical error bars show the variances across 100 QPM mocks. A slight horizontal offset between multipoles is included for visualization and both plots show the same results but with different $k$-ranges and vertical scales. Beyond $k\sim 0.4h\,\mathrm{Mpc}^{-1}$ the errors arising from the truncation radius become negligible compared to the statistical errors, with the difference consistent with zero. This is of a similar form to the idealized power ratios of Fig.\,\ref{fig: aniso_power_broadening_log}, but we here elect to plot only the difference in power spectra to avoid large deviations from unity when $P_2(k)\approx 0$, due the the FoG effect. The differences between datasets are expected to arise due to the convolution of the true power with Fourier-space pair-separation windows, $\widetilde{W}(k;R_0)$, of different widths.}
    \label{fig: qpm_R0_comparison}
\end{minipage}
\end{figure}

Fig.\,\ref{fig: qpm_power_spectrum} shows the small-scale power spectrum multipoles for the first QPM mock, as well as fractional errors obtained from 200 independent mocks for both aforementioned truncation radii. We note excellent agreement between the two datasets, which is explored further in Fig.\,\ref{fig: qpm_R0_comparison}. The monopole is seen to decrease to small levels with increasing $k$ as expected, and we do \textit{not} find any evidence for a shot-noise plateau (which would give a positive linear asymptote in $kP_0(k)$ for large $k$), as discussed in Sec.\,\ref{subsec: shot-noise}. In addition, the monopole is measured to extremely high precision, at the sub-percent level for much of the range of $k$, with the error bars in the top plot being too small to see. We note that the quadrupole becomes negative for $k\gtrsim0.3h\,\mathrm{Mpc}^{-1}$, which is a result of the FoG effect which dominates at high $k$. FoG has the effect of convolving the true power spectrum with a smoothing kernel along the LoS, which boosts the $k_\parallel$ power, giving the opposite effect to the Kaiser phenomenon, hence leading to $P_2(k)<0$. In this sample, we note the hexadecapole to be small, as expected, with relatively strong constraints obtained on small scales.

In the ideal survey limit with isotropic (and Gaussian) $P(\vec k)$, the error bars should scale as $\sigma_{P_\ell}\propto\sqrt{2\ell+1}P_0(k)$; in the lower part of Fig.\,\ref{fig: qpm_power_spectrum}, the scaling is found to be roughly accurate, though we note larger errors at high-$k$, likely due to the growing importance of higher $P_\ell(k)$ multipoles. We also note a small increase in the variance of the power spectrum when moving from $R_0 = 50h^{-1}\mathrm{Mpc}$ to $R_0 = 100h^{-1}\mathrm{Mpc}$ which may seem counter-intuitive, since we are utilizing more pairs and obtaining a less well constrained result. This may be rationalized by noting that the bin-spacing ($\Delta k$) is increased by a factor of two between the two datasets (cf.\,Sec.\,\ref{subsec: data-binning}) in order to minimize the correlations between neighbouring $k$-bins. Larger $\Delta k$ implies an average over a larger range of $k$-modes, which, in the Gaussian limit, are independent, leading to a reduced variance at larger $R_0$. If the bin-size were kept constant, we would still expect reduced variance, since the diagonal power would be distributed into nearby bins due to the convolution with the pair-separation window $\widetilde{W}(k;R_0)$, whose breadth increases with $R_0$.

\begin{figure}
    \centering
\includegraphics[width=0.5\textwidth]{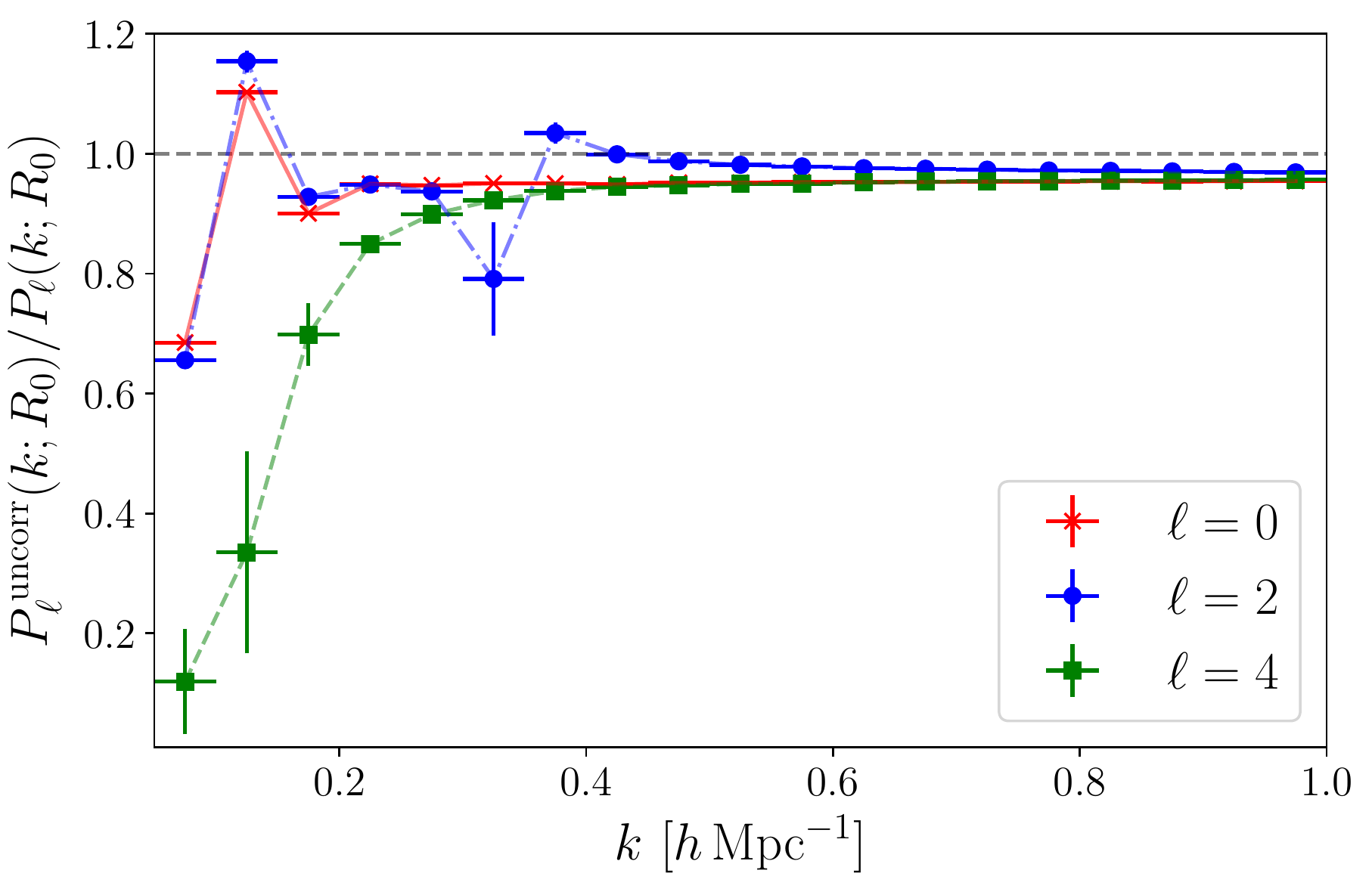}
  \caption{Ratio of small-scale power spectrum multipoles (a) uncorrected for the survey geometry and (b) corrected, incorporating the survey correction function $\Phi$ (Eq.\,\ref{eq: phi_r_mu_defn}), as used in the rest of this paper. We plot the mean and variance of the ratios computed from 20 QPM mocks utilizing a pair-separation truncation radius of $R_0 = 100h^{-1}\mathrm{Mpc}$. Neither set of multipoles have contributions from shot-noise, since we exclude self-counts in our configuration-space estimator. The uncorrected power monopole $P_0^\mathrm{\,uncorr}(k;R_0)$ is equivalent to the FKP power spectrum estimator for large $R_0$ and the higher multipoles are equal to the \citet{2015MNRAS.453L..11B} estimator, except without the \citet{2006PASJ...58...93Y} angle approximations (which are negligible at high-$k$). We note significant survey-window-function effects at small $k$, but no significant effects at $k\gtrsim0.4h\,\mathrm{Mpc}^{-1}$ except for a slight renormalization (with the ratio tending to $\approx0.954$). This is analogous to Fig.\,\ref{fig: phi-correction-plot} except utilizing full simulated data appropriate for BOSS DR12 (and includes additional effects such as the FoG giving different hexadecapole behavior, resulting in the larger deviations between datasets at small $k$).}
  \label{fig: qpm_no_phi_comparison}
\end{figure}

The effect of $R_0$ on the power spectrum multipoles is considered in Fig.\,\ref{fig: qpm_R0_comparison}, where we consider the difference between $P_\ell(k;R_0)$ at the two values of $R_0$, normalizing by the statistical error $\sigma_{P_\ell}(k)$ at the larger $R_0$. This utilizes $\Delta k = 0.05h\,\mathrm{Mpc}^{-1}$ binning for both choices of $R_0$ to allow comparison. The differences between datasets are heuristically similar to those found in the simple forecasts used in Fig.\,\ref{fig: aniso_power_broadening_log}, though we now plot the difference in $P_\ell(k)$ rather than the ratio, to avoid errors when $P_2(k)\approx 0$. On large scales, there is significant differences between the datasets which swamp the statistical error, implying that $R_0 = 50h\,\mathrm{Mpc}^{-1}$ should not be used to measure the $k\lesssim0.3h\,\mathrm{Mpc}^{-1}$ power spectrum. For large $k$ however, we note excellent agreement between the two datasets, with deviations consistent with zero (given the statistical error) similar to that found in Fig.\,\ref{fig: aniso_power_broadening_log}. This clearly demonstrates that $R_0 = 50h^{-1}\mathrm{Mpc}$ is a fine (and fast) approximation for measuring very small-scale power. Differences between the simple forecasts and the behavior of Fig.\,\ref{fig: qpm_R0_comparison} result from the simplifications in the physical model of the former case (which did not include FoG effects) and the binning in $k$-space. 

\subsubsection{Effects of the Survey-Correction Function $\Phi$}
Using the above estimators, we may assess the dependence of the power spectrum multipoles on the survey-correction function $\Phi$, analogous to Sec.\,\ref{sec: survey-correction-integrals}. This (defined as a ratio of $RR$ pair-counts in Eq.\,\ref{eq: phi_r_mu_defn}) was introduced to correct for the non-uniform finite survey geometry and in Fig.\,\ref{fig: phi-correction-plot} was forecasted to significantly effect the mid-$k$ power spectrum of higher multipoles. To fully investigate this, we estimate $P_\ell(k)$ from QPM mocks with $R_0 = 100h^{-1}\mathrm{Mpc}$ in the same manner as before, but do not include the correction function (i.e. set $\Phi(r,\mu) = 1$ for all $r$ and $\mu$). The ratios of these powers are shown in Fig.\,\ref{fig: qpm_no_phi_comparison}, and we note heuristically similar behavior to Fig.\,\ref{fig: phi-correction-plot}, with a large underestimate observed for the hexadecapole, and oscillatory behavior for the lower multipoles whose amplitude falls at large $k$. Here, we note a weaker dependence of the hexadecapole on $\Phi$ than before; this is likely due to the lack of inclusion of FoG effects in the previous forecasting. At large $k$ all ratios are seen to converge to $\approx 0.95$, with the survey-correction function seen to cause an additional slight change in the normalization. Additionally, the error bars show little variation in the uncorrected-to-corrected power ratios with different QPM mocks (except for the small and poorly constrained quadrupole) indicating that this is a true effect of the survey geometry.

\subsubsection{\resub{Comparison with Simple Models of Anisotropy}}\label{subsec: FoG-model}
For this analysis we will adopt the simple model of \citet[Eq.\,21]{2009MNRAS.393..297P} incorporating both Kaiser and FoG effects;
\beq\label{eq: FoG-model}
    P(k,\mu) &=& b^2(1+\beta\mu^2)^2P_\mathrm{NL}(k)F(k,\mu^2)\\\nonumber
    F(k,\mu^2) &=& \left[1+(k\mu\sigma_\mathrm{FoG})^2\right]^{-1}
\eeq
for (local) bias parameter $b$, Kaiser parameter $\beta = f/b$ and FoG velocity dispersion $\sigma_\mathrm{FoG}$. The factor $F(k,\mu^2)$ adds Lorentzian FoG effects \citep{1995MNRAS.275..515C,1998ASSL..231..185H} to the standard Kaiser power spectrum \citep{1987MNRAS.227....1K,2009MNRAS.393..297P}, using a non-linear matter power spectrum $P_\mathrm{NL}(k)$, assuming only linear local galaxy bias and no velocity bias. \resub{We stress that this is included only to give context and show the approximate impacts of the various RSD effects on the power spectrum. They do not represent our best models, which would incorporate velocity biases and higher order loop corrections. For a redshift-space power spectrum model accurate to $k=0.4h\,\mathrm{Mpc}^{-1}$, see \citet{2017JCAP...07..002H}.}

The non-linear power spectrum is computed via the `halofit' prescription of \citet{2015MNRAS.454.1958M} using the cosmology $\{\Omega_b = 0.048, \Omega_\Lambda = 0.71, h = 0.7\}$ (matching that used by the \citet{2014MNRAS.437.2594W} Quick Particle Mesh (QPM) BOSS DR12 mock catalogs). In the analysis of \citet{2016MNRAS.460.4188G} using the CMASS QPM mocks, the values of $b,\beta,\sigma_{FoG}$ were constrained to $b = 1.74\pm 0.03$, $\beta = 0.64\pm 0.05$ and $\sigma_\mathrm{FoG} = 3.35\pm 0.32$ in a somewhat more complex model; we adopt these parameters here as a rough estimate. From Eq.\,\ref{eq: FoG-model}, we may compute the Legendre moments of the model spectrum via the standard integral
\beq
    P_\ell(k) = \frac{2\ell+1}{2}\int_{-1}^1d\mu\,P(k,\mu)L_\ell(\mu);
\eeq 
here we use the full forms tabulated in \citet{2018JCAP...02..039L}.

\begin{figure}
    \centering
    \includegraphics[width=\textwidth]{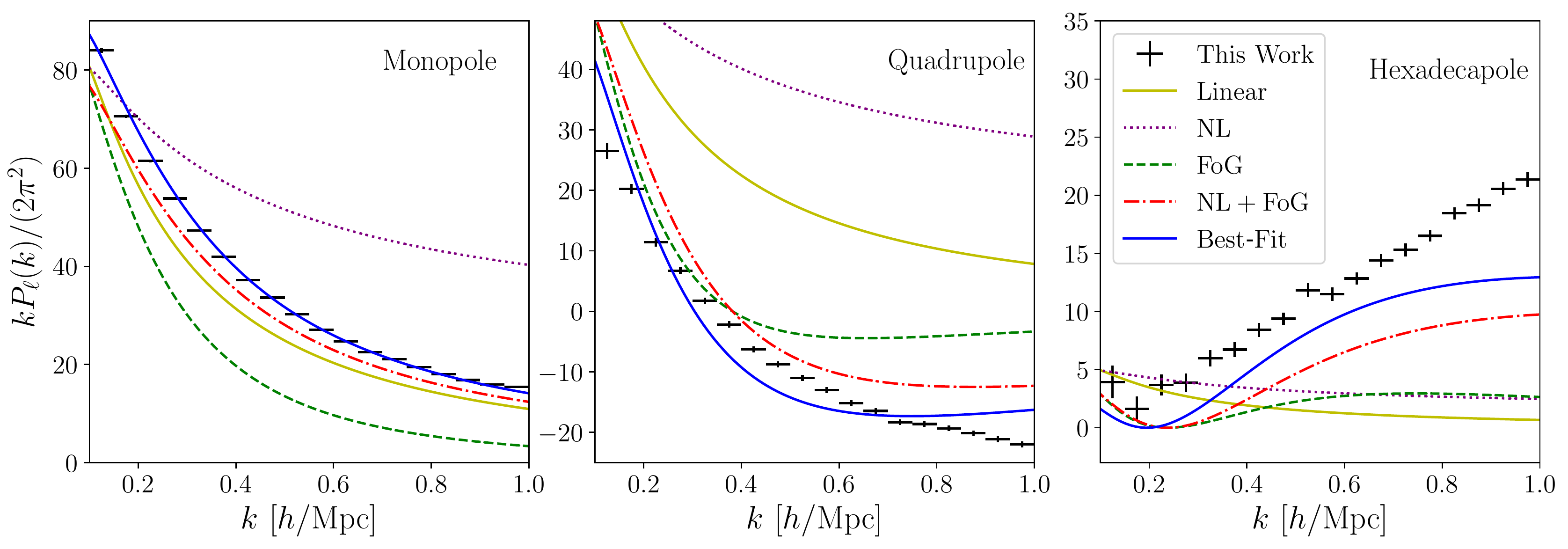}
\caption{Comparison of observed (crosses) and model (lines) small-scale power spectrum multipoles (for $\ell\in\{0,2,4\}$) with a variety of simple theoretical RSD models. \resub{These models are presented only for context; other models likely provide more accuracy.} Observational multipoles are from the QPM results shown in Fig.\,\ref{fig: qpm_power_spectrum} with $R_0 = 100h^{-1}\mathrm{Mpc}$, \resub{as discussed in Sec.\,\ref{subsec: aniso-power-dat}}. The linear power spectrum is computed from \texttt{CAMB} \citep{Lewis:2002ah} using cosmology appropriate for the QPM mocks at redshift $z = 0.57$. Redshift space distortions (RSD) are added to all models via the Kaiser prescription for (local) bias and $\beta$ parameters 1.74 and 0.64 respectively. The `NL' models include a non-linear power spectrum from `halofit' \citep{2015MNRAS.454.1958M} at the same redshift and `FoG' models add a Lorentzian Finger of God kernel function (Eq.\,\ref{eq: FoG-model}) with $\sigma_\mathrm{FoG} = 3.35$. All parameters are taken from the DR12 QPM analysis of \citet{2016MNRAS.460.4188G}. The `best-fit' model includes both non-linear and FoG effects using a new parameter fit. We note that our simple model does not exactly match that used in the DR12 analysis and is clearly inadequate at higher multipoles (as expected).}
    \label{fig: qpm_FoG_theory}
\end{figure}

In Fig.\,\ref{fig: qpm_FoG_theory} we plot the $R_0 = 100h^{-1}\mathrm{Mpc}$ power spectrum estimates for a single QPM mock (as above) alongside a variety of models based on Eq.\,\ref{eq: FoG-model}. These use the aforementioned DR12 analysis parameters, assessing the effects of power spectrum non-linearities and the FoG phenomenon (simply tested by using a linear power spectrum and setting $\sigma_\mathrm{FoG}=0$ respectively). For the monopole spectrum we note surprisingly good agreement between the data and the `fiducial' spectrum (simply a linear Kaiser spectrum), with non-linearities and FoG having opposing effects on $P_0(k)$ which almost cancel. This highlights the use of the higher multipoles in constraining cosmological parameters. In the quadrupole, models without FoG are seen to significantly overestimate the measured power (additionally constraining it to be everywhere positive), with the best approximation given by the inclusion of both FoG and non-linearity. Similar effects are seen for the hexadecapole, with a characteristic minimum in $kP_4(k)$ around $0.2h\,\mathrm{Mpc}^{-1}$ caused by FoG.

Note that the best-fit parameters from the DR12 analysis may not match those required by our somewhat simpler model here; to fully assess how the power spectrum can be fitted by Eq.\,\ref{eq: FoG-model}, we must optimize for the parameters $\chi = \{b, \beta, \sigma_\mathrm{FoG}\}$. This is done by minimizing the negative log likelihood
\beq
    -\log\mathcal{L}(\chi) = \frac{1}{2}\left(\mathbf{P}^T_\mathrm{model}(\chi)\mathbf{\Psi}_\mathrm{QPM}\mathbf{P}_\mathrm{model}(\chi)\right) + \mathrm{const.}
\eeq
where $\mathbf{P}_\mathrm{model}$ is the (stacked) vector of power spectrum multipoles and $\mathbf{\Psi}_\mathrm{QPM}$ is the QPM precision matrix (see Sec.\,\ref{subsec: QPM-cov-matrices}) which is independent of the modeling parameters $\chi$. The errors on the derived parameters $\chi^*$ may be estimated via a standard Fisher forecast giving error on the $i$-th parameter
\beq
    \sigma_{\chi_i} \approx \sqrt{\left(\mathbb{F}^{-1}\right)_{ii}}
\eeq
for Fisher matrix
\beq
    \mathbb{F}_{ij} = \left.\frac{\partial \mathbf{P}^T_\mathrm{model}}{\partial \chi_i}\mathbf{\Psi}_\mathrm{QPM}\frac{\partial \mathbf{P}_\mathrm{model}}{\partial \chi_j}\right|_{\chi = \chi^*}
\eeq
which has a simple form since the precision matrix is independent of the model parameters in this case. Using the $R_0 = 100h^{-1}\mathrm{Mpc}$ dataset and the (noisy) QPM precision matrix discussed in Sec.\,\ref{subsec: QPM-cov-matrices}, we obtain the optimal parameters $b^* = 1.95\pm 0.01$, $\beta^* = 0.48\pm 0.02$ and $\sigma_\mathrm{FoG}^* = 3.54\pm 0.02$, significantly different from the DR12 analysis due to the simpler model applied here. This model is also shown in Fig.\,\ref{fig: qpm_FoG_theory} and gives excellent agreement with the monopole power spectrum up to $k = 1h\,\mathrm{Mpc}^{-1}$. The quadrupole spectrum is broadly in agreement up to $k\sim 0.5h\,\mathrm{Mpc}^{-1}$ and we note an underestimation of the hexadecapole at all scales. This indicates that we require a more sophisticated model to fit the higher-order power multipoles, such as Effective Field Theory. The quoted errors above are very small; this is a result of the very small statistical errors on the monopole and quadrupole. Note that (a) this does \textit{not} imply that our model is correct, and these are an underestimate since we have not included systematic errors arising from, for example, the pair-separation window function bias, which is non-negligible at small $k$, and (b) the statistical error bars are slightly biased by noise in the precision matrix.

\subsection{Covariance Matrix Estimates}\label{subsec: QPM-cov-matrices}
The covariance matrix of the above anisotropic power estimates is here computed in two manners; (a) by numerically computing the covariance across the 200 individual QPM power estimates and (b) using the shot-noise rescaled Gaussian prescription of Sec.\,\ref{sec: power-cov}. Whilst it is possible to get far less noisy estimates of the covariance by the latter approach, it is not certain \textit{a priori} how strongly we are biased by the exclusion of non-Gaussian terms at high-$k$.

\subsubsection{QPM Covariance Matrix}
\begin{figure}%
    \centering
    \subfloat[$R_0 = 50h^{-1}\mathrm{Mpc}$, $\Delta k = 0.1h\,\mathrm{Mpc}^{-1}$]{{\includegraphics[width=0.4\textwidth]{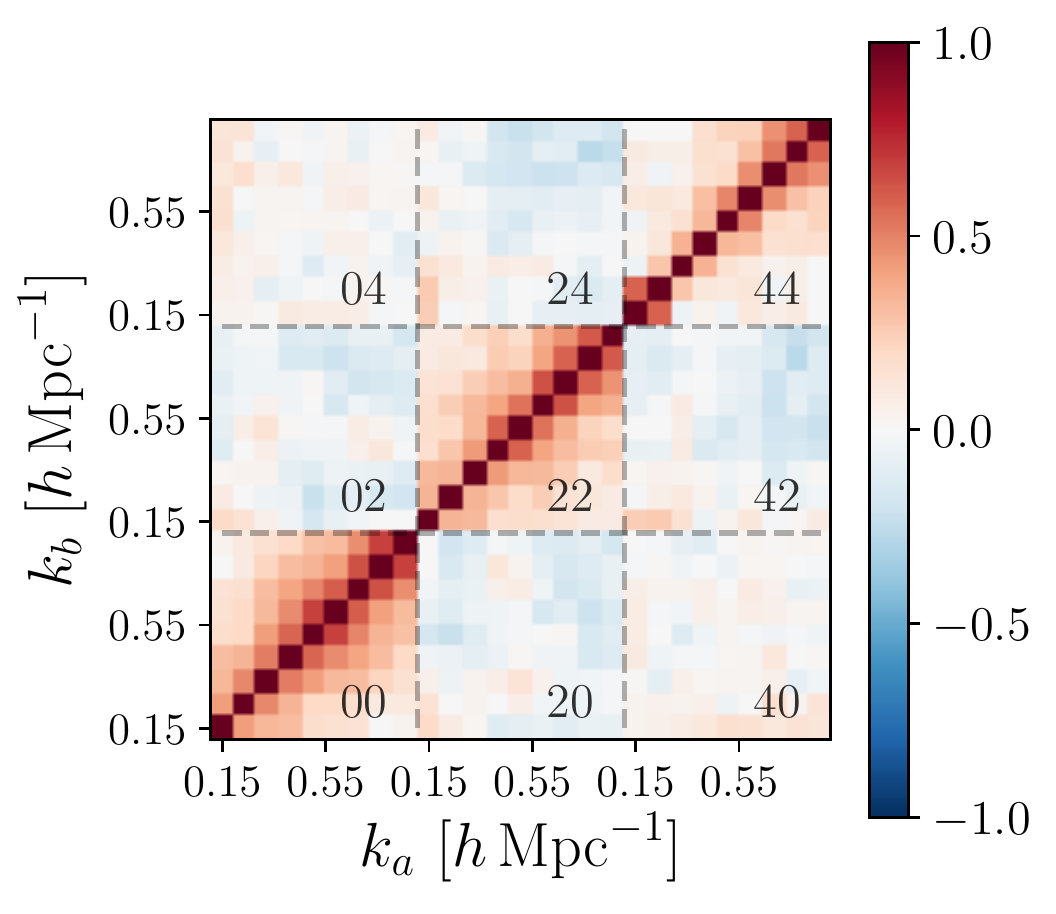} }}%
    \subfloat[$R_0 = 100h^{-1}\mathrm{Mpc}$, $\Delta k = 0.05h\,\mathrm{Mpc}^{-1}$]{{\includegraphics[width=0.4\textwidth]{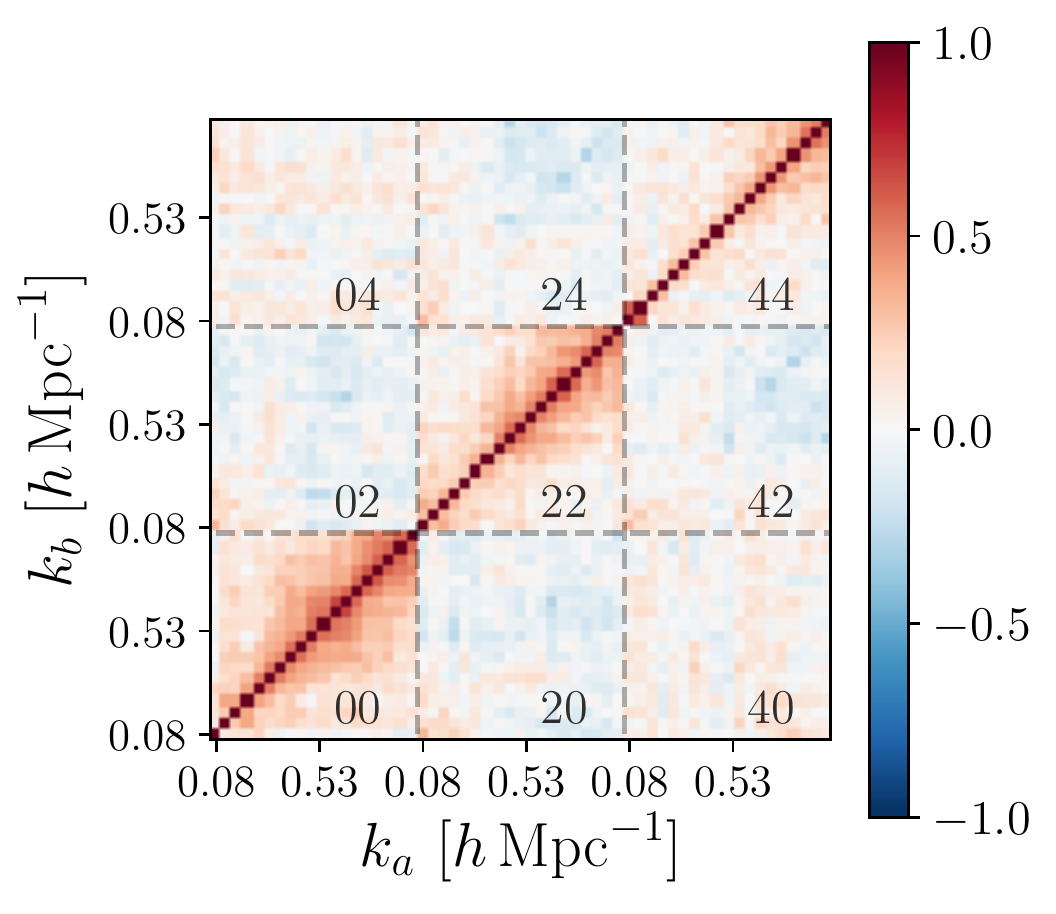} }}%
    \caption{Correlation matrix for small-scale power spectrum multipoles measured from 200 QPM mocks using the configuration-space power estimator (Eq.\,\ref{eq: windowed_pair_counts}) with pair-counts truncated at two radii $R_0$ (with associated powers shown in Fig.\,\ref{fig: qpm_power_spectrum}). The correlation matrix is defined by Eq.\,\ref{eq: corr_matrix_def} and we display all combinations of Legendre multipoles in each (symmetric) figure. The dotted lines demarcate submatrices for different pairs of Legendre multipoles (e.g. $(02)$ indicates the cross-covariance between $\ell = 0$ and $\ell = 2$) with a total of 9 (19) radial bins in each for $R_0 = 50$ ($100$) $h^{-1}\mathrm{Mpc}$. We note negative correlations with the $\ell = 2$ multipoles and significant off-diagonal correlations increasing to larger $k$ due to the survey selection function, the pairwise window function and non-Gaussianity.}%
    \label{fig: corr_matrices}%
\end{figure}

Denoting an estimate of the $p$-th multipole power in $k$-space bin $a$ from mock $i$ as $\hat P^a_{p,(i)}$, the QPM covariance matrix in radial bins $a,b$ and Legendre moments $p,q$ obtained from $N_\mathrm{mocks}=200$ mocks is defined as
\beq
    \mathbf{C}_{pq}^{ab} = \frac{1}{N_\mathrm{mocks}-1}\sum_{i=1}^{N_\mathrm{mocks}}\left[\hat P^a_{p,(i)}-\overline{P}^a_p\right]\left[\hat P^b_{q,(i)}-\overline{P}^b_q\right]
\eeq
for power spectrum mean
\beq
    \overline{P}^a_p = \frac{1}{N_\mathrm{mocks}}\sum_{i=1}^{N_\mathrm{mocks}}\hat P^a_{p,(i)}.
\eeq
This may be used to find the \textit{correlation matrix} $\mathbf{R}^{ab}_{pq}$ which is defined by 
\beq\label{eq: corr_matrix_def}
    \mathbf{R}_{pq}^{ab} = \frac{\mathbf{C}_{pq}^{ab}}{\sqrt{\mathbf{C}_{pp}^{aa}\mathbf{C}_{qq}^{bb}}}
\eeq
with unity along the $p=q$, $a=b$ leading diagonal by construction. This is shown for the two choice of $R_0$ in Fig.\,\ref{fig: corr_matrices}, using the $\Delta k = 5/R_0$ binning as before, to ensure minimal cross-talk between $k$-bins induced by the pair-separation window function. Notably we observe a significant positive correlation between all $k$-bins at the same Legendre multipole especially at larger $k$; this may result from non-Gaussian terms (cf. trispectrum terms discussed in Sec.\,\,\ref{eq: ideal-covariance}), the effects of the non-trivial survey geometry or additional mixing between $k$-bins induced by the pair-separation window $\widetilde{W}(k;R_0)$. In addition, we note small negative correlations between the quadrupole and other multipoles and very weak correlations between the hexadecapole and monopole.

\begin{figure}%
    \centering
    \subfloat[$R_0 = 50h^{-1}\mathrm{Mpc}$]{{\includegraphics[width=0.4\textwidth]{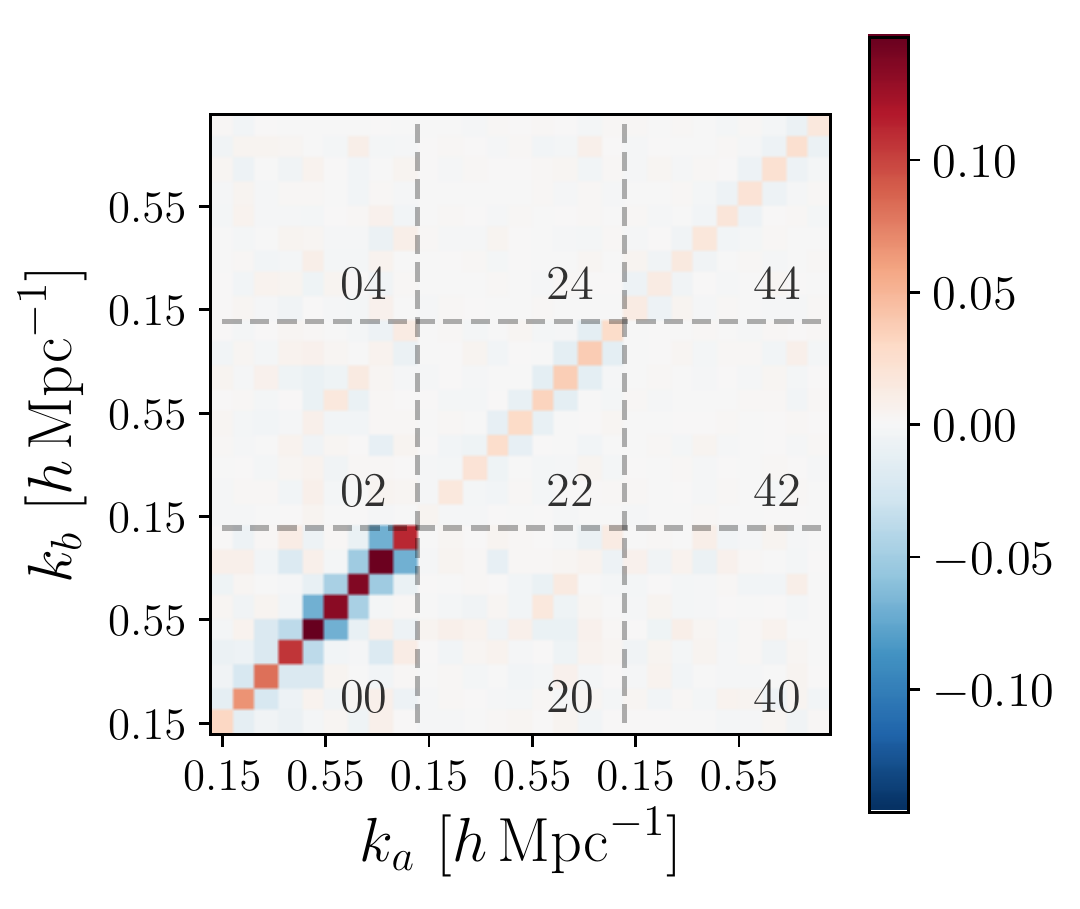} }}%
    \subfloat[$R_0 = 100h^{-1}\mathrm{Mpc}$]{{\includegraphics[width=0.4\textwidth]{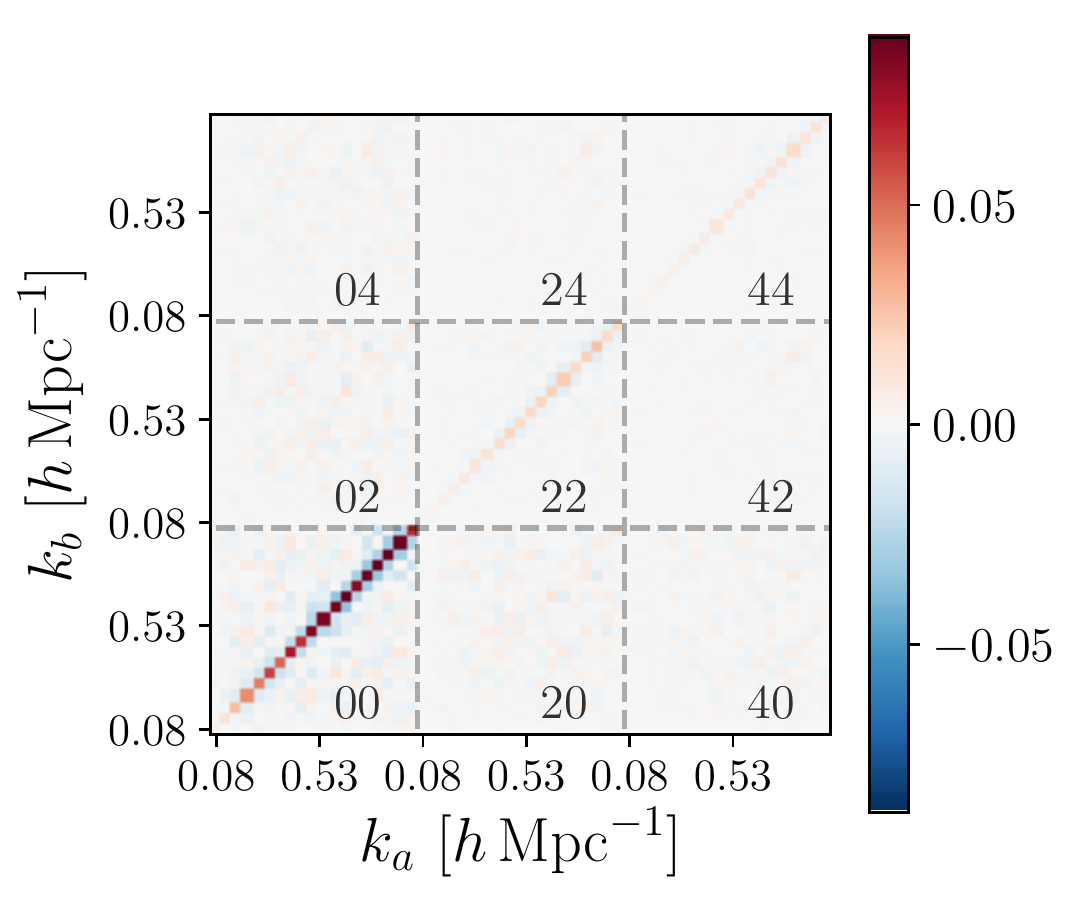} }}%
    \caption{Sample QPM precision matrices $\mathbf{\Psi}_{pq}^{ab}$ (defined by Eq.\,\ref{eq: precision_matrix_def}) for the two covariance matrices shown in Fig.\,\ref{fig: corr_matrices}. These are divided by $k_1k_2$ for ease of visualization.These use 200 mocks as before and we note a strong diagonal and next-to-diagonal term for $p=q$ but little other power. These can be used in Fisher forecasting to find the constraints on derived model parameters such as RSD parameters.}
    \label{fig: prec_matrices}%
\end{figure}

The precision matrix is formally defined as the inverse of the covariance matrix, yet, in the limit of finite mocks, this introduces a bias to the estimate which causes bias in the derived parameter variances. Assuming Wishart noise \citep{wishart28}, we apply the \citet{2007A&A...464..399H} correction factor to obtain the QPM precision matrix estimate
\beq\label{eq: precision_matrix_def}
    \mathbf{\Psi}_{pq}^{ab} = \frac{N_\mathrm{mocks}-N_\mathrm{bins}-2}{N_\mathrm{mocks}-1}\left(\mathbf{C}_{pq}^{ab}\right)^{-1}
\eeq
where $N_\mathrm{bins}$ is the total number of bins across all multipoles. This is shown in Fig.\,\ref{fig: prec_matrices} in the same fashion as above, and has an approximately tridiagonal form, with strong positive diagonal contributions for $p = q$ and significant negative next-to-diagonal terms but little additional power. As expected, the precision matrix elements are largest for the monopole as this is known most precisely (cf.\,Fig.\,\ref{fig: qpm_power_spectrum}). This precision matrix may be used to compute Fisher forecasts, as in Sec.\,\ref{subsec: FoG-model}.

\subsubsection{Theoretical Covariance Matrix}
\begin{figure}%
    \centering
    \subfloat[Two-point, ${}^2\Omega_{pq}(r)$]{{\includegraphics[width=0.4\textwidth]{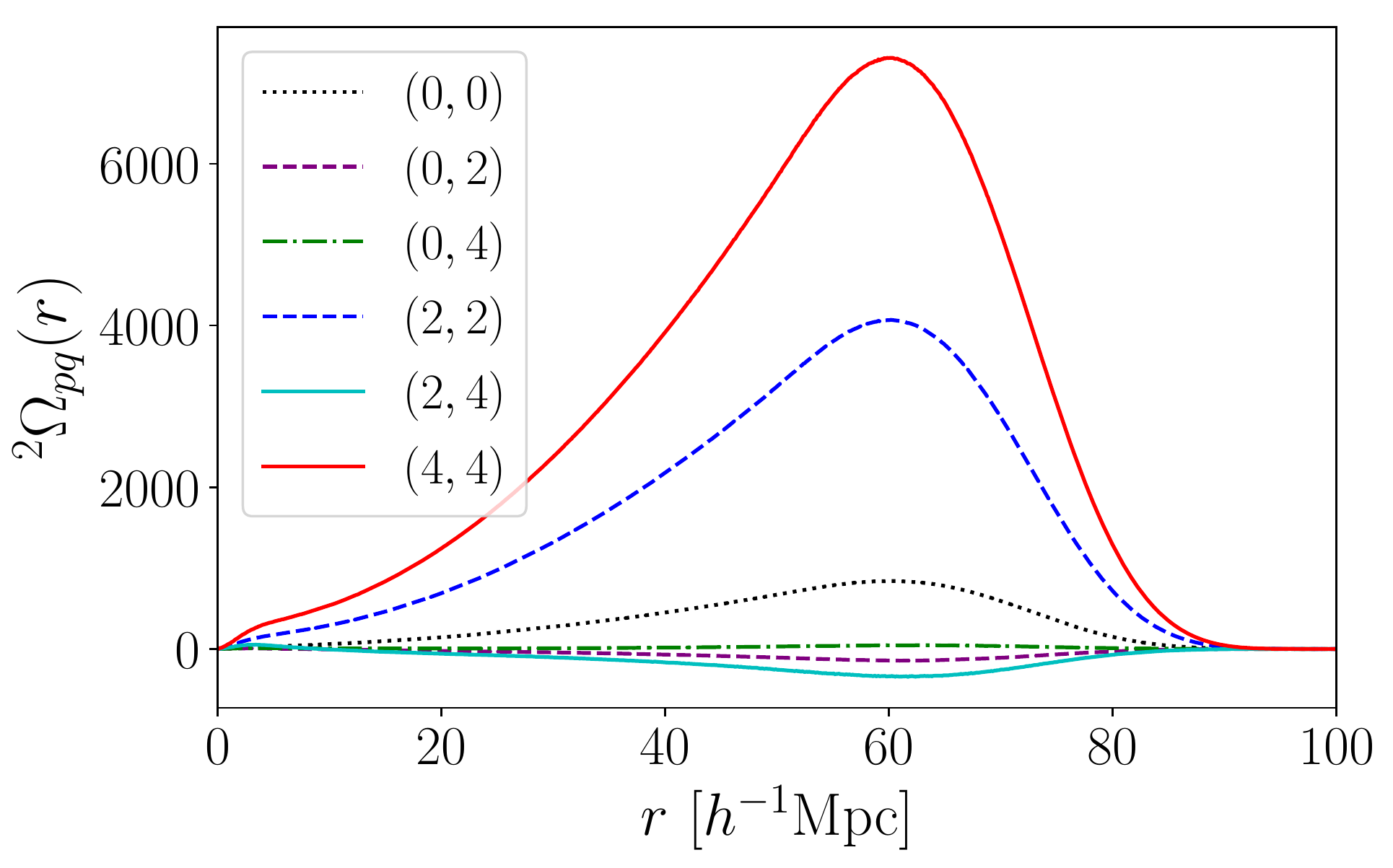} }}%
    \subfloat[Three-point, ${}^3\Omega_{pq}(r_1,r_2)$]{{\includegraphics[width=0.3\textwidth]{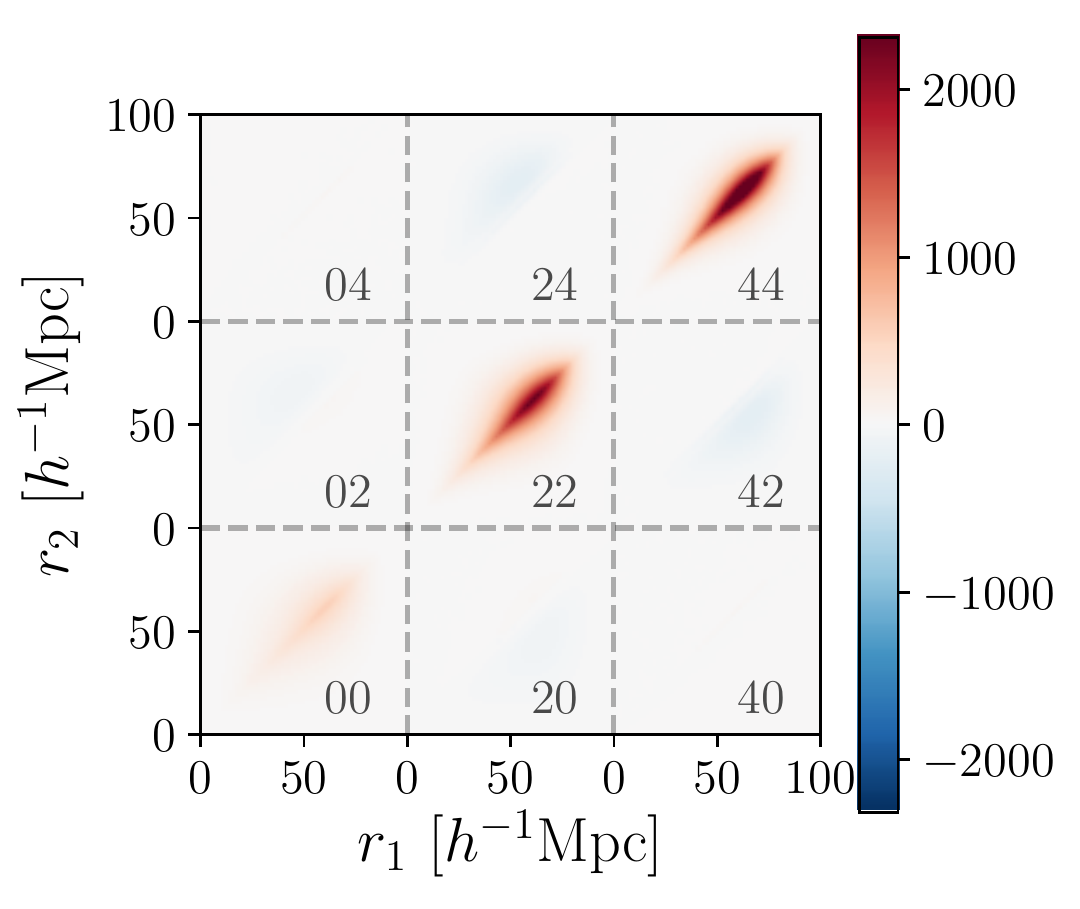} }}%
    \subfloat[Four-point, ${}^4\Omega_{pq}(r_1,r_2)$]{{\includegraphics[width=0.3\textwidth]{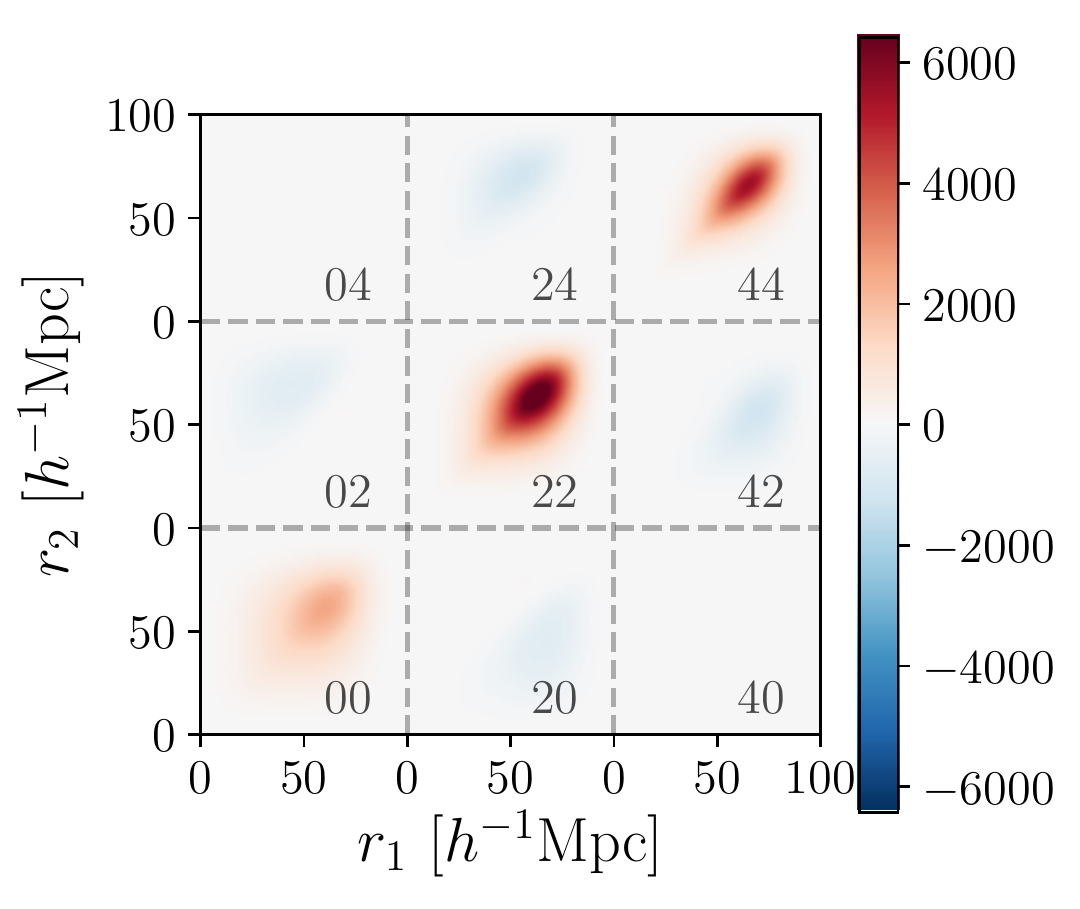} }}%
    \caption{Gaussian covariance unnormalized probability density functions (PDFs) for the two-, three- and four-point power spectrum covariance matrices at $R_0 = 100h^{-1}\mathrm{Mpc}$, as defined in Eqs.\,\ref{eq: omega2-def}\,\&\,\ref{eq: Omega34def}, here using $\Delta k = 0.1h^{-1}\mathrm{Mpc}$ radial bins. These are used to compute the expected rescaled-Gaussian power spectrum covariance matrices via Eqs.\,\ref{eq: cov2-pdf-expression}\,\&\,\ref{eq: cov34-pdf-expression}, shown in Fig.\,\ref{fig: gaussian_pdf_plots}. Physically these are the distribution functions of pair-wise separations, weighted by the survey correction functions, pair-separation windows, number densities and correlation functions, such that the covariance matrices can be defined as a integral over these with a $k$-dependent kernel. The two-point integral is shown in the leftmost plot, which is diagonal for each set of Legendre indices with lines labelled by $(p,q)$ representing the ${}^2\Omega_{pq}(r)$ function. The ${}^3\Omega_{pq}$ and ${}^4\Omega_{pq}$ matrices are plotted in the same style as Figs.\,\ref{fig: corr_matrices}\,\&\,\ref{fig: prec_matrices} except in configuration-space.}
    \label{fig: gaussian_pdf_plots}%
\end{figure}

To compute the theoretical rescaled-Gaussian covariances, we adopt the implementation described in Sec.\,\ref{subsec: power-cov-implementation}, first computing the weighting functions $\left\{{}^2\Omega_{pq}(u),{}^3\Omega_{pq}(u_1,u_2),{}^4\Omega_{pq}(u_1,u_2)\right\}$ across the domain $[0,R_0]$. This is performed using a modified version of the \texttt{RascalC} as previously noted, and we produce estimates of the $\Omega_{pq}$ functions (Eqs.\,\ref{eq: omega2-def}\,\&\,\ref{eq: Omega34def}) in 1000 radial bins with $\Delta r = 0.1h^{-1}\mathrm{Mpc}$ regularly spaced in $[0,R_0]$, computed for $\sim 10^{13}$ quads of particles in $\sim 200$ CPU-hours. The two-, three- and four-point weighting functions for all combinations of Legendre moments $(p,q)$ are shown in Fig.\,\ref{fig: gaussian_pdf_plots} (and we note that these appear to be smooth, with $\Omega_{pq}(u)\rightarrow0$ as $u \rightarrow R_0$ for all functions, as expected. Notably, the $\Omega$ functions are large and positive for $p=q$ elements, with negative terms arising for combinations of the quadrupole with other multipoles.

\begin{figure}%
    \centering
\subfloat[Two-point, ${}^2C_{pq}^{ab}\times k_ak_b$]{{\includegraphics[width=0.32\textwidth]{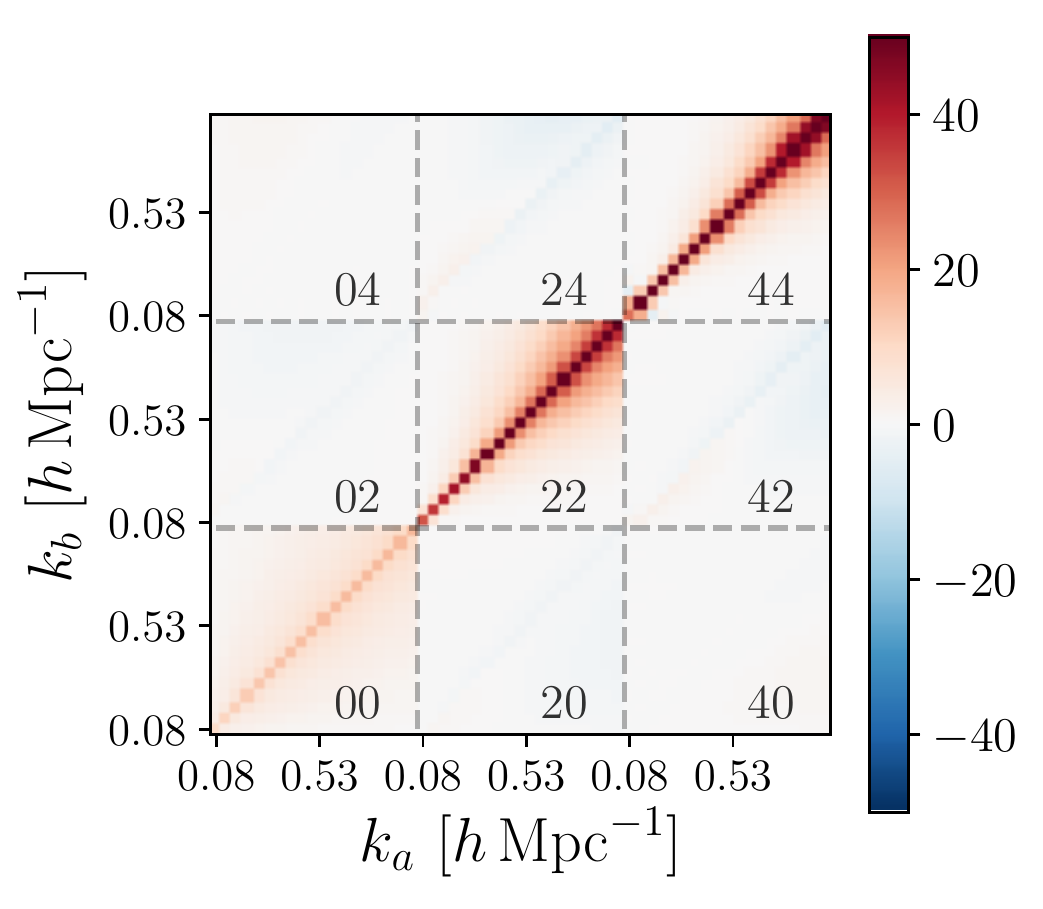} }}%
    \subfloat[Three-point, ${}^3C_{pq}^{ab}\times k_ak_b$]{{\includegraphics[width=0.32\textwidth]{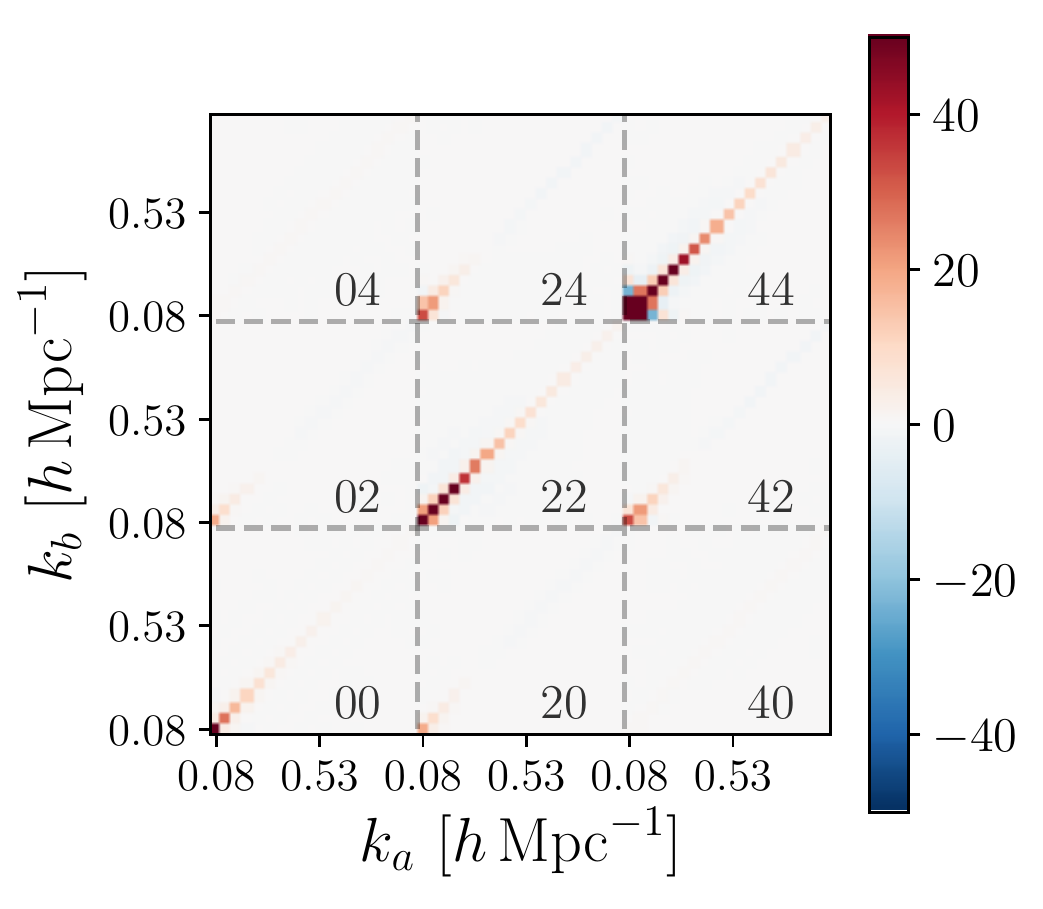} }}%
    \subfloat[Four-point, ${}^4C_{pq}^{ab}\times k_ak_b$]{{\includegraphics[width=0.32\textwidth]{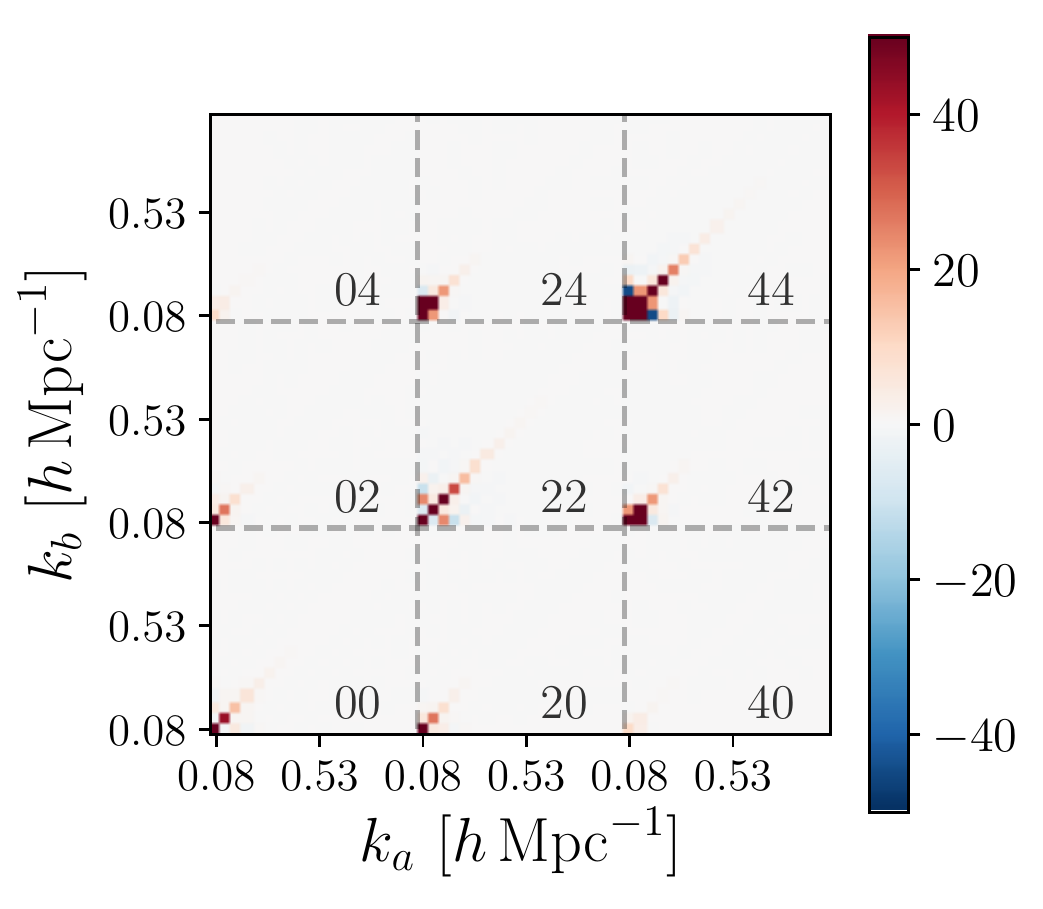} }}%
\caption{Two-, three- and four-point theoretical Gaussian covariance matrices, as defined in Eq.\,\ref{eq: power-cov-matrix-full}, here using $R_0 = 100h^{-1}\mathrm{Mpc}$ and 19 $k$-bins linearly spaced in $[0.15,1]$ (matching that of Fig.\,\ref{fig: corr_matrices}). Matrices are multiplied by $k_ak_b$ for visibility. The individual terms are given by integrals of the 2PCF $\xi$ and kernel functions $A^a_p(r)$ over two to four copies of the survey. Two- and three-point terms arise from shot-noise contributions to the covariance. Here, they are computed from integrating $k$-space kernel functions over the smooth distributions functions of Fig.\,\ref{fig: gaussian_pdf_plots}, according to Eqs.\,\ref{eq: cov2-pdf-expression}\,\&\,\ref{eq: cov34-pdf-expression}. These may be summed via Eq.\,\ref{eq: power-cov-matrix-gaussian} to form a full estimate of the covariance matrix, with shot-noise rescaling parameter $\alpha$ encapsulating some non-Gaussianity. The three- and four-point covariance terms are seen to dominate at small $k$, with largest contributions on very small scales from the two-point term. We note clear off-diagonal terms here resulting from the non-uniform survey geometry and the pair-separation window function.}
    \label{fig: gaussian_cov_plots}%
\end{figure}

These are combined via Eqs.\,\ref{eq: cov2-pdf-expression}\,\&\,\ref{eq: cov34-pdf-expression} to form the full Gaussian covariance matrix components, which are displayed in Fig.\,\ref{fig: gaussian_cov_plots}. Since the $\Omega_{pq}$ matrices are well converged, the integrals over $r$ are approximated simply by summations; for less well converged matrices one could fit $\Omega_{pq}$ to smooth functions before performing the $k$-space kernel weighted integration. Although the four-point terms are dominant on large scales (small $k$), the two- and three-point covariance matrix terms are the leading contributions at large $k$. The latter terms are sourced by shot-noise contractions (appearing in the covariances not the power spectra) which are expected to dominate on small-scales when the power is subdominant. In addition, we note significant off-diagonal power in the two-point matrix and (at small $k$) the higher-point matrices. Since these do not include non-Gaussian terms, this is expected to arise from $k$-bin mixing or survey geometry effects, not from any intrinsic 4PCF terms.

The full theoretical covariance matrix is found from Eq.\,\ref{eq: power-cov-matrix-gaussian}, with a shot-noise rescaling parameter $\alpha$ allowing inclusion of some non-Gaussianity. Here the parameter is constrained by fitting the theoretical to QPM covariance matrix; one could also compute this from a single survey via jackknife approaches \citep[cf.\,][]{2019MNRAS.487.2701O,rascalC}. The two matrices are compared via the $\mathcal{L}_1$ likelihood
\beq\label{eq: L1-likelihood}
    -\log\mathcal{L}_1(\alpha) = 2 D_{KL}\left(\Psi(\alpha),\mathbf{C}_\mathrm{QPM}\right) = \operatorname{trace}[\Psi(\alpha)\mathbf{C}_\mathrm{QPM}] - \log\det \mathbf{C}_\mathrm{QPM} - \log\det \Psi(\alpha) - N_\mathrm{bins}
\eeq
where $D_{KL}$ is the Kullback-Leibler (KL) divergence \citep{kullback1951}. Note that this depends on the theoretical precision matrix; since this is computed at low noise, we can assume $\Psi(\alpha) = C^{-1}(\alpha)$ with minimal error. Numerical minimization of this likelihood gives an optimal shot-noise rescaling parameter $\alpha^* = 1.13$.

\subsubsection{Covariance Matrix Comparison}
A simple method by which to compare the QPM and theory covariance matrices is the KL divergence used in Eq.\,\ref{eq: L1-likelihood}. As shown in \citet[Appendix D]{rascalC}, the expected KL divergence between the two matrices is approximately given by
\beq
    \av{D_{KL}(\Psi,\mathbf{C}_\mathrm{QPM})} \approx \frac{N_\mathrm{bins}(N_\mathrm{bins}+1)}{4N_\mathrm{mocks}} 
\eeq
(assuming a smooth precision matrix). Here, the true KL divergence is found to be $5.3$ compared to an expected value of $4.3$. Since this only becomes exact in the limit of $N_\mathrm{mocks}\gg N_\mathrm{bins}$ we conclude that the matrices appear broadly consistent at this level.

\begin{figure}
\centering
\begin{minipage}[t]{.48\textwidth}
\vspace{0pt}
  \centering
  \includegraphics[width=0.95\textwidth]{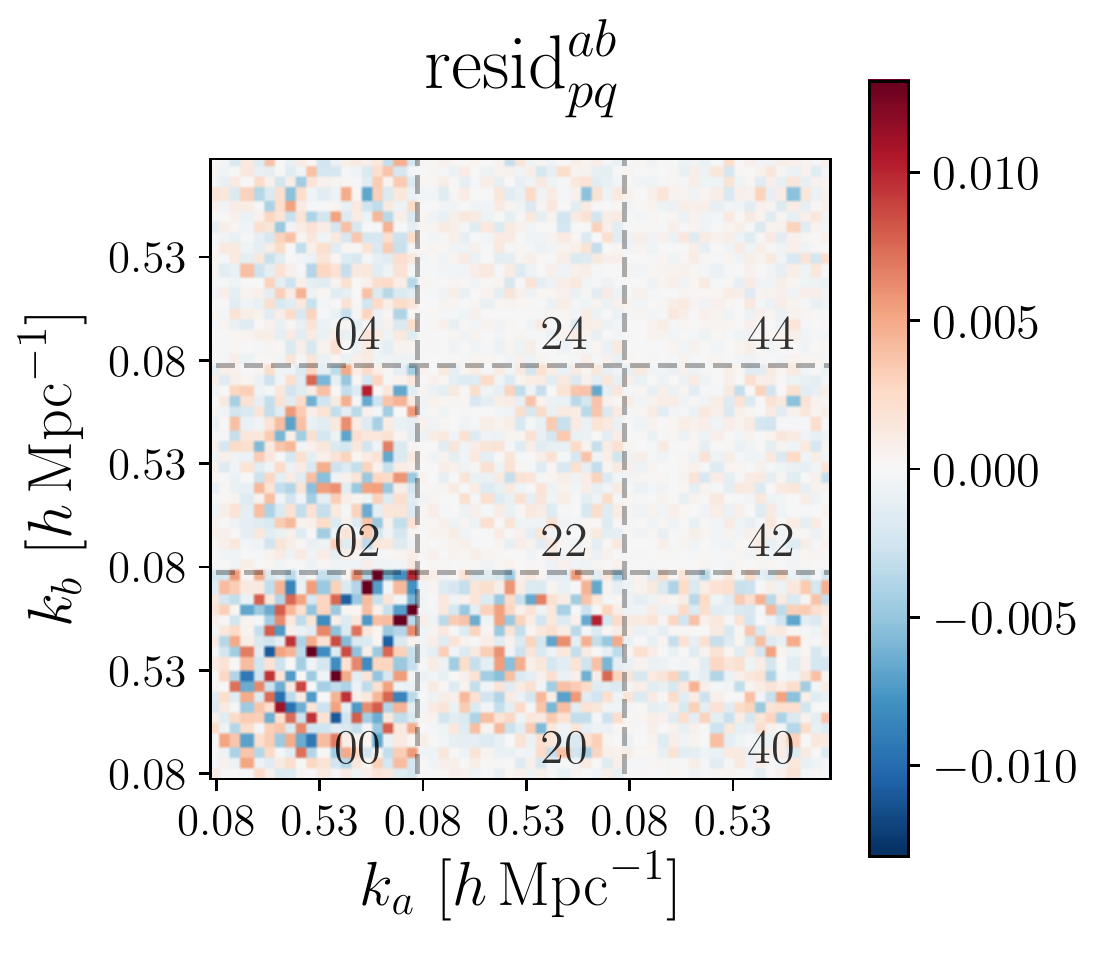}
  \caption{Residual matrix between the QPM precision matrix (Fig.\,\ref{fig: prec_matrices}) and the theoretical precision matrix computed from the shot-noise-rescaled Gaussian terms of Fig.\,\ref{fig: gaussian_cov_plots}, i.e. Eq.\,\ref{eq: residual-prec-matrix}, using shot-noise rescaling $\alpha = 1.13$. The residual is normalized by $k_ak_b$ for visualization and uses $R_0 = 100h^{-1}\mathrm{Mpc}$ and $\Delta k = 0.05h\,\mathrm{Mpc}^{-1}$, displaying all cross-covariances for $\ell\in\{0,2,4\}$ as before. The residual matrix appears mostly consistent with noise here. A more obvious comparison of these covariances is presented in Fig.\,\ref{fig: gaussian_disc_matrix}.}
    \label{fig: prec_matrix_difference}
\end{minipage}%
\hfill
\begin{minipage}[t]{.48\textwidth}
\vspace{0pt}
  \centering
  \includegraphics[width=0.95\textwidth]{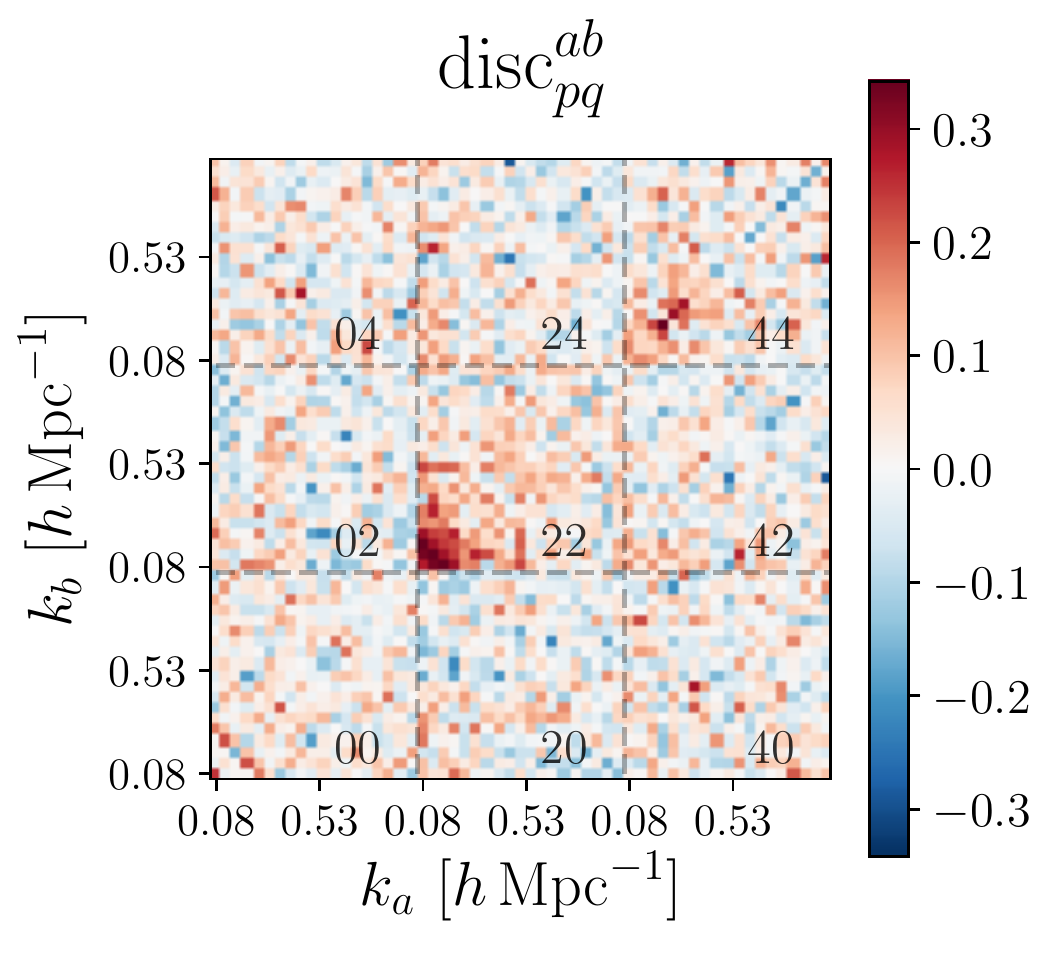}
  \caption{Discriminant matrix encapsulating the differences between theory (Fig.\,\ref{fig: gaussian_cov_plots} with $\alpha = 1.13$) and simulation (Figs.\,\ref{fig: corr_matrices}\,\&\,\ref{fig: prec_matrices}) covariance matrices for QPM mocks, as defined in Eq.\,\ref{eq: disc-matrix-def}. For identical covariance matrices, the discriminant matrix will be equal to zero in the limit of zero noise; systematic differences from zero indicate biases in the theoretical model. We note no obvious biases in the monopole matrices, with a slight positive bias in the two-point matrix at small $k$ and a small diagonal bias at large $k$ in the $p=q=4$ matrix. This could be removed using an improved model for non-Gaussianity in the covariance matrix, for example hierarchical models.}
  \label{fig: gaussian_disc_matrix}
\end{minipage}
\end{figure}

A more graphic method is to consider the matrix residual, given by the rescaled difference between the precision matrices;
\beq\label{eq: residual-prec-matrix}
    \mathrm{resid}^{ab}_{pq} = \frac{\left(\Psi(\alpha^*) - \mathbf{\Psi}_\mathrm{QPM}\right)^{ab}_{pq}}{k_ak_b}.
\eeq
Any significant departures of this from zero indicate systematic differences between the matrices. Note that we here compare the precision matrices rather than the covariances since these are more relevant cosmologically. Fig.\,\ref{fig: prec_matrix_difference} displays this matrix and the behavior appears largely consistent with noise, with larger fluctuations observed for the $\ell = 0$ multipoles as expected, since these include the dominant power in the precision matrices. We note a slight underestimation of the $p=q=4$ matrix diagonal in the theoretical matrix which is likely due to the lack of inclusion of non-Gaussianity in the model. At this level of noise, it is difficult to see any major systematic differences between matrices.

An additional test uses the `discriminant' matrix which we define as
\beq\label{eq: disc-matrix-def}
    \mathrm{disc}^{ab}_{pq} = \left(\sqrt{\Psi(\alpha^*)}^T\mathbf{C}_\mathrm{QPM}\sqrt{\Psi(\alpha^*)}-\mathbb{I}\right)^{ab}_{pq}
\eeq
where $\sqrt{\Psi(\alpha^*)}$ indicates the Cholesky factorization of the theoretical precision matrix. For identical covariance matrices in the zero-noise limit, we expect $\Psi(\alpha^*) = \mathbf{C}_\mathrm{QPM}^{-1}$ thus $\mathrm{disc}^{ab}_{pq} = \mathbf{0}$. Any systematic deviations from zero indicate shortcomings in our theoretical matrices. This is plotted in Fig.\,\ref{fig: gaussian_disc_matrix} for the same matrices as before, and we note no significant departures from zero for the monopole covariances implying that these terms are well estimated. For the $p=q=2$ quadrupole autocovariance there is a clear bias for small $k$, indicating that our model is insufficient there. This likely arises from non-Gaussianity since the non-Gaussian terms only appear in the three- and four-point theoretical covariance matrices which dominate at small $k$. We additionally note a slight negative diagonal bias for $p=q=4$. Overall, we note remarkable agreement between the rescaled-Gaussian theory and QPM mock covariance matrices for most monopoles, with the exception of the $k\lesssim 0.3h\,\mathrm{Mpc}^{-1}$ quadrupole autocovariance. Increasing the model complexity is expected to further improve the fit; a simple approach to this is via hierarchical 3PCF and 4PCF models \citep[e.g.][]{1975ApJ...196....1P,1977ApJ...217..385G} that add an additional free hierarchical parameter $Q$ which could be fit to mocks or jackknifes.

\section{Bispectrum Estimators}\label{sec: bispectrum-estimators}
We conclude by outlining estimators for the isotropic bispectrum in a similar fashion to the above, with computation possible via weighted triple-counting. We defer considerations of the bispectrum covariance and power spectrum - bispectrum cross-covariance to future work, although note that this may be derived as for the power spectrum and estimated using small modifications to the 3PCF covariance algorithm of \citet{3pcfCov}.

\subsection{Idealized Estimator}
We start with the definition of the bispectrum as the transform of the three-point correlation function (3PCF) $\zeta$;
\beq\label{eq: bispectrum-from-3pcf}
    \delta^{(D)}(\vec k_1+\vec k_2+\vec k_3)\,B(\vec k_1,\vec k_2,\vec k_3) \equiv B(\vec k_1,\vec k_2) =  \int d^3\vec x_1\,d^3\vec x_2\,e^{i[\vec k_1\cdot\vec x_1+\vec k_2\cdot\vec x_2]}\zeta(\vec x_1,\vec x_2)
\eeq
for Dirac delta function $\delta^{(D)}$ enforcing the triangle condition on the wavevector $\vec k_i$. On the right hand side, this condition has been used, and we write the 3PCF as a function of two triangle sides, $\vec x_1$ and $\vec x_2$. Using the \citet{1998ApJ...494L..41S} 3PCF estimator, we may proceed analogously to the power spectrum case, writing  
\beq
    B(\vec k_1,\vec k_2) = \int d^3\vec x_1\,d^3\vec x_2\,e^{i[\vec k_1\cdot\vec x_1+\vec k_2\cdot\vec x_2]}\frac{NNN(\vec x_1,\vec x_2)}{RRR(\vec x_1,\vec x_2)}
\eeq
for triple-counts $NNN$ and $RRR$. These are defined as
\beq
    RRR(\vec x_1,\vec x_2) &\equiv& \int  \left\{\prod_{i=1}^3 d^3\vec r_i\,n(\vec r_i)w(\vec r_i)\right\} \left[\delta^{(D)}(\vec x_1 - (\vec r_1-\vec r_3))\delta^{(D)}(\vec x_2-(\vec r_2-\vec r_3)) + \left(\text{5 perms. of $\{\vec r_1, \vec r_2, \vec r_3\}$}\right) \right]\\\nonumber
    NNN(\vec x_1,\vec x_2) &\equiv& \int \left\{\prod_{i=1}^3 d^3\vec r_i\,n(\vec r_i)w(\vec r_i)\delta(\vec r_i)\right\}\left[\delta^{(D)}(\vec x_1 - (\vec r_1-\vec r_3))\delta^{(D)}(\vec x_2-(\vec r_2-\vec r_3)) + \left(\text{5 perms. of $\{\vec r_1, \vec r_2, \vec r_3\}$}\right) \right].
\eeq
For thin $k$-space bins $(a,b)$ with center $(\vec x_a, \vec x_b)$ and size $\delta \vec x$, in the ideal survey limit (with uniform $n$, $w$) we obtain;
\beq
    RRR^{ab}&\approx&RRR(\vec x_a,\vec x_b)\left(\delta\vec x\right)^2 =(nw)^3\int d^3\vec r_1\,d^3\vec r_2\,d^3\vec r_3\,\left[\Theta^{a}(\vec r_1-\vec r_3)\Theta^{b}(\vec r_2-\vec r_3) + \left(\text{5 perms.}\right) \right]\\\nonumber
    &=& 6V(nw)^3 \left(\delta \vec x\right)^2
\eeq
using binning functions $\Theta^c(\vec x)$ which are unity if $\vec x$ is in bin $c$ and zero else. As before, this permits the modeling of a general $RRR$ count as
\beq
    RRR(\vec x_1,\vec x_2) = \frac{6V\overline{(nw)^3}}{\Phi(\vec x_1,\vec x_2)}
\eeq
for survey correction factor $\Phi(\vec x_1,\vec x_2)$ now depending on three points in space (parametrized by two separation vectors). Analogous to the power spectrum definition, we may write the bispectrum as 
\beq\label{eq: bispectrum}
    B(\vec k_1,\vec k_2) &=& \frac{1}{6V\overline{(nw)^3}}\int \left\{\prod_{i=1}^3 d^3\vec r_i\,n(\vec r_i)w(\vec r_i)\delta(\vec r_i)\right\}\left[\Phi(\vec r_1-\vec r_3,\vec r_2-\vec r_3)e^{i\left[\vec k_1\cdot(\vec r_1-\vec r_3)+\vec k_2\cdot(\vec r_2-\vec r_3)\right]} + \left(\text{5 perms.}\right)\right]\\\nonumber
    &=& \frac{1}{V\overline{(nw)^3}}\int \left\{\prod_{i=1}^3 d^3\vec r_i\,n(\vec r_i)w(\vec r_i)\delta(\vec r_i)\right\}\Phi(\vec r_1-\vec r_3,\vec r_2-\vec r_3)e^{i\left[\vec k_1\cdot(\vec r_1-\vec r_3)+\vec k_2\cdot(\vec r_2-\vec r_3)\right]}
\eeq
integrating over the Dirac delta functions in the $NNN$ definition to set $\vec x_1, \vec x_2$ in $\Phi$, and noting that the second line follows since $\Phi(\vec r_i-\vec r_j,\vec r_i-\vec r_k)$ is symmetric under any permutation of $\{i,j,k\}$. Here we note that, in the limit of an ideal survey, where $\Phi=1$ everywhere, this is identical to the standard estimator, which can be written as
\beq\label{eq: bispectrum_conventional}
    \delta^{(D)}(\vec k_1+\vec k_2+\vec k_3)\,B(\vec k_1, \vec k_2) &=& \widetilde{F}_3(\vec k_1)\widetilde{F}_3(\vec k_2)\widetilde{F}_3(\vec k_3)\\\nonumber
    F_3(\vec r) &=& \left(I_3\right)^{-1/3}n(\vec r)w(\vec r)\delta(\vec r)\\\nonumber
    I_3 &=& \int d^3\vec r\,n^3(\vec r)w^3(\vec r)\equiv V\overline{(nw)^3}.
\eeq
\citep[e.g.][]{2012PhRvD..86f3511F,2013PhRvD..88f3512S,2015PhRvD..92h3532S}. The two vectors $\vec k_1$ and $\vec k_2$ fully describe the bispectrum geometry; for the isotropic bispectrum estimator we parametrize simply by the lengths $k_1, k_2$ and the Legendre polynomial of the angle between them. For $k$-bins $a,b$ (of volume $v_a$, $v_b$) and Legendre moment $\ell$ the estimator becomes
\beq
    B_\ell^{ab} &=& \frac{1}{v_av_b}\int d^3\vec k_1\,d^3\vec k_2\,B(\vec k_1,\vec k_2)\Theta^a(|\vec k_1|)\Theta^b(|\vec k_2|)L_\ell(\hat{\vec k}_1\cdot\hat{\vec k}_2)\\\nonumber 
    &=& \frac{1}{V\overline{(nw)^3}v_av_b}\int d^3\vec k_1\,d^3\vec k_2\,\Theta^a(|\vec k_1|)\Theta^b(|\vec k_2|)L_\ell(\hat{\vec k}_1\cdot\hat{\vec k}_2)\int \prod_{i=1}^{3}\left[d^3\vec r_i\,n(\vec r_i)w(\vec r_i )\delta(\vec r_i)\right]\Phi(\vec r_1-\vec r_3,\vec r_2-\vec r_3)e^{i\vec k_1\cdot \vec r_1}e^{i\vec k_2\cdot \vec r_2}e^{-i(\vec k_1+\vec k_2)\cdot\vec r_3}.
\eeq
inserting Eq.\,\ref{eq: bispectrum} in the second line. Analogously to the anisotropic power spectrum estimator (Sec.\,\ref{sec: ideal-power-estimators}), we consider the kernel function
\beq\label{eq: bispectrum_kernel_init}
    A^{ab}_\ell(\vec r_1,\vec r_2;\vec r_3) &\equiv& \frac{1}{v_av_b}\int d^3\vec k_1\,d^3\vec k_2 \Theta^a(|\vec k_1|)\Theta^b(|\vec k_2|)L_\ell(\hat{\vec k}_1\cdot\hat{\vec k}_2)\\\nonumber
    &\approx& (-1)^\ell P_\ell(\hat{\vec x}\cdot\hat{\vec y}) j_\ell(k_ax)j_\ell(k_by)
\eeq
using the approximate form derived in appendix \ref{appen: bispectrum-deriv} for $\vec x = \vec r_1-\vec r_3$ and $\vec y = \vec r_2-\vec r_3$. The full form is also given in appendix \ref{appen: bispectrum-deriv}, depending again on generalized hypergeometric functions or Sine integrals. 

We thus arrive at a closed form for the bispectrum integral depending only on the particle separations and bins;
\beq\label{eq: bispectrum_integral}
    B^{ab}_\ell = \frac{1}{V\overline{(nw)^3}}\int \prod_{i=1}^{3}\left[d^3\vec r_i\,n(\vec r_i)w(\vec r_i )\delta(\vec r_i)\right]\Phi(\vec r_1-\vec r_3,\vec r_2-\vec r_3)A^{ab}_\ell(\vec r_1,\vec r_2;\vec r_3)\Phi(\vec r_1-\vec r_3,\vec r_2-\vec r_3)
\eeq
which may be computed by triple-counting data and randoms analogous to the \citet{1998ApJ...494L..41S} 3PCF estimator. The bispectrum estimator is thus
\beq
    \hat B^{ab}_\ell = \frac{\widetilde{NNN}^{ab}_\ell - 3\widetilde{NNR}^{ab}_\ell + 3\widetilde{NRR}^{ab}_\ell - \widetilde{RRR}^{ab}_\ell}{V\overline{(nw)^3}},
\eeq
defining the generalized triple-count $\widetilde{XYZ}^{ab}_\ell$ over fields $X, Y, Z \in \{D,R\}$ as
\beq\label{eq: generalized triple-count}
    \widetilde{XYZ}^{ab}_\ell = \sum_{i\in X}\sum_{j \in Y^*}\sum_{k\in Z^*} w_iw_jw_kA_\ell^{ab}(\vec r_i,\vec r_j;\vec r_k)\Phi(\vec r_i-\vec r_k,\vec r_j-\vec r_k)
\eeq
where the asterisks indicate that we exclude self-counts for identical fields (i.e. $i\neq j$ if $X=Y$ etc.). Note that we can easily distribute the integral over the \citet{1998ApJ...494L..41S} estimator since the denominator is independent of spatial coordinates, following our definition of $\Phi$. 

\subsection{Window Functions}
In practice, we must include a kernel function to allow for efficient triple-counting for the large-scale bispectrum as for the power spectrum (Sec.\,\ref{sec: windows}). To require that all 3 sides of the bispectrum triangle be small, we simply require that the pair-separation between two pairs of points in the triangle be small; the third side is constrained via the triangle inequality. Using the asymmetric triple-count integral of Eq.\,\ref{eq: bispectrum_integral}, we simply insert the Kaiser window function $W_{R_0}$ between the two sides $\vec r_1-\vec r_3$ and $\vec r_2-\vec r_3$ constrained from the $A_\ell^{ab}(\vec r_1,\vec r_2;\vec r_3)$ kernel, giving a windowed form
\beq
    B^{ab}_\ell = \frac{1}{V\overline{(nw)^3}}\int \prod_{i=1}^{3}\left[d^3\vec r_i\,n(\vec r_i)w(\vec r_i )\delta(\vec r_i)\right]\Phi(\vec r_1-\vec r_3,\vec r_2-\vec r_3)A^{ab}_\ell(\vec r_1,\vec r_2;\vec r_3)\Phi(\vec r_1-\vec r_3,\vec r_2-\vec r_3)W(\vec r_i-\vec r_k;R_0)W(\vec r_j-\vec r_k;R_0).
\eeq
This can be naturally be incorporated into the triple-count estimator by inserting the factor $W(\vec r_i-\vec r_k,R_0)W(\vec r_j-\vec r_k;R_0)$ into Eq.\,\ref{eq: generalized triple-count}.

To see the effect on the measured bispectrum, consider the windowed estimator for $B(\vec k_1,\vec k_2;R_0)$;
\beq
    B(\vec k_1,\vec k_2;R_0) &=& \int \prod_{i=1}^3\left[d^3\vec r_i n(\vec r_i)w(\vec r_i)\delta(\vec r_i)\right]\frac{d^3\vec p_1\,d^3\vec p_2}{(2\pi)^6}\,e^{i(\vec k_1-\vec p_1)\cdot(\vec r_1-\vec r_3)}e^{i(\vec k_2-\vec p_2)\cdot(\vec r_2-\vec r_3)}\Phi(\vec r_1-\vec r_3,\vec r_2-\vec r_3)\widetilde{W}(\vec p_1;R_0)\widetilde{W}(\vec p_2;R_0)\\\nonumber
    &=& \int \frac{d^3\vec p_1\,d^3\vec p_2}{(2\pi)^6}\, B(\vec k_1-\vec p_1, \vec k_2-\vec p_2)\widetilde{W}(\vec p_1;R_0)\widetilde{W}(\vec p_2;R_0),
\eeq
which is simply a double convolution of $B$ with $\widetilde{W}_{R_0}$. As before, the windowed bispectrum Legendre moments, which are simply the Legendre moments of the double convolution of $B$ with $\widetilde{W}$. As for the power spectrum, the inclusion of the pair-separation window function allows for fast computation of the high-$k$ isotropic bispectrum multipoles and we note that the algorithm has complexity $\mathcal{O}(Nn^2R_0^6)$, since we must count all $N$ primary particles but only the secondary and tertiary particles in volume $4\pi R_0^3/3$. 

\section{Summary and Outlook}\label{sec: conclusion}
In this work we have derived a new estimator for the anisotropic small-scale power spectrum, based on computing weighted pair-counts for data and random catalogs in configuration-space, with no need for explicit Fourier transforms. Truncating the pair-counts at radius $R_0$ for speed, our estimator has complexity $\mathcal{O}(NnR_0^3)$ and may be applied analogously to standard 2PCF estimators. We showed that the truncation error in the monopole (higher multipoles) at $R_0\sim 100h^{-1}\mathrm{Mpc}$ with optimal $k$-space binning is negligible for $k\gtrsim0.2h\,\mathrm{Mpc}^{-1}$ ($0.4h\,\mathrm{Mpc}^{-1}$). The main benefits of our estimator, which has been made publicly available,\footnote{HIPSTER: HIgh-k Power SpecTrum EstimatoR (\href{https://HIPSTER.readthedocs.io}{HIPSTER.readthedocs.io})} are as follows;
\begin{itemize}
    \item \textbf{Efficiency}: The computation time for FFT-based small-scale power increases with the size of the wavevector $\vec k$. Our algorithm requires fewer pairs to be counted at larger $k$, thus its efficiency \textit{increases} at smaller scales. For logarithmic binning in $k$, we may use a truncation radius $R_0\sim 1/k$ and the computation time scales as $\mathcal{O}(R_0^3) = \mathcal{O}(k^{-3})$. By combining an FFT-based approach with the methods presented herein (with some overlap at moderate $k$) we can efficiently compute power across all wavenumbers, \resub{obtaining much faster sampling at high-$k$ than FFT-based approaches}.
    \item \textbf{Boundary-Correction}: Unlike the standard power spectrum estimators, our approach accounts for the non-uniform survey geometry (as in the \citet{1993ApJ...412...64L} 2PCF estimator) via a survey-correction function $\Phi$, defined as the ratio of expected and true $RR$ pair-counts. This allows power spectra to be estimated without the need for deconvolution in post-processing. Although surveys are large, the window functions typically include significant small-scale power due to excluded regions (e.g. stars and fiber collisions) thus this remains important at high $k$.
    \item \textbf{Self-counts}: Since the power spectra are computed in configuration-space we may exclude all galaxy (and random) self-counts, \resub{allowing us to correctly compute the Fourier transform of the correlation function, avoiding any high-$k$ plateau.} This is seen to be true in simulations, and simplifies analysis, since any non-Poissonian shot-noise is not well understood.
\end{itemize}

The estimator has been tested on mock galaxy simulations appropriate to the BOSS DR12 dataset and the dependencies on the survey geometry and truncation radius explored. Notably the survey geometry has strong effects on the intermediate- and large-scale multipoles, thus our correction function is important to include. Fitting to established simple (tree-level) models of RSD shows excellent agreement with our monopole power, with higher multipoles requiring more complex models. Additionally, we have considered the shot-noise-rescaled Gaussian theoretical covariance matrix (analogous to \citealt{2016MNRAS.462.2681O}) which is found to be in good agreement with the sample covariance for large $k$. We have also described the extension to the bispectrum, which is computed in a similar fashion, requiring triple-counts at complexity $\mathcal{O}(Nn^2R_0^6)$. We note a variety of avenues of future work based on this method;
\begin{itemize}
    \item \textbf{Application to Survey Data}: In this paper, we have focused solely on BOSS DR12-like simulations, which do not include complexities such as systematic weights from fiber-collisions etc. Their inclusion will result in an updated weighting scheme but little further modification.
    \item \textbf{Comparison with Detailed Models}: The power spectrum estimates can be used to test extended Perturbation Theory models on the smallest scales and rigidly constrain RSD and expansion parameters. This will additionally require consideration of the systematic errors of our approach.
    \item \textbf{N-body Simulations}: Our $P_\ell(k)$ estimator can be well applied to cubic simulations, computing spectra free from discretization errors without using large, memory-intensive meshes. It may also be extended to large scales (and incorporated into one of the many efficient pair-counting routines), using sub-sampling to ensure manageable computation times. This will be discussed further in upcoming work.
    \item \textbf{Bispectrum Application and Covariances}: The presented bispectrum estimators may be applied to data as for the power spectrum, and full theoretical autocovariance matrices, as well as power spectrum - bispectrum cross-covariances may be computed in the same manner as above. 
    \item \textbf{Anisotropic Bispectra}: We may similarly extend the methods to the anisotropic bispectrum. This may also be computed via triple-counting, using a modified kernel yet little additional computational effort. The algorithm retains complexity of $\mathcal{O}(Nn^2R_0^6)$, although we note that the anisotropic bispectrum depends on five parameters rather than three.
    \item \textbf{Joint Statistical Analyses}: To extract maximal information from survey data we should use multiple statistics in concert. In the simplest case, we can use the 2PCF and power spectrum jointly, with analysis made possible via the cross-covariances discussed above. These may be similarly extended to higher order statistics, such as cross-covariances between the 3PCF and power spectrum or the 2PCF and bispectrum.
\end{itemize}
The methods presented above thus provide an exciting avenue into a wide variety of applications, allowing small-scale spectra to be estimated robustly in a fraction of the previous computational time.

\section*{Acknowledgements}
We thank Lehman Garrison \resub{and the anonymous referee} for insightful comments. OHEP acknowledges funding from the Herchel-Smith foundation. DJE is supported by U.S. Department of Energy grant DE-SC0013718 and as a Simons Foundation Investigator.

Some of the computations in this paper were run on the Odyssey cluster supported by the FAS Division of Science, Research Computing Group at Harvard University. Funding for SDSS-III has been provided by the Alfred P. Sloan Foundation, the Participating Institutions, the National Science Foundation, and the U.S. Department of Energy Office of Science. The SDSS-III web site is http://www.sdss3.org/.

SDSS-III is managed by the Astrophysical Research Consortium for the Participating Institutions of the SDSS-III Collaboration including the University of Arizona, the Brazilian Participation Group, Brookhaven National Laboratory, Carnegie Mellon University, University of Florida, the French Participation Group, the German Participation Group, Harvard University, the Instituto de Astrofisica de Canarias, the Michigan State/Notre Dame/JINA Participation Group, Johns Hopkins University, Lawrence Berkeley National Laboratory, Max Planck Institute for Astrophysics, Max Planck Institute for Extraterrestrial Physics, New Mexico State University, New York University, Ohio State University, Pennsylvania State University, University of Portsmouth, Princeton University, the Spanish Participation Group, University of Tokyo, University of Utah, Vanderbilt University, University of Virginia, University of Washington, and Yale University.

\bibliographystyle{mnras}
\bibliography{adslib,otherlib} 



\appendix

\section{Derivation of the Anisotropic Pair-Count Kernel}\label{appen: aniso-power-deriv}
Below, we outline the derivation of the pair-count kernel, $A^a_\ell$, for the configuration-space anisotropic power estimator. Starting from Eq.\,\ref{eq: aniso-power-kernel-init};
\beq\label{eq: aniso-power-kernel-init-appen}
    A_\ell^a(\vec r_i\,,\vec r_j) \equiv \frac{2\ell+1}{V_\mathrm{shell}}\int d^3\vec k\,\Theta^a(|\vec k|)e^{i\vec k\cdot\vec u}L_\ell(\measuredangle[\vec k, \vec x]),
\eeq
we may use the plane-wave expansion $e^{i\vec x\cdot \vec y} = \sum_{\ell=0}^\infty (2\ell+1)\,i^{\ell}j_\ell(xy)L_\ell(\hat{\vec x}\cdot \hat{\vec y})$ \citep[Eq.\,16.63]{arfken2013mathematical} and the Legendre polynomial decomposition $L_\ell(\hat{\vec x}\cdot\hat{\vec y}) = {4\pi}(2\ell+1)^{-1}\sum_{m=-\ell}^\ell Y^{}_{\ell m}(\hat{\vec x})Y_{\ell m}^*(\hat{\vec y})$ for spherical harmonics $Y_{\ell m}$ \citep[Eq.\,14.30.9]{nist_dlmf} to give
\beq
    A_\ell^a(\vec r_i\,,\vec r_j) &=& \frac{2\ell+1}{V_\mathrm{shell}} \int k^2dk \Theta^a(k)\sum_{\ell'}(2\ell'+1)\,i^{\ell'}j_{\ell'}(ku)L_{\ell'}(\hat{\vec k}\cdot\hat{\vec u})L_\ell(\hat{\vec k}\cdot\hat{\vec x})\\\nonumber
    &=& \frac{(4\pi)^2}{V_\mathrm{shell}} \int k^2dk\,d\Omega_k \Theta^a(k)\sum_{\ell'}\sum_{m'}\sum_{m}i^{\ell'}j_{\ell'}(ku)Y^{}_{\ell'm'}(\hat{\vec k})Y^{*}_{\ell' m'}(\hat{\vec u})Y^{*}_{\ell m}(\hat{\vec k})Y^{}_{\ell m}(\hat{\vec x}).
\eeq
Via spherical harmonic orthonormality, $\int d\Omega_k\,Y^{}_{\ell'm'}(\hat{\vec k})Y^{*}_{\ell m}(\hat{\vec k}) = \delta_{\ell \ell'}\delta_{mm'}$ \citep[Eq.\,14.30.8]{nist_dlmf}, thus
\beq
    A_\ell^a(\vec r_i\,,\vec r_j) &=& i^\ell\frac{(4\pi)^2}{V_\mathrm{shell}} \int k^2dk\, j_\ell(ku)\Theta^a(k)\sum_{m} Y^{*}_{\ell m}(\hat{\vec u})Y^{}_{\ell m}(\hat{\vec x}) = 4\pi i^\ell L_\ell(\hat{\vec x}\cdot \hat{\vec u})\frac{(2\ell+1)}{V_\mathrm{shell}} \int k^2dk\,\Theta^a(k)j_\ell(ku)
\eeq
utilizing the Legendre polynomial decomposition in reverse. In the narrow bin limit $V_\mathrm{shell}\approx4\pi k_a^2\Delta k$ and the kernel simplifies to
\beq\label{eq: simple_aniso_kernel2}
    A_\ell^a(\vec r_i\,,\vec r_j) \approx (-1)^{\ell/2} (2\ell+1) L_\ell(\vec x\cdot \vec u) j_\ell(k_au),
\eeq
(for even $\ell$). In general, our expression is more complex, and a general form may be found in terms of generalized hypergeometric functions (e.g. using \texttt{Mathematica}), via the definition
\beq\label{eq: hypergeom_indefinite_integral}
    D_\ell(ku) \equiv u^3\int k^2dk\,j_\ell(ku) = \frac{\sqrt{\pi}}{2^{2+\ell}}(ku)^{3+\ell}\Gamma\left(\frac{3+\ell}{2}\right){}_1\widetilde{F}_2\left[\left(\tfrac{3+\ell}{2}\right),\left(\tfrac{3}{2}+\ell,\tfrac{5+\ell}{2}\right),-\tfrac{1}{4}(ku)^2\right]
\eeq
where $\Gamma(x)$ is the Gamma function and ${}_1\widetilde{F}_2$ is a regularized generalized hypergeometric function, given by 
\beq\label{eq: D_function_general}
    {}_1\widetilde{F}_2\left[(a)\,,(b_1,b_2),x\right] \equiv \frac{1}{\Gamma(b_1)\Gamma(b_2)}\sum_{n=0}^\infty \frac{(a)_n}{(b_1)_n(b_2)_n}\frac{x^n}{n!} = \sum_{n=0}^\infty \frac{(a)_n}{\Gamma(b_1+n)\Gamma(b_2+n)}\frac{x^n}{n!}
\eeq
\citep[Eq.\,16.2.1\,\&\,16.2.5]{nist_dlmf}, where $(a)_n \equiv a(a-1)(a-2)...(a+n-1)$ for $n\geq 1$, with $(a)_0=1$. This gives
\beq\label{eq: full-aniso-kernel}
    A_\ell^a(\vec r_i\,,\vec r_j) &=& 3(-1)^{l/2}\frac{(2\ell+1)}{k_{a,\mathrm{max}}^3-k_{a,\mathrm{min}}^3} \frac{L_\ell(\hat{\vec x}\cdot \hat{\vec u})}{|\vec r_i-\vec r_j|^3} \left[D_\ell(k_{a,\mathrm{max}}|\vec r_i-\vec r_j|)-D_\ell(k_{a,\mathrm{min}}|\vec r_i-\vec r_j|)\right]\\\nonumber
\eeq
The $D_\ell$ functions may seem cumbersome to compute (though possible with C\texttt{++} packages such as \texttt{arblib}\footnote{\href{http://arblib.org}{arblib.org}}) but we note that the ${}_1\widetilde{F}_2$ term is simply a power series in $-\tfrac{1}{4}(k|\vec r_i-\vec r_j|)^2$ with coefficients dependent only on $\ell$ which may be pre-computed. By Taylor expanding the term in square brackets in Eq.\,\ref{eq: full-aniso-kernel}, one can show that the thin-bin assumption error is $\mathcal{O}\left(\left(\tfrac{\Delta k}{k}\right)^2\right)$, and thus only truly important for $k_a\sim \Delta k$.

An alternative representation (which is used in this paper) may be found by expanding the functions $D_\ell$ purely in terms of trigonometric functions and the well-defined Sine integral $\operatorname{Si}(x) = \int_0^x \sin{t} dt/t$, using the recursion relations of \citet{2017arXiv170306428B} and spherical Bessel function definitions. For the first few even $\ell$ we obtain the results
\beq\label{eq: D_function_simple}
    D_0(ku) &\equiv& u^3\int k^2dk\,j_0(ku) = -\eta\cos\eta+\sin\eta = \eta^2j_1(\eta)\\\nonumber
    D_2(ku) &\equiv& u^3\int k^2dk\,j_2(ku) = \eta\cos\eta-4\sin\eta+3\operatorname{Si}(\eta)\\\nonumber
    D_4(ku) &\equiv& u^3\int k^2dk\,j_4(ku) = \frac{1}{2}\left[\left(\frac{105}{\eta}-2\eta\right)\cos\eta + \left(22-\frac{105}{\eta^2}\right)\sin\eta+15\operatorname{Si}(\eta)\right]
\eeq
defining $\eta\equiv ku\equiv k|\vec r_i-\vec r_j|$. This can be continued to arbitrary high Legendre multipoles $\ell$, involving only polynomials in $\eta$ up to order $\ell$ and the Sine integral. Notably, all multipoles involve the same $\operatorname{Si}(\eta)$ function, thus we need only compute $\sin\eta$, $\cos\eta$ and $\operatorname{Si}(\eta)$ once per pair of particles drawn and $k$-bin. The $\ell=0$ case matches the isotropic $P(k)$ form (Eq.\,\ref{eq: iso-power-pair-count-kernel}) as expected.

\section{Relating Configuration- and Fourier-space Multipoles}\label{appen: power-to-2pcf-multipoles}
The multipoles of a function $A(\vec r)$ can be simply related to those of its Fourier transform $\widetilde{A}(\vec k)$, as described below. We here define multipoles to be measured with respect to a LoS vector $\vec x$ (carried by each field) which is averaged over. This avoids having to make the flat-sky approximation, where we would assume a uniform LoS for the survey. In this paper, the LoS vector is simply the vector joining the observer to the midpoint of a pair of particles. Fourier multipoles are thus defined as 
\beq
    \widetilde{A}_\ell(k) &=& (2\ell+1)\int \frac{d\Omega_k}{4\pi}\int \frac{d^3\vec x}{V}\,L_\ell(\hat{\vec k}\cdot\hat{\vec x})\widetilde{A}(\vec k; \vec x) = \frac{(2\ell+1)}{4\pi V}\int d\Omega_k \int d^3\vec x\,L_\ell(\hat{\vec k}\cdot\hat{\vec x})\int d^3\vec r\,e^{i\vec k\cdot\vec r}A(\vec r; \vec x)\\\nonumber
    &=& \frac{2\ell+1}{4\pi V}\int d\Omega_k\,d^3\vec x\,d^3\vec r\,L_\ell(\hat{\vec k}\cdot\hat{\vec x})\sum_{\ell_1}(2\ell_1+1)i^{\ell_1}j_{\ell_1}(kr)L_{\ell_1}(\hat{\vec k}\cdot\hat{\vec r})\sum_{\ell_2}A_{\ell_2}(r)L_{\ell_2}(\hat{\vec r}\cdot\hat{\vec x})\\\nonumber
    &=& \frac{(4\pi)^2}{(2\ell_2+1)V}\sum_m\sum_{\ell_1m_1}\sum_{\ell_2m_2}i^{\ell_1}\int x^2dx\,r^2dr\,j_{\ell_1}(kr)A_{\ell_2}(r)\int d\Omega_k\,d\Omega_x\,d\Omega_r\,\left(Y^{}_{\ell m}(\hat{\vec k})Y^{*}_{\ell_1m_1}(\hat{\vec k})\right)\left(Y^{}_{\ell_1m_1}(\hat{\vec r})Y^{*}_{\ell_2m_2}(\hat{\vec r}\right)\left(Y^{}_{\ell_2m_2}(\hat{\vec x})Y^{*}_{\ell m}(\hat{\vec x})\right)
\eeq
where we express $\widetilde{A}(\vec k)$ in terms of $A(\vec r)$ in the first line. In the second line we expand the exponential using the plane-wave expansion \citep[Eq.\,16.63]{arfken2013mathematical} and express $A$ in terms of its Legendre moments, before expanding all Legendre polynomials in terms of spherical harmonics in the third line \citep[Eq.\,14.30.9]{nist_dlmf}. Via spherical harmonic orthonormality \citep[Eq.\,14.30.8]{nist_dlmf}, the angular integrals enforce $\ell=\ell_1=\ell_2$ and $m=m_1=m_2$ which yields
\beq\label{eq: fourier-multipoles-from-real}
    \widetilde{A}_\ell(k) &=& \frac{(4\pi)^2}{2\ell+1}\sum_{m}i^\ell\int r^2dr\,A_\ell(r) j_{\ell}(kr)\int \frac{x^2dx}{V} = 4\pi i^{\ell}\int r^2dr\,A_\ell(r)j_\ell(kr).
\eeq
The final equality follows from noting that $\int x^2dx = \int d^3\vec x/(4\pi) = V/(4\pi)$ and that there are $2\ell+1$ possible values of $m$.

\section{Legendre Multipoles of Transformed Power Spectra}\label{appen: transformed-multipole-power}
Here, we derive a useful result relating the Legendre multipoles of some transformed power spectrum, $\mathcal{P}_\ell(k)$ to the 2PCF multipoles $\xi_\ell(r)$. This is a generalization of the results of appendix \ref{appen: power-to-2pcf-multipoles}. We begin by assuming the following form for the local power spectrum at LoS vector $\vec x$;
\beq
    \mathcal{P}(\vec k;\vec x) \equiv\mathcal{F}\left[\xi(\vec r; \vec x)\omega(\vec r; \vec x)\right] = \int d^3\vec r\,e^{i\vec k\cdot\vec r}\xi(\vec r;\vec x)\omega(\vec r;\vec x),
\eeq
where $\xi(\vec r; \vec x)$ is the 2PCF for (midpoint) LoS $\vec x$ and $\omega$ is an arbitrary function. Averaging over the LoS over a volume $V$ (denoted by angle brackets) gives the full power spectrum $\mathcal{P}(\vec k) = \av{\mathcal{P}(\vec k; \vec x)} = \mathcal{F}\left[\xi(\vec r)\omega(\vec r)\right](\vec k)$. The multipoles are defined as
\beq
    \mathcal{P}_\ell(k) &=& (2\ell+1)\int \frac{d\Omega_k}{4\pi}\int \frac{d^3\vec x}{V}\int d^3\vec r\,e^{i\vec k\cdot\vec r}\,\xi(\vec r; \vec x)\omega(\vec r; \vec x)L_\ell(\hat{\vec k}\cdot\hat{\vec x})\\\nonumber
    &=& \frac{(2\ell+1)}{4\pi V}\int d\Omega_k\,d^3\vec x\,d^3\vec r\,\sum_L i^L(2L+1)j_L(kr)L_L(\hat{\vec k}\cdot\hat{\vec r})\times \sum_{\ell_1}\xi_{\ell_1}(r)L_{\ell_1}(\hat{\vec r}\cdot\hat{\vec x})\times \sum_{\ell_2}\omega_{\ell_2}(r)L_{\ell_2}(\hat{\vec r}\cdot\hat{\vec x})\times L_\ell(\hat{\vec k}\cdot\hat{\vec x})  
\eeq
using the plane wave expansion \citep[Eq.\,16.63]{arfken2013mathematical} and decomposing $\xi$ and $\omega$ into spherical harmonic coefficients in the second line. Next, note that
\beq
    \int d\Omega_k L_L(\hat{\vec k}\cdot\hat{\vec r})L_\ell(\hat{\vec k}\cdot\hat{\vec x}) &=& \frac{(4\pi)^2}{(2\ell+1)(2L+1)}\sum_{m,M}\int d\Omega_k Y^{}_{LM}(\hat{\vec k})Y^{*}_{LM}(\hat{\vec r})Y^{*}_{\ell m}(\hat{\vec k})Y^{}_{\ell m}(\hat{\vec x})\\\nonumber
    &=& \frac{(4\pi)^2}{(2\ell+1)(2L+1)}\delta_{\ell L}\sum_m Y_{\ell m}^{*}(\hat{\vec r})Y_{\ell m}^{}(\hat{\vec x}) = \frac{4\pi}{2\ell+1}\delta_{\ell L}L_{\ell}(\hat{\vec r}\cdot\hat{\vec x})
\eeq
via Legendre polynomial decomposition and orthonormality \citep[Eq.\,14.30.8\,\&\,14.30.9]{nist_dlmf}, which gives
\beq
    \mathcal{P}_\ell(k) &=& i^\ell\frac{2\ell+1}{V}\int d^3\vec x\,d^3\vec r\,j_\ell(kr)L_\ell(\hat{\vec r}\cdot\hat{\vec x})\sum_{\ell_1}\xi_{\ell_1}(r)\sum_{\ell_2}\omega_{\ell_2}(r)L_{\ell_1}(\hat{\vec r}\cdot\hat{\vec x})L_{\ell_2}(\hat{\vec r}\cdot\hat{\vec x})\\\nonumber
    &=& i^\ell\frac{2\ell+1}{V}\int d^3\vec x\,d^3\vec r\,j_\ell(kr)L_\ell(\hat{\vec r}\cdot\hat{\vec x})\sum_{\ell_1,\ell_2}\xi_{\ell_1}(r)\omega_{\ell_2}(r)\sum_{L}(2L+1)\left(\begin{array}{ccc}
         \ell_1 & \ell_2 & L \\
          0 & 0 & 0
    \end{array}\right)^2L_L(\hat{\vec r}\cdot\hat{\vec x}).
\eeq
In the second line we have used \citealt[Eq.\,34.3.19]{nist_dlmf} to replace the product of two Legendre polynomials with a sum over a single function, weighted by Wigner $3j$ symbols. We further note that 
\beq
    \int d\Omega_r d\Omega_x\,L_\ell(\hat{\vec r}\cdot\hat{\vec x})L_L(\hat{\vec r}\cdot\hat{\vec x}) &=& \frac{(4\pi)^2}{(2\ell+1)(2L+1)}\int d\Omega_r d\Omega_x\,\sum_{m,M}Y^{}_{LM}(\hat{\vec r})Y^{*}_{LM}(\hat{\vec x})Y^{*}_{\ell m}(\hat{\vec r})Y^{}_{\ell m}(\hat{\vec x})\\\nonumber
    &=& \frac{(4\pi)^2}{(2\ell+1)(2L+1)}\sum_{m,M}\delta_{\ell L}\delta_{mM} = \frac{(4\pi)^2}{2\ell+1}
\eeq
noting that there are $2\ell+1$ possible values of $m$. This gives
\beq
    \mathcal{P}_\ell(k) &=& (4\pi)^2i^\ell\frac{2\ell+1}{V}\int x^2dx\,r^2dr\,\,j_\ell(kr)\sum_{\ell_1,\ell_2}\xi_{\ell_1}(r)\omega_{\ell_2}(r)\left(\begin{array}{ccc}
         \ell_1 & \ell_2 & \ell \\
          0 & 0 & 0
    \end{array}\right)^2\\\nonumber
    &=& 4\pi i^\ell(2\ell+1)\int r^2dr\,\,j_\ell(kr)\sum_{\ell_1,\ell_2}\xi_{\ell_1}(r)\omega_{\ell_2}(r)\left(\begin{array}{ccc}
         \ell_1 & \ell_2 & \ell \\
          0 & 0 & 0
    \end{array}\right)^2\\\nonumber
\eeq
evaluating the integral over $x$ as $V/(4\pi)$ as before. In the limit of isotropic $\omega$ (where $\omega_\ell(r) = 0\,$ for all $l>0$), this reduces to the standard expression
\beq
    \mathcal{P}_\ell(k) \rightarrow 4\pi i^\ell \int r^2dr\,j_\ell(kr)\xi_\ell(r)\omega(r)
\eeq
(cf.\,Eq.\,\ref{eq: fourier-multipoles-from-real}) noting that, for $\ell_2 = 0$, the $3j$ symbol is equal to $\delta_{\ell \ell_1}/\sqrt{2\ell+1}$.

\section{Derivation of the Bispectrum Triple-Count Kernel}\label{appen: bispectrum-deriv}
The bispectrum kernel function is derived analogously to that of the anisotropic power spectrum (appendix \ref{appen: aniso-power-deriv}). Starting from Eq.\,\ref{eq: bispectrum_kernel_init}, we can apply the spherical harmonic addition theorem and separate integrals to yield
\beq\label{eq: bispectrum_kernel}
    A^{ab}_\ell(\vec r_1,\vec r_2;\vec r_3) &\equiv& \frac{1}{v_av_b}\int d^3\vec k_1\,d^3\vec k_2 \Theta^a(|\vec k_1|)\Theta^b(|\vec k_2|)L_\ell(\hat{\vec k}_1\cdot\hat{\vec k}_2)e^{i\vec k_1\cdot(\vec r_1-\vec r_3)}\,e^{i\vec k_2\cdot(\vec r_2-\vec r_3)}\\\nonumber
    &=& \frac{4\pi}{2\ell+1}\sum_{m=-\ell}^\ell\int\frac{d^3\vec k_1}{v_a}\,\Theta^a(|\vec k_1|)\,e^{i\vec k_1\cdot(\vec r_1-\vec r_3)}Y^{}_{\ell m}(\hat{\vec k}_1)\times \int \frac{d^3\vec k_2}{v_b}\,\Theta^b(|\vec k_2|)\,e^{i\vec k_2\cdot(\vec r_2-\vec r_3)}Y_{\ell m}^*(\hat{\vec k}_2).
\eeq
Each Fourier-space integral may be solved as
\beq
    \int \frac{d^3\vec k}{v_a}\,\Theta^a(|\vec k|)e^{i\vec k\cdot\vec x} &=& \frac{4\pi}{v_a}\sum_{\ell'=0}^\infty\sum_{m'=-\ell'}^{\ell'}i^{\ell'}Y^{}_{\ell'm'}(\hat{\vec x})\int k^2dk\,j_{\ell'}(kx)\,\Theta^a(k)\int d\Omega_k Y^{*}_{\ell'm'}(\hat{\vec k})\,Y^{}_{\ell'm'}(\hat{\vec k})\\\nonumber
    &=& \frac{4\pi}{v_a} i^\ell Y_{\ell m}(\hat{\vec x})\int k^2dk\,j_\ell(kx)\,\Theta^a(k)\\\nonumber
    &=& \frac{4\pi i^\ell}{x^3v_a} Y_{\ell m}(\hat{\vec x})\left[D_\ell(k_{a,\mathrm{max}}x)-D_\ell(k_{a,\mathrm{min}}x)\right]
\eeq
via spherical harmonic completeness, using the $D_\ell$ definitions of Eq.\,\ref{eq: D_function_general} or Eq.\,\ref{eq: D_function_simple}. Inserting into Eq.\,\ref{eq: bispectrum_kernel} gives
\beq
    A_\ell^{ab}(\vec r_1,\vec r_2;\vec r_3) &=& \frac{(4\pi)^3}{2\ell+1}\frac{(-1)^\ell}{(xy)^3v_av_b}\sum_{m=-\ell}^\ell Y^{}_{\ell m}(\hat{\vec x})Y^{*}_{\ell m}(\hat{\vec y})\left[D_\ell(k_{a,\mathrm{max}}x)-D_\ell(k_{a,\mathrm{min}}x)\right]\left[D_\ell(k_{b,\mathrm{max}}y)-D_\ell(k_{b,\mathrm{min}}y)\right]\\\nonumber
    &=& \frac{(4\pi)^2(-1)^\ell}{(xy)^3v_av_b} P_\ell(\hat{\vec x}\cdot\hat{\vec y})\left[D_\ell(k_{a,\mathrm{max}}x)-D_\ell(k_{a,\mathrm{min}}x)\right]\left[D_\ell(k_{b,\mathrm{max}}y)-D_\ell(k_{b,\mathrm{min}}y)\right]\\\nonumber
    &\approx& (-1)^\ell P_\ell(\hat{\vec x}\cdot\hat{\vec y}) j_\ell(k_ax)j_\ell(k_by)
\eeq
for $\vec x = \vec r_1 - \vec r_3$ and $\vec y = \vec r_2 - \vec r_3$, assuming the $n$-th bin to lie in the range $[k_{n,\mathrm{min}},k_{n,\mathrm{max}}]$. Although this appears to be asymmetric in $\vec r_1$, $\vec r_2$ and $\vec r_3$, a symmetric expression can be wrought simply by averaging over all possible permutations of the three positions in the above expression (which all give identical results). The final line is derived assuming the thin-bin limit, where $\Delta k\ll k_a$. This gives an analytic form for the kernel function in a particular bin depending only on the lengths of the triangle side-lengths and $\ell$.

\bsp	
\label{lastpage}
\end{document}